\documentclass[12pt]{article}
\usepackage[dvips]{epsfig}
\usepackage{amsmath,amssymb,amsfonts,enumerate,bbm}
\usepackage{caption}
\usepackage{url}
\usepackage{fullpage}

\usepackage{ifthen}
\usepackage{graphicx}

\usepackage{subfigure}

\usepackage{color,soul}

\long\def\commabs #1\commabsend{}
\long\def\commful #1\commfulend{#1}
\long\def\comment #1\commentend{}
\newtheorem{theorem}{Theorem}[section]
\newtheorem{lemma}[theorem]{Lemma}
\newtheorem{observation}[theorem]{Observation}
\newtheorem{corollary}[theorem]{Corollary}

\newtheorem{claim}[theorem]{Claim}
\newtheorem{fact}[theorem]{Fact}
\newtheorem{Definition}[theorem]{Definition}

\newcommand{\REAL}{\mathbb R}
\def\deg{\mbox{\tt deg}}
\def\depth{\mbox{\tt depth}}
\def\LCA{\mbox{\tt LCA}}

\def\Cost{\mbox{\tt Cost}}
\def\Value{\mbox{\tt Val}}
\newcommand{\dist}{\mbox{\rm dist}}

\def\CLUSTER{\mbox{\sf Cluster}}

\def\inline#1:{\par\vskip 7pt\noindent{\bf #1:}\hskip 10pt}
\def\Proof{\par\noindent{\bf Proof:~}}
\def\blackslug{\hbox{\hskip 1pt \vrule width 4pt height 8pt
    depth 1.5pt \hskip 1pt}}
\def\QED{\quad\blackslug\lower 8.5pt\null\par}

\def\cI{{\cal I}}
\newcommand{\cS}[0]{{\cal S}}
\newcommand{\Sensitive}[0]{{\mathcal{S}}}
\newcommand{\New}[0]{\mbox{\tt New}}
\def\LAB{\mbox{\tt Label}}

\def\LastE{\mbox{\tt LastE}}
\def\FirstEN{\mbox{\tt FirstNewE}}

\def\RepOne{P^1}
\def\RepTwo{P^2}

\def\Depth{\mbox{\tt Depth}}

\def\FTBFS{\mbox{\tt FT-BFS}}
\def\FTSPANNERBFS{\mbox{\tt FT-ABFS}}

\def\Weight{\omega}

\def\Pairs{\Gamma}
\def\LEdgeDelta{E}

\def\CurrSpanner{\mathcal{F}}

\newcommand{\FBFS}[0]{\mbox{\tt ConsFT-BFS}}
\newcommand{\APPBFS}[0]{\mbox{\tt ConsFT-AddSpan}}
\newcommand{\MSPAANER}[0]{\mbox{\tt ConsSpan}}

\begin{document}



\title{Fault Tolerant Approximate BFS Structures}

\author{
Merav Parter
\thanks{Department of Computer Science and Applied Mathematics.
The Weizmann Institute of Science, Rehovot, Israel.
E-mail: {\tt \{merav.parter,david.peleg\}@ weizmann.ac.il}.
Supported in part by the Israel Science Foundation (grant 894/09),
the United States-Israel Binational Science Foundation
(grant 2008348),
the I-CORE program of the Israel PBC and ISF (grant 4/11),
the Israel Ministry of Science and Technology
(infrastructures grant), and the Citi Foundation.}
\thanks{Recipient of the Google European Fellowship in distributed computing;
research is supported in part by this Fellowship.}
\and
David Peleg $^*$
\footnote{An extended abstract of this paper has appeared in the proceedings of the 2014 ACM-SIAM Symposium on Discrete Algorithms.}
}

\maketitle


\begin{abstract}
A {\em fault-tolerant} structure for a network is required to continue
functioning following the failure of some of the network's edges or vertices.
This paper addresses the problem of designing a {\em fault-tolerant}
$(\alpha, \beta)$ approximate BFS structure
(or {\em \FTSPANNERBFS\ structure} for short), namely,
a subgraph $H$ of the network $G$
such that subsequent to the failure of some subset $F$ of edges or vertices,
the surviving part
of $H$ still contains
an \emph{approximate} BFS spanning tree for (the surviving part of) $G$,
satisfying
$\dist(s,v,H\setminus F) \leq \alpha \cdot \dist(s,v,G\setminus F)+\beta$
for every $v \in V$.

We first consider {\em multiplicative} $(\alpha,0)$ \FTSPANNERBFS\ structures
resilient to a failure of a single edge
and present an algorithm that given an $n$-vertex unweighted undirected graph
$G$ and a source $s$ constructs a $(3,0)$ \FTSPANNERBFS\ structure rooted at $s$
with at most $4n$ edges (improving by an $O(\log n)$ factor on the near-tight
result of \cite{BS10} for the special case of edge failures). Assuming at most $f$ edge failures, for constant integer $f>1$,
we prove that there exists a (poly-time constructible)
$(3(f+1), (f+1) \log n)$ \FTSPANNERBFS\ structure with $O(f n)$ edges.

We then consider {\em additive} $(1,\beta)$ \FTSPANNERBFS\ structures.
In contrast to the linear size of $(\alpha,0)$ \FTSPANNERBFS\ structures,
we show that for every $\beta \in [1, O(\log n)]$ there exists an $n$-vertex
graph $G$ with a source $s$ for which any $(1,\beta)$ \FTSPANNERBFS\ structure
rooted at $s$ has $\Omega(n^{1+\epsilon(\beta)})$ edges, for some function
$\epsilon(\beta) \in (0,1)$. In particular, $(1,3)$ \FTSPANNERBFS\ structures admit
a lower bound of $\Omega(n^{5/4})$ edges.
These lower bounds demonstrate
an interesting dichotomy between multiplicative and additive spanners;
whereas $(\alpha,0)$ \FTSPANNERBFS\ structures of size $O(n)$ exist
(for $\alpha\geq 3$), their additive counterparts, $(1,\beta)$ \FTSPANNERBFS\ structures, are of super-linear size.
Our lower bounds are complemented by an upper bound, showing that there exists
a poly-time algorithm that for every $n$-vertex unweighted undirected graph $G$ and source $s$ constructs a $(1,4)$ \FTSPANNERBFS\ structure
rooted at $s$ with at most $O(n^{4/3})$ edges.
%
\end{abstract}


\section{Introduction}

\paragraph{Background and Motivation.}
Fault-tolerant subgraphs are subgraphs designed to maintain a certain desirable
property in the presence of edge or vertex failures. This paper focuses on
the property of containing a BFS tree with respect to some source $s$.
A {\em fault tolerant BFS structure} (or {\em \FTBFS\ structure}) resistant to a single
edge failure is a subgraph $H \subseteq G$ satisfying that
$\dist(s, v, H \setminus \{e\})=\dist(s, v, G \setminus \{e\})$
for every vertex $v \in V$ and  edge $e \in E$.

To motivate our interest in such structures, consider a situation where it
is required to lease a subnetwork of a given network, which will provide
short routes from a source $s$ to all other vertices. In a failure-free
environment one can simply lease a BFS tree rooted at $s$. However, if
links might disconnect, then one must prepare by leasing a larger set of links,
and specifically an \FTBFS\ structure. Moreover, taking costs into account,
this example also motivates our interest in constructing
{\em sparse} \FTBFS\ structure.

This question has recently been studied by us in \cite{PPFTBFS13}.
Formally, a spanning graph $H \subseteq G$ is an
{\em $f$ edge (resp., vertex) fault-tolerant BFS (\FTBFS) structure} for $G$
with respect to the source $s \in V$ iff for every $v \in V$ and every set $F \subseteq E(G)$ (resp., $F\subseteq V$), $|F| \leq f$,  it holds that
$\dist(s, v, H \setminus F)=\dist(s, v, G \setminus F).$
It is shown in \cite{PPFTBFS13} that for every graph $G$ and source $s$
there exists a (poly-time constructible) 1-edge  \FTBFS\ structure $H$ with
$O(n^{3/2})$ edges. This result is complemented by a matching lower bound
showing that for every sufficiently large integer $n$, there exist
an $n$-vertex graph $G$ and a source $s \in V$, for which every 1-edge \FTBFS\
structure is of size $\Omega(n^{3/2})$.
Hence {\em exact} \FTBFS\ structures may be rather expensive.

This last observation motivates the approach of resorting to \emph{approximate}
distances, in order to allow the design of a sparse subgraph with properties
resembling those of an \FTBFS\ structure.
The current paper aims at exploring
this approach, focusing mainly on subgraphs that contain {\em approximate} BFS
structures and are resistant to a single edge failure.
Formally, given an unweighted undirected $n$-vertex graph $G=(V,E)$
and a source $s \in V$, the subgraph $H \subseteq G$ is
an $f$-edge (resp., vertex) $(\alpha, \beta)$ \FTSPANNERBFS\ structure
with respect to $s$ if for every vertex $v \in V$ and every set
$F \subseteq E(G)$ (resp., $F\subseteq V$), $|F| \leq f$,
$$\dist(s, v, H \setminus F) \leq
\alpha \cdot \dist(s, v, G \setminus F)+\beta~.$$
(An $(\alpha, \beta)$ \FTSPANNERBFS\ structure is a
fault-tolerant BFS (\FTBFS) structure if $\alpha=1$ and $\beta=0$.)
We show that this relaxed requirement
allows structures that are sparser than their exact counterparts.

Approximate BFS tree structures can also be compared against a different
type of structures, namely, fault-tolerant {\em spanners}.
Given an $n$-vertex graph $G=(V,E)$, the subgraph $H \subseteq G$
is an {\em $f$-edge fault-tolerant $(\alpha, \beta)$ spanner} of $G$
if for every two vertices $v,w \in V$ and every set $F \subseteq E(G)$,
$|F| \leq f$, we have
$\dist(v,w, H \setminus F) \leq \alpha \cdot \dist(v, w, G \setminus F)+\beta$.
Observe that the union of $(\alpha, \beta)$ \FTSPANNERBFS\ structures with respect to every source $s \in V$ forms an (all-pairs) fault tolerant $(\alpha, \beta)$ spanner for $G$. In fact,
\FTSPANNERBFS\ structures can be viewed as \emph{single source spanners}.
Algorithms for constructing an $f$-vertex fault tolerant $(2k-1)$ spanner
of size $O(f^2 k^{f+1} \cdot n^{1+1/k}\log^{1-1/k}n)$ and an $f$-edge fault tolerant
$2k-1$ spanner of size $O(f\cdot n^{1+1/k})$ for a given $n$-vertex graph $G$ were presented in \cite{CLPR09-span}.
A randomized construction attaining an improved tradeoff for vertex fault-tolerant spanners was then presented in \cite{DK11}.

For the case of $f$ edge failures for constant $f\geq 1$, we show
(in Sec. \ref{subsec:spanner_ub}) that
there exists a poly-time algorithm that for every $n$-vertex graph constructs
a $(3(f+1), (f+1) \log n)$ \FTSPANNERBFS\ structure $H$ with
$O(f n)$ edges overcoming up to $f$ edge faults.
For the special case of a single edge failure ($f=1$),
we get a somewhat stronger result, namely, that
for every $n$-vertex graph $G$ and source $s$,
there is a (poly-time constructible) $(3,0)$ \FTSPANNERBFS\ structure with
at most $4n$ edges, thus improving on the near-tight construction of \cite{BS10}
by a $O(\log n)$ factor for the special case of $\alpha=3$ and edge failures.

This result is to be contrasted with two different structures:
the (single-source) fault tolerant exact \FTBFS\ structure of \cite{PPFTBFS13}, and
the (all-pairs) fault tolerant  $(3,0)$ spanner of \cite{CLPR09-span},
which both contain $\Theta(n^{3/2})$ edges.
This implies that using \FTSPANNERBFS\ structures is more efficient than using
fault-tolerant spanners
even if it is necessary to handle not a single source $s$ but
a set $S \subseteq V$ of sources where $|S|=\Omega(n^{\epsilon})$
for $\epsilon < 1/2$; a collection of approximate
$(\alpha, \beta)$ \FTSPANNERBFS\ structures rooted at each of the sources $s \in S$
will still be cheaper than a fault-tolerant spanner.

\emph{Additive} fault tolerant $(1, \beta)$ spanners were recently defined
and studied by \cite{FTAdd12}, establishing the following general result.
For a given $n$-vertex graph $G$, let $H_1$ be an ordinary additive
$(1, \beta)$ spanner for $G$ and $H_2$ be a fault tolerant $(\alpha,0)$ spanner
for $G$ resilient against up to $f$ edge faults. Then $H=H_1 \cup H_2$ is a
$(1, \beta(f))$ additive fault tolerant spanner for $G$
(for up to $f$ edge faults) for $\beta(f)=O(f(\alpha+\beta))$.
In particular, fixing the number of $H$ edges to be $O(n^{4/3})$ and the number
of faults to $f=1$ yields an additive stretch of $38$
(See \cite{FTAdd12}; Cor. 1).

When considering \FTBFS\ structures with an {\em additive} stretch, namely,
$(1, \beta)$ \FTSPANNERBFS\ structures, the improvement is less dramatic
compared to the size of the single-source exact or the all-pairs approximate
variants. In Sec. \ref{sec:lowerbound_add}, we show that for every additive
stretch $\beta \in [1, \log n]$, there exists a superlinear lower bound
on the size of the \FTSPANNERBFS\ structure with additive stretch $\beta$,
i.e., $\Omega(n^{1+\epsilon(\beta)})$.
These new lower bound constructions are independent of the correctness
of Erd\"{o}s conjecture.
Importantly, our results reveal an interesting dichotomy between
{\em multiplicative} $(\alpha, 0)$ \FTSPANNERBFS\
and {\em additive} $(1, \beta)$ \FTSPANNERBFS\ structures:
whereas every graph $G$ contains a (poly-time constructible)
$(3,0)$ \FTSPANNERBFS\ structure rooted at $s \in V$ of size $\Theta(n)$,
there exist an $n$-vertex graph $G$ and a source $s \in V$ for which
every $(1,\beta)$ \FTSPANNERBFS\ structure contains a \emph{super-linear}
number of edges. For example, for additive stretch $\beta=3$,
we have a lower bound construction with $\Omega(n^{5/4})$ edges.

On the positive side, in Sec. \ref{sec:upperbound_add} we complement
those results
by presenting a (rather involved) poly-time algorithm that
for any given $n$-vertex graph $G$ and source $s$ constructs a
$(1,4)$ \FTSPANNERBFS\ structure with $O(n^{4/3})$ edges
(hence improving the additive stretch of the (all-pairs) fault tolerant
additive spanner with $O(n^{4/3})$ edges of \cite{FTAdd12} from 38 to 4).
This algorithm is inspired by the (non-fault-tolerant) additive spanner
constructions of \cite{BSADD10,ChechikAdd13,CGK13}.
The main technical contribution of our algorithm is in adapting the path-buying
strategy used therein to failure-prone settings.
So far, the correctness and size analysis of this strategy
heavily relied on having a fault-free input graph $G$.
We show that by a proper construction of the sourcewise replacement paths,
the path-buying technique can be extended to support the construction
even in the presence of failures.

\paragraph{Related work.}
\FTBFS\ structures are closely related to the notion of
\emph{replacement paths}.
For a source $s$, a target vertex $v$ and an edge $e\in G$,
a \emph{replacement path} is the shortest $s-v$ path $P_{s,v,e}$ that does not
go through $e$.
An \FTBFS\ structure is composed of a collection consisting of a replacement
path $P_{s,v,e}$ for every target $v \in V$ and edge $e \in E$.
Analogously, the notion of \FTSPANNERBFS\ structures is closely related to the
problem of constructing \emph{approximate replacement paths}
\cite{BS06,CLPR09-do,Bern10},
and in particular to its {\em single source} variant studied in \cite{BS10}.
That problem requires to compute a collection $\mathcal{P}_{s}$ consisting of
an approximate $s-t$ replacement path $P_{s,t,e}$ for every $t \in V$
and every failed edge $e$ that appears on the $s-t$ shortest-path in $G$,
such that $|P_{s,t,e}|\leq \alpha \cdot \dist(s,t, G \setminus \{e\})$.
In the resulting {\em fault tolerant distance oracle}, in response to a query
$(s,t,F)$ consisting of an $s-t$ pair and a set $F$ of failed edges or vertices
(or both), the oracle $\cS$ must return the distance between $s$ and $t$
in $G'=G\setminus F$.
Such a structure is sometimes called an {\em $F$-sensitivity distance oracle}.
The focus is on both fast preprocessing time, fast query time and low space.
An  approximate {\em single source fault tolerant distance oracle} has been
first studied at \cite{BS10}, which proposed an $O(n \log n /\epsilon^3)$ space
data structure that can report a $(1+\epsilon)$ approximate shortest path
for any $\epsilon >0$. An additional by-product of the data structure of
\cite{BS10} is the construction of an $(1+\epsilon,0)$ \FTSPANNERBFS\ structure
with $O(n/\epsilon^3+n \log n)$ edges. Setting $\epsilon=2$, this yields
a $(3,0)$ \FTSPANNERBFS\ structure with $O(n \log n)$ edges. Hence our
$(3,0)$ \FTSPANNERBFS\ structure construction with at most $3n$ edges
improves that construction by a factor of $O(\log n)$ for the case of single edge failure (the construction of \cite{BS10} supports the case of vertex failures as well).

It is important to note that the literature on
approximate replacement paths (cf. \cite{BS06,Bern10}) mainly focuses on \emph{time-efficient} computation of the these paths, as well as their efficient maintenance within distance oracles.
In contrast, the main concern in the current paper is with optimizing
the \emph{size} of the resulting fault tolerant structure that contains the collection of approximate replacement paths.

Moreover, this paper considers both multiplicative and additive stretch,
whereas the long line of existing approximate distance oracles concerned mostly
multiplicative (and not additive) stretch,
with the exception of \cite{PatrascuByondT}.
To illustrate the dichotomy between the additive and multiplicative setting,
consider the issue of lower bounds for additive \FTSPANNERBFS\ structures.
In the all-pairs fault-free setting,
the best known lower bound for additive spanners is based on the
{\em girth conjecture} of Erd\"{o}s \cite{Erdos:book}, stating that there exist
$n$-vertex graphs with $\Omega(n^{1+1/k})$ edges and girth
(minimum cycle length) $2k+2$ for any integer $k$.
Removing any edge in such a graph increases the distance between its endpoints
from $1$ to $2k+1$, hence any $(1, \beta)$ spanner with
$\beta \leq 2k - 1$ must have $\Omega(n^{1+1/k})$ edges.
This conjecture is settled only for $k = 1, 2, 3,5$ (see \cite{Wenger91}).
In  \cite{Woodruff06}, Woodruff presented a lower bound for additive spanners
matching the girth conjecture bounds but independent of the correctness
of the conjecture. More precisely, he showed the existence of graphs
for which any spanner of size $O(k^{-1}n^{1+1/k})$ has an additive stretch
of at least $2k-1$, hence establishing a lower bound of
$\Omega(k^{-1}n^{1+1/k})$ on the size of additive spanners.
The lower bound constructions of \cite{Woodruff06} are formed by appropriately
gluing together certain complete bipartite graphs.
Since for every $n$-vertex graph $G$ there exists a (poly-time constructible)
multiplicative spanner of size $O(n^{1+1/k})$ and stretch $\alpha=2k-1$,
so far there has been no theoretical indication for a dichotomy between
additive and multiplicative spanners.
Such a dichotomy is believed to exist mainly based on the existing gap
between the current upper and lower bounds for additive spanners
(the current additive lower bounds match the
lower bounds of its multiplicative counterpart).
Perhaps surprisingly, such a dichotomy is revealed by our current results,
obtained for the most basic setting of fault tolerance, namely,
single edge fault and sourcewise distances.

Upper bounds for constant stretch (non-fault-tolerant) additive spanners
are currently known for but a few stretch values.
A $(1,2)$ spanner with $O(n^{3/2})$ edges is presented in \cite{Aing99},
a $(1,6)$ spanner with $O(n^{4/3})$ edges is presented in \cite{BSADD10}, and
a $(1,4)$ spanner with $O(n^{7/5})$ edges is presented in \cite{ChechikAdd13}.
The latter two constructions use the {\em path-buying} strategy, which is
adopted in our additive upper bound in Sec. \ref{sec:upperbound_add}.
Recently, the path-buying strategy was employed in the context of pairwise spanners, where the objective is to construct a subgraph
$H \subseteq G$ that satisfies the bounded additive stretch requirement only for a \emph{subset} of pairs \cite{CGK13}.
\paragraph{Preliminaries.}
Given a graph $G=(V,E)$ and a source $s$, let $T_0(s) \subseteq G$ be
a shortest paths (or BFS) tree rooted at $s$.
Let $\pi(x, y)$ be the (unique) $x-y$ path in $T_0(s)$.
Let $E(v,G)=\{(u,v) \in E(G)\}$ be the set of edges incident to $v$
in the graph $G$ and let $\deg(v,G)=|E(v,G)|$ denote the degree of vertex $v$
in $G$. When the graph $G$ is clear from the context,
we may omit it and simply write $\deg(v)$.
Let $\depth(s, v) = \dist(s,v,G)$ denote the {\em depth} of $v$
in the BFS tree $T_0(s)$. When the source $s$ is clear from the context,
we may omit it and simply write $\depth(v)$ and $T_0$.
Let $\Depth(s)=\max_{u \in V} \{ \depth(s, u) \}$ be the {\em depth}
of $T_0(s)$.
For a subgraph $G'=(V', E') \subseteq G$
(where $V' \subseteq V$ and $E' \subseteq E$)
and a pair of vertices $u,v \in V$, let $\dist(u,v, G')$ denote the
shortest-path distance in edges between $u$ and $v$ in $G'$.
For a path $P=[u_1, \ldots, u_k]$, let $\LastE(P)$ denote the last edge of $P$,
let $|P|$ denote the length of $P$ and let $P[u_i, u_j]$ be the subpath of $P$
from $u_i$ to $u_j$. For paths $P_1$ and $P_2$
where the last vertex of $P_1$ equals the first vertex of $P_2$,
let $P_1 \circ P_2$ denote
the path obtained by concatenating $P_2$ to $P_1$.
Assuming an edge weight function $W: E(G)\to \REAL^{+}$, let $SP(s, u_i, G, W)$
be the set of $s-u_i$ shortest-paths in $G$ according to the edge weights
of $W$. (When the graph is unweighted, the parameter $W$ is omitted.)
Throughout, the edges of these paths are considered to be directed
away from the source $s$. Given an $s-t$ path $P$ and an edge
$e=(u,v) \in P$, let $\dist(s, e, P)$ be the distance (in edges) between $s$
and $e$ on $P$.
In addition, for an edge $e=(u,v)\in T_0(s)$, define
$\dist(s,e)=i$ if $\depth(u)=i-1$ and $\depth(v)=i$.
For a subset $V' \subseteq V$, let $G(V')$ be the induced subgraph on $V'$.
Let $\LCA(V')$ be the least common ancestor of all the vertices in $V'$.
A \emph{replacement path} $P^*_{i,j}$ is a shortest path in
$SP(s, u_i, G \setminus \{e_j\})$. Note that if $e_{j} \notin \pi(s,u_i)$
then the replacement path $P^*_{i,j}$ is simply the shortest-path $\pi(s,u_i)$.
%
\comment
Let
$$\mathcal{T}(s, G, \alpha, \beta) ~=~
\{H  \subseteq G \mid H \mbox{~is an~}
(\alpha, \beta) \FTSPANNERBFS\ structure \mbox{~with respect to~} s\}.$$
Define $\Cost^*(s, G, \alpha, \beta)=\min\{|E(T')| \mid T' \in \mathcal{T}(s, G, \alpha, \beta)$ as the minimum over all $(\alpha, \beta)$ \FTSPANNERBFS\
structure $T' \in \mathcal{T}(s, G, \alpha, \beta)$ for $s$.
\commentend

Fix the source $s \in V$. For an edge $e=(u,v)\in T_0$,
denote the set of vertices in $T_0(v)$, the subtree of $T_0$ rooted at $v$, by
$$\Sensitive(e) ~=~ V(T_0[v]).$$
Note that the vertices of $\Sensitive(e)$ are precisely those sensitive to the
failure of the edge $e$, i.e.,
the vertices $w$ having $e$ on $\pi(s, w)$, their $s-w$ path in $T_0(s)$,
hence also $\Sensitive(e) ~=~ \{w \mid e \in \pi(s, w)\}.$

\section{Multiplicative FT-ABFS Structures}
\label{subsec:spanner_ub}

This section describes algorithms for constructing \FTSPANNERBFS\ structures
for unweighted undirected graphs.

\subsection{Single edge fault}
We establish the following.

\begin{theorem}
\label{thm:multbfs}
There exists a poly-time algorithm that for every $n$-vertex graph $G$
and source $s$ constructs a 1-edge $(3,0)$ \FTSPANNERBFS\ structure
with $O(n)$ edges.
\end{theorem}

We begin by providing an informal intuition for the algorithm.
The construction is based on starting from a BFS tree $T_0$ and adding edges to it
until it satisfies the requirement.
Specifically, the algorithm constructs a collection of replacement paths,
$P^*_{i,j} \in SP(s, u_i, G \setminus \{e_j\})$ for every vertex-edge pair
$(i,j)$, satisfying that $e_j \in \pi(s, u_i)$. From each such path $P^*_{i,j}$,
only the first new edge $e=(x,y) \in P^*_{i,j} \setminus T_0$, i.e.,
the new edge closest to $s$, is taken into the spanner.

The correctness analysis shows that the construction of the $P^*_{i,j}$ collection guarantees that the endpoint $y$ of the first new edge $e$, as well as the path endpoint $u_i$, are both sensitive to the failure of the edge $e_j=(x',y')$, namely, $u_i,y \in T_0(y')$. Therefore, the $y-u_i$ path $\pi(y, u_i) \subseteq T_0(y') \subseteq T_0 \setminus \{e_j\}$ in the BFS tree $T_0$ is free of the failing edge $e_j$ and hence provides a safe alternative path to the segment $P^*_{i,j}[y,u_i]$, which possibly might contain many edges that are missing in $T_0$. Then, by employing the triangle inequality, we also get that the alternative $s-u_i$ path $P^*_{i,j}[s,y] \circ \pi(y, u_i)$ is not much longer than the optimal counterpart $P^*_{i,j}$.
Perhaps the more surprising part is the \emph{size} analysis, where we show that every vertex $y$ can appear as the endpoint of the \emph{first} new edge of at most three replacement paths.
This should be contrasted with \cite{PPFTBFS13}, where it is shown that
a vertex can be the endpoint of the \emph{last} new edge of $\Omega(\sqrt{n})$
replacement paths. Hence, taking the last edge of every replacement path
results in an exact \FTBFS\ structure with $\Theta(n^{3/2})$ edges,
while taking the first new edge of every replacement path results in
an approximate $(3,0)$ \FTSPANNERBFS\ structure with at most $3n$ edges.

We now provide some intuition explaining why a vertex  might be the endpoint
of the \emph{first} new edge of at most \emph{three} replacement paths.
Let $P^*_{i_1,j_1}, P^*_{i_2,j_2}, \ldots, P^*_{i_k,j_k}$ be the replacement paths in which the endpoint of the first new edge is $y$, i.e., $\widehat{e}_{\ell}=\LastE(P^*_{i_\ell,j_\ell}[s,y])\notin T_0$ and $P^*_{i_\ell,j_\ell}[s,y]\setminus \{\widehat{e}_{\ell}\}\subseteq T_0$ for every $\ell \in  \{1, \ldots, k\}$.
We then show that upon a proper construction of the replacement paths,
the fact that $\widehat{e}_{\ell} \notin T_0$ implies that
$e_{j_1},\ldots, e_{j_k} \subseteq \pi(s, y)$
(otherwise the original shortest path $\pi(s, y)$ could be used in
$P^*_{i_{\ell},j_{\ell}}$ instead of the segment $P^*_{i_\ell,j_\ell}[s,y]$)
and letting $e_{j_\ell}$ be sorted in increasing distance from $s$
on $\pi(s, y)$, it also holds that the truncated replacement paths are
monotonically decreasing, i.e.,
$|P^*_{i_1,j_1}[s, y]|>\ldots >|P^*_{i_k,j_k}[s, y]|$. Let $x_\ell$ be such that
$\widehat{e}_{\ell}=(x_{\ell}, y)$.
Since $P^*_{i_\ell,j_\ell}[s, x_{\ell}]=\pi(s, x_{\ell})$, we have that $y$ is
connected by an edge to $k$ vertices $x_1,\ldots, x_k$ of distinct distances
from $s$,
$\dist(s, x_1,G)>\ldots>\dist(s, x_k,G)$, hence by the triangle inequality,
necessarily $k \leq 3$.

\begin{figure}[htbp]
\begin{center}
\includegraphics[scale=0.27]{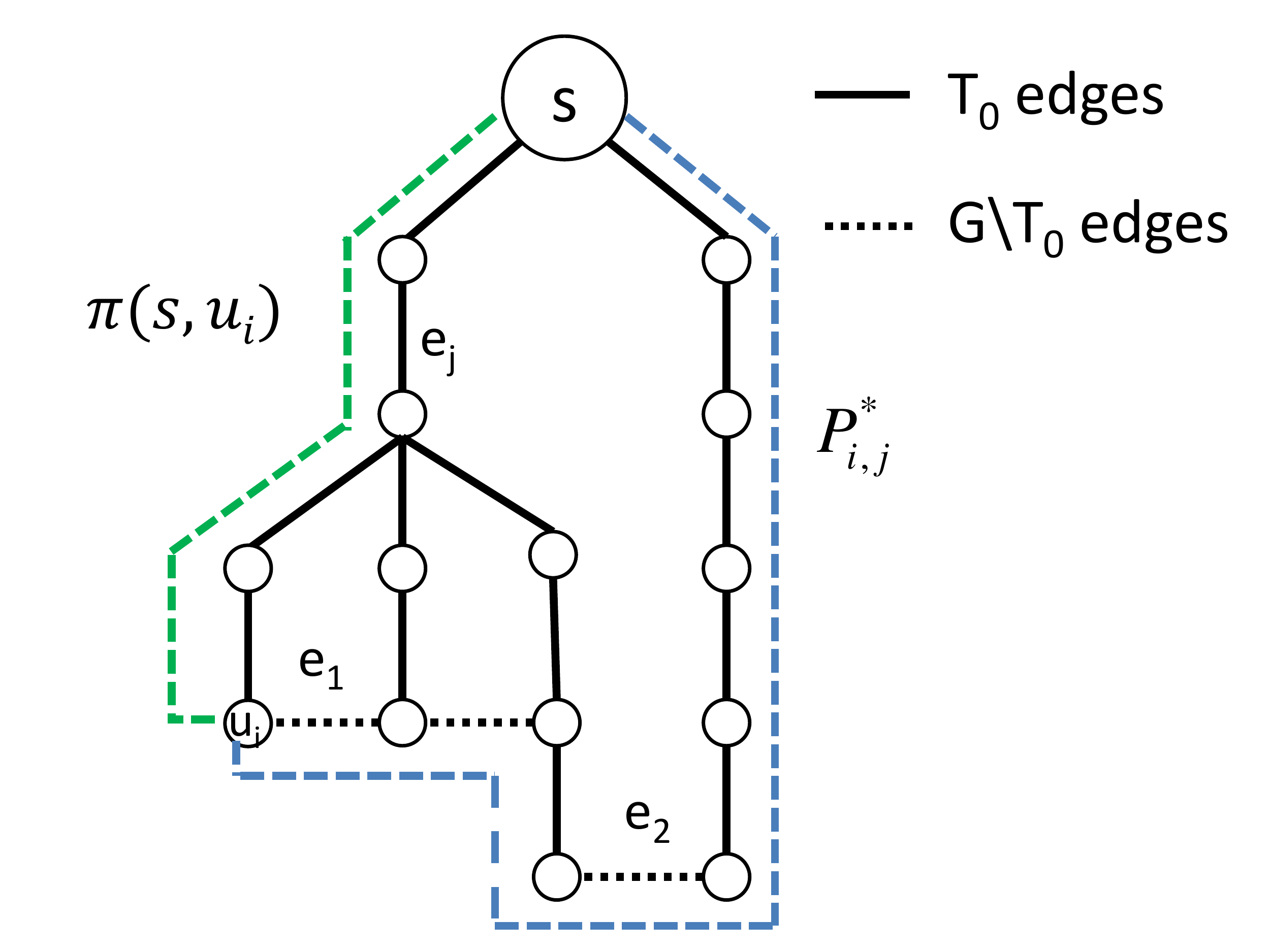}
\caption{The path $P^*_{i,j}$ protects $u_i$ against the failure of the edge $e_j$. Here $e_1=\LastE(P^*_{i,j})$ and $e_2=\FirstEN(P^*_{i,j})$. The algorithm will add $e_2$ to $E^*_{i}$ and subsequently to the output structure $H$ because $e_1$ is also a new edge, so $P^*_{i,j}$ is new-ending. \label{fig:figprotecting}}
\end{center}
\end{figure}

\paragraph{Algorithm Description.}
We next formally describe the algorithm.
For a path $P$ in the constructed structure, let $\New(P)=E(P) \setminus E(T_0)$ be the set of new edges in $P$, namely, edges that were added to $T_0$ during the construction process. Let $\FirstEN(P^*_{i,j})$ be the first (from $s$) new edge on $P^*_{i,j}$ that is not in $T_0$. See Fig. \ref{fig:figprotecting} for an illustration of these definitions.

The $(3,0)$ \FTSPANNERBFS\ structure $H$ is constructed by adding to $T_0$ only new edges that appear as the first edges on \emph{some} of the replacement paths. The algorithm operates as follows.
\par Fix an ordering on the edges $E=\{e_1, \ldots, e_{m}\}$ and on the vertices $u_1, \ldots, u_n$. In round $i$, the vertex $u_i$ is considered. The round consists of $\depth(u_i)$ iterations. In iteration $j$, consider $e_j \in \pi(s, u_i)$ and define the path protecting $u_i$ against the failure of $e_j$ to be $P^*_{i,j} \in SP(s, u_i, G \setminus \{e_j\}, W_{i})$, where $W_{i}: E \to \mathbb{R}_{>0}$ is a weight assignment for the edges of $E$ defined by
\begin{equation}
\label{eq:well_multis_onef}
W_{i}(e_\ell) =
\begin{cases}
\Weight_\ell, & \text{if $e_\ell \in E(T_0) \setminus E(\pi(s,u_i))$},\\
\Weight_\ell+\epsilon_2, & \text{if $e_\ell \in \pi(s,u_i)$,}\\
\Weight_\ell+\epsilon_1, & \text{otherwise,}
\end{cases}
\end{equation}
where $\Weight_\ell = n^6 \cdot 2^{m+1} + 2^\ell$,
$\epsilon_1=n^3 \cdot 2^{m+1}$ and $\epsilon_2= 2^{m+1}$.
Call a replacement path $P^*_{i,j}$  \emph{new-ending} if its last edge is new,
namely, $\LastE(P^*_{i,j}) \notin E(T_0)$. For every vertex $u_i \in V$, define
$$E^*_i ~=~ \{\FirstEN(P^*_{i,j}) \mid ~ e_j \in \pi(s, u_i) \mbox{~and~}
P^*_{i,j} \mbox{~is new-ending}\}.$$
Let $E^*_{new}=\bigcup_{u_i \in V}E^*_i$ ~~~ and ~~~ $H=T_0 \cup E^*_{new}$.

\paragraph{Correctness.}
We now prove the correctness of the algorithm
and establish Thm. \ref{thm:multbfs},
by showing that taking into the constructed $H$ merely the first new edge from
each new-ending replacement path $P^*_{i,j}$ is sufficient in order to guarantee
the existence of an approximate $s-u_i$ replacement path in the surviving
structure $H \setminus \{e_j\}$, for every $u_i$ and $e_j \in \pi(s, u_i)$.

%
%
Let us start by explaining the specific weight assignment chosen.
The role of $W_{i}(e_\ell)$ is to enforce a unique $s-u_i$ shortest-path
in $SP(s, u_i, G \setminus \{e_j\})$. This is important for both the
correctness and the size analysis of the $(3,0)$ \FTSPANNERBFS\ structure.
(In other words, carelessly taking the first edge of an arbitrary replacement
path might result in a dense subgraph which is also not a
$(3,0)$ \FTSPANNERBFS\ structure.)
The weight assignment achieves this as follows.
For every $u_i \in V$, let $\Cost_{i}(P)=\sum_{e \in P} W_{i}(e)$ be the weighted
cost of $P$, i.e., the sum of its edge weights.
Then given paths $P_1, P_2$, the weight assignment $W_{i}$ has the following
properties, implying that $\Cost_{i}(P)$ can in some sense be viewed as based
on $|P|$,  $|\New(P)|$ and $|E(P) \cap \pi(s,u_i)|$ lexicographically.
\begin{fact}
\label{obs:weigh:prop}
For every two paths $P_1, P_2$ and $u_i \in V$,
\begin{description}
\label{desc:weight}
\item{(a)}\label{desc:q1}
If $|P_1| < |P_2|$, then $\Cost_{i}(P_1)<\Cost_{i}(P_2)$.
\item{(b)} \label{desc:q2}
If $|P_1|=|P_2|$ and $|\New(P_1)| < |\New(P_2)|$,
then $\Cost_{i}(P_1)< \Cost_{i}(P_2)$.
\item{(c)}\label{desc:q3}
If $|P_1|=|P_2|$, $|\New(P_1)|= |\New(P_2)|$ and
$|E(P_1) \cap E(\pi(s,u_i))| < |E(P_2) \cap E(\pi(s,u_i))|$,
then $\Cost_{i}(P_1)< \Cost_{i}(P_2)$.
\item{(d)}\label{desc:q4}
If $|P_1|=|P_2|$, $|\New(P_1)|= |\New(P_2)|$ and
$|E(P_1) \cap E(\pi(s,u_i))| = |E(P_2) \cap E(\pi(s,u_i))|$,
then $\Cost_{i}(P_1)< \Cost_{i}(P_2)$ iff
$\sum_{e_k \in P_1}\Weight_k<\sum_{e_k \in P_2}\Weight_k$.
\end{description}
\end{fact}
Conversely we also have the following.
\begin{fact}
If $\Cost_{i}(P_1)<\Cost_{i}(P_2)$, then necessarily one of the following four
conditions holds:
\begin{description}
\item{(a)} $|P_1|<|P_2|$,
\item{(b)} $|\New(P_1)| < |\New(P_2)|$,
\item{(c)} $|E(P_1) \cap E(\pi(s,u_i))| < |E(P_2) \cap E(\pi(s,u_i))|$,
\item{(d)} $\sum_{e_k \in P_1}\Weight_k<\sum_{e_k \in P_2}\Weight_k$.
\end{description}
\end{fact}

The following key observation is used repeatedly in what follows.
\begin{observation}
\label{cl:sp_edge}
For every replacement path $P^{*}_{i,j}$ and every new edge $e=(x, y) \in \New(P^{*}_{i,j})$ on it,
$y \in \Sensitive(e_{j})$ (or $e_{j} \in\pi(s,y)$).
\end{observation}
\Proof
Assume, towards contradiction, that $e=(x, y) \in \New(P^{*}_{i,j})$ and yet $e_{j} \notin\pi(s,y)$. Since $e=(x, y) \in \New(P^{*}_{i,j})$ is a new edge, it holds that $P^{*}_{i,j}[s,y] \neq \pi(s,y)$.
Consider an alternative $s-u_i$ replacement path $P'=\pi(s,y) \circ P^{*}_{i,j}[y, u_{i}]$. Since $|\pi(s,y)| \leq |P^{*}_{i,j}[s,y]|$ but $|\New( \pi(s,y))|=1$ and $|\New(P^{*}_{i,j}[s,y])|\geq 1$ (since $T_0$ contains only one $s-y$ path, $\pi(s,y)$). It follows by Obs. \ref{desc:weight}, that
$\Cost_{i}(\pi(s,y))< \Cost_{i}(P^{*}_{i,j}[s,y])$ and thus also
$\Cost_{i}(P')< \Cost_{i}(P^{*}_{i,j})$, in contradiction to the fact that
$P^{*}_{i,j} \in SP(s, u_i, G \setminus \{e_j\}, W_i)$.
The observation follows.
\QED
We next provide the following claim, showing that if a BFS edge $e_j \in \pi(s,u_i)$ fails, then adding the first new edge $e'$ of a new-ending replacement path $P^*_{i,j}$ to the BFS tree $T_0$,  recovers its connectivity. In order words, $T'=T_0 \setminus \{e_j\} \cup \{e'\}$ is a connected spanning tree of the graph $G$. Moreover, we show that the $s-u_i$ path in $T'$ has low stretch compared to the $s-u_i$ shortest-path in $G \setminus \{e_j\}$.

\begin{figure}[htbp]
\begin{center}
\includegraphics[scale=0.3]{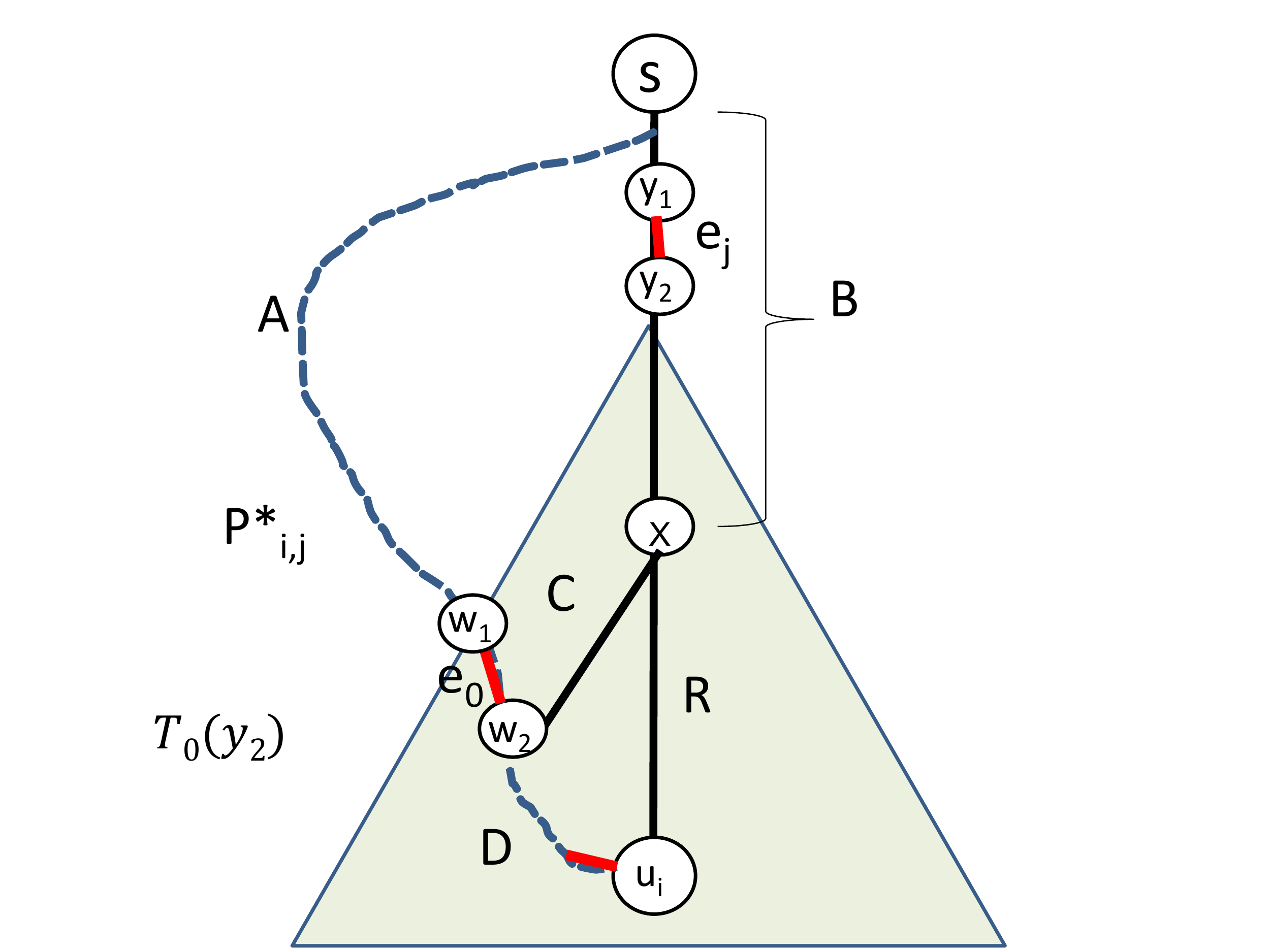}
\caption{Illustration of the approximate replacement path.
Solid lines represent tree edges.
\label{fig:figsp}}
\end{center}
\end{figure}

\begin{lemma}
\label{cl:first_edge}
Let $P^*_{i,j}$ be a new-ending replacement path, let $e_0=(w_1, w_2)=\FirstEN(P^*_{i,j})$ and let
$T'=T_0 \cup \{e_0\}$. Then $\dist(s, u_i, T' \setminus \{e_j\}) \leq 3 \cdot |P^*_{i,j}|$.
\end{lemma}
\Proof
Let $e_j=(y_1,y_2)$; see Fig. \ref{fig:figsp} for illustration.
Note that by Obs. \ref{cl:sp_edge}, $w_2$ is in $\Sensitive(e_j)$, hence
both $u_i, w_2\in T_0(y_2)$, the subtree of $T_0$ rooted at $y_2$.
Therefore the path between $w_2$ and $u_i$, $\pi(w_2,u_i)$, does not use $e_j$,
hence it exists in $T_0 \setminus \{e_j\}$.

Let $x=\LCA(u_i,w_2)$ be the least common ancestor of $u_i$ and $w_2$ in
$T_0$. Then $x \in \pi(y_2, u_i)$.
Let $A=P^*_{i,j}[s, w_2], B = \pi(s, x), C=\pi(x,w_2), D=P^*_{i,j}[w_2, u_i]$ and
$R=\pi(x, u_i)$.
Consider an alternative $s-u_i$ replacement path $P'=A \circ C \circ R$
that uses the $w_2-u_i$ path in $T_0$ (see Fig. \ref{fig:figsp}).
Note that since $e_0=(w_1,w_2)$ is the first new edge on $P^*_{i,j}$,
it follows that $P' \subseteq T'=T_0 \cup \{e_0\}$.
Since $P'$ is a replacement path for $u_i$ in $T'$, it remains to bound
its length. Note that $|P'|=|A|+|C|+|R|$ and $|P^*_{i,j}|=|A|+|D|$.
First, since $A$ is a shortest $s-w_2$ path in $G \setminus \{e_j\}$ but
$B \circ C$ is an $s-w_2$ shortest-path in $G$,
it follows that $|B|+|C| \leq |A|$. Next, consider the two $x-u_i$ paths
$R$ and $C \circ D$. Since $R$ is a shortest $x-u_i$ path,
it follows that $|R| \leq |C|+|D|$.
Therefore $|P'| \leq |A|+2|C|+|D| \leq 3|A|+|D| \leq 3|P^*_{i,j}|$.
%
The lemma follows.
\QED

\begin{lemma}
\label{lemma:correctness}
$H$ is a $(3,0)$ \FTSPANNERBFS\ structure.
\end{lemma}
\Proof
Assume, towards contradiction, that $H$ is not a $(3,0)$ \FTSPANNERBFS\ structure. Let $BP=\{(i,j) \mid
\dist(s,u_i, H\setminus \{e_j\}) >  3 |P^*_{i,j}|\}$
be the set of ``bad pairs," namely, vertex-edge pairs $(i,j)$ for which
the length of the replacement $s-u_i$ path in $H \setminus \{e_j\}$ is greater than $3 \cdot \dist(s,u_i, G\setminus \{e_j\})$.
(By the contradictory assumption, $BP\ne \emptyset$.)
For each bad pair $(i,j) \in BP$, define
$BE(i,j)=P^{*}_{i,j} \setminus E(H)$ to be the set of ``bad edges,''
namely, the set of $P^{*}_{i,j}$ edges that are missing in $H$.
By definition, $BE(i,j) \neq \emptyset$ for every bad pair $(i,j) \in BP$.
Let $d(i,j)=\max_{e \in BE(i,j)}\{\dist(s,e,P^{*}_{i,j})\}$ be the maximal depth of a missing edge in $BE(i,j)$, and let $DM(i,j)$ denote that ``deepest
missing edge'', i.e., the edge $e$ on $P^{*}_{i,j}$ satisfying
$d(i,j) = \dist(s,e,P^{*}_{i,j})$.
Finally, let $(i',j') \in BP$ be the pair that minimizes $d(i,j)$,
and let $e_1=(v_{\ell_1}, u_{i_1}) \in BE(i',j')$ be the deepest missing edge on $P^{*}_{i',j'}$, namely, $e_1=DM(i',j')$. Note that $e_1$ is the {\em shallowest} ``deepest missing edge'' over all bad pairs $(i,j) \in BP$.
By Obs. \ref{cl:sp_edge}, $e_{j'} \in \pi(s, u_{i_1})$.

Consider the $s-u_{i_1}$ replacement path $P^*_{i_1,j'}$.
Note that there are two $s-u_{i_1}$ replacement paths,
$P_1=P^{*}_{i',j'}[s,u_{i_1}]$ and $P_2=P^*_{i_1,j'} \in G \setminus \{e_{j'}\}$,
and by their optimality
we have that $|P_1|=|P_2|$
(these paths might - but do not have to - be the same).

We distinguish between two cases:
(C1) $\LastE(P_2) \notin T_0$ and  (C2) $\LastE(P_2) \in T_0$.
Begin with case (C1).
By construction, $\FirstEN(P_2) \in E^*_{i_1}$, so $\FirstEN(P_2) \in H$.
By Lemma \ref{cl:first_edge}, there exists an $s-u_{i_1}$ replacement path $P'$ in $G \setminus \{e_{j'}\}$ such that $|P'| \leq 3 \cdot |P_2|$ and $P' \subseteq \left(T_0 \cup \{\FirstEN(P_2)\}\right) \subseteq H$.
Consider the $s-u_{i'}$ replacement path
$$P'' ~=~ P' \circ P^{*}_{i',j'}[u_{i_1},u_{i'}].$$
Note that since $e_1$ is the deepest missing edge in $P^*_{i',j'}$, it holds that $P^*_{i',j'}[u_{i_1},u_{i'}] \subseteq H$ and by the previous argument $P' \subseteq H\setminus \{e_{j'}\}$, concluding that $P''$ is an $s-u_{i'}$ replacement path in $ H\setminus \{e_{j'}\}$ Moreover, its length is bounded by
\begin{eqnarray*}
|P''| &=& |P'|+|P^{*}_{i',j'}[u_{i_1},u_{i'}]|
\leq 3|P_1|+|P^{*}_{i',j'}[u_{i_1},u_{i'}]| ~\leq~ 3|P^*_{i_1,j'}|,
\end{eqnarray*}
contradicting the fact that $(i',j') \in BP$ is a bad pair.

Now consider case (C2) where $\LastE(P_2) \in T_0$.
We show that in this case $(i_1, j') \notin BP$. Assume, towards contradiction, that $(i_1, j')$ is a bad pair. This implies that $P_2 \nsubseteq H$. Since $|P_1|=|P_2|$, $\LastE(P_1)\notin T_0$ but $\LastE(P_2)\in T_0$, it holds that the ``deepest missing edge'' $e''$ in $P_2$ is such that $\dist(s,e'',P_2)<\dist(s,e_1,P_1)$ (or $d(i_1,j')<d(i',j')$) in contradiction to the selection of $(i',j')$. Hence, we conclude that $(i_1, j') \notin BP$, which guarantees the existence of an $s-u_{i_1}$ replacement path $P' \in G \setminus \{e_{j'}\}$ such that $|P'|\leq |P_2|$.
Finally, the $s-u_{i'}$ path $P''=P' \circ P^*_{i',j'}[u_{i_1},u_{i'}]$ exists in $H\setminus \{e_{j'}\}$ and $|P''|\leq 3|P_1|+|P^*_{i',j'}[u_{i_1},u_{i'}]|\leq 3|P^*_{i',j'}|$,
in contradiction to the fact that $(i',j') \in BP$. The lemma follows.
\QED

\begin{figure}[htbp]
\begin{center}
\includegraphics[scale=0.4]{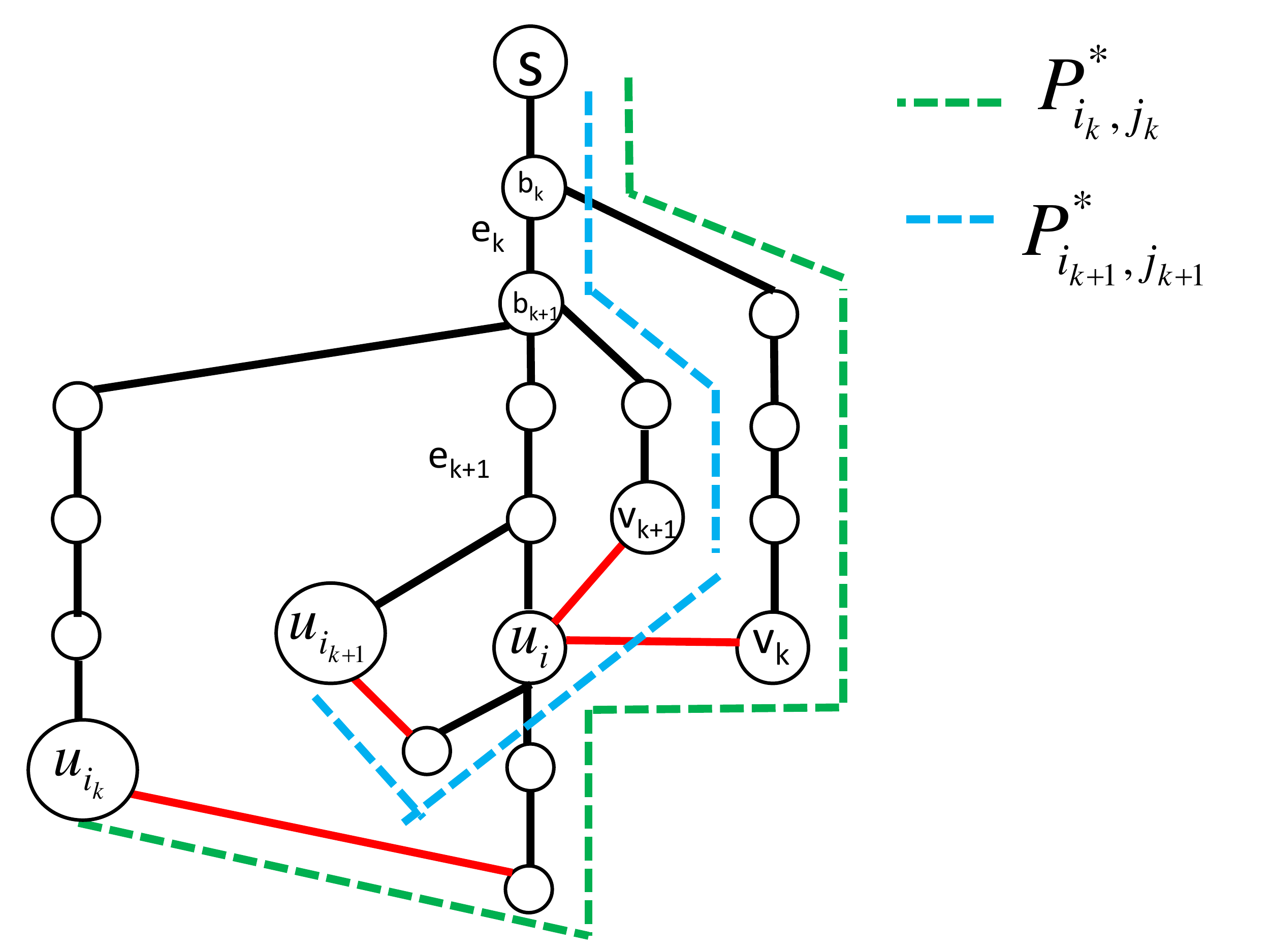}
\caption{Illustration of the replacement paths in which edge of $u_i$ are
the first new edges. New edges are represented in red.
\label{fig:sizebound}}
\end{center}
\end{figure}

\paragraph{Size analysis.}
\begin{lemma}
\label{lem:bfsspanner_sizebound}
$|E(H) \setminus E(T_0)|=|E^*_{new}| \leq 3n$.
\end{lemma}
\Proof
We show that every vertex $u_i$ can have at most 3 of its incident edges in $E^*_{new}$.
Assume, towards contradiction, that there exists some $u_i$ with (at least) 4 edges in $E^*_{new}$, $e_k=(v_k,u_i)$  for $k \in \{1,\ldots, 4\}$, that appear as first new edges in the replacement paths $P^*_{i_1, j_1}, P^*_{i_2, j_2}, P^*_{i_3, j_3}, P^*_{i_4, j_4}$ respectively.
By Obs. \ref{cl:sp_edge}, it holds that the 4 failed edges $e_{j_k}\in \pi(s, u_i)$, $k \in \{1,\ldots, 4\}$ appear on $\pi(s, u_i)$ and by definition, $e_{j_k}\in \pi(s, u_{i_k})$, for every $k \in \{1,\ldots, 4\}$.
Without loss of generality, assume that $\dist(s, e_{j_k}, \pi(s, u_i)) \leq
\dist(s, e_{j_{k+1}}, \pi(s, u_i))$ for every $k \in \{1,2,3\}$, namely, that the edges $e_{j_1}, e_{j_2},e_{j_3}, e_{j_4}$ occur on $\pi(s,u_i)$ in that order.
For illustration see Fig. \ref{fig:sizebound}.

\begin{figure}[htbp]
\begin{center}
\includegraphics[scale=0.4]{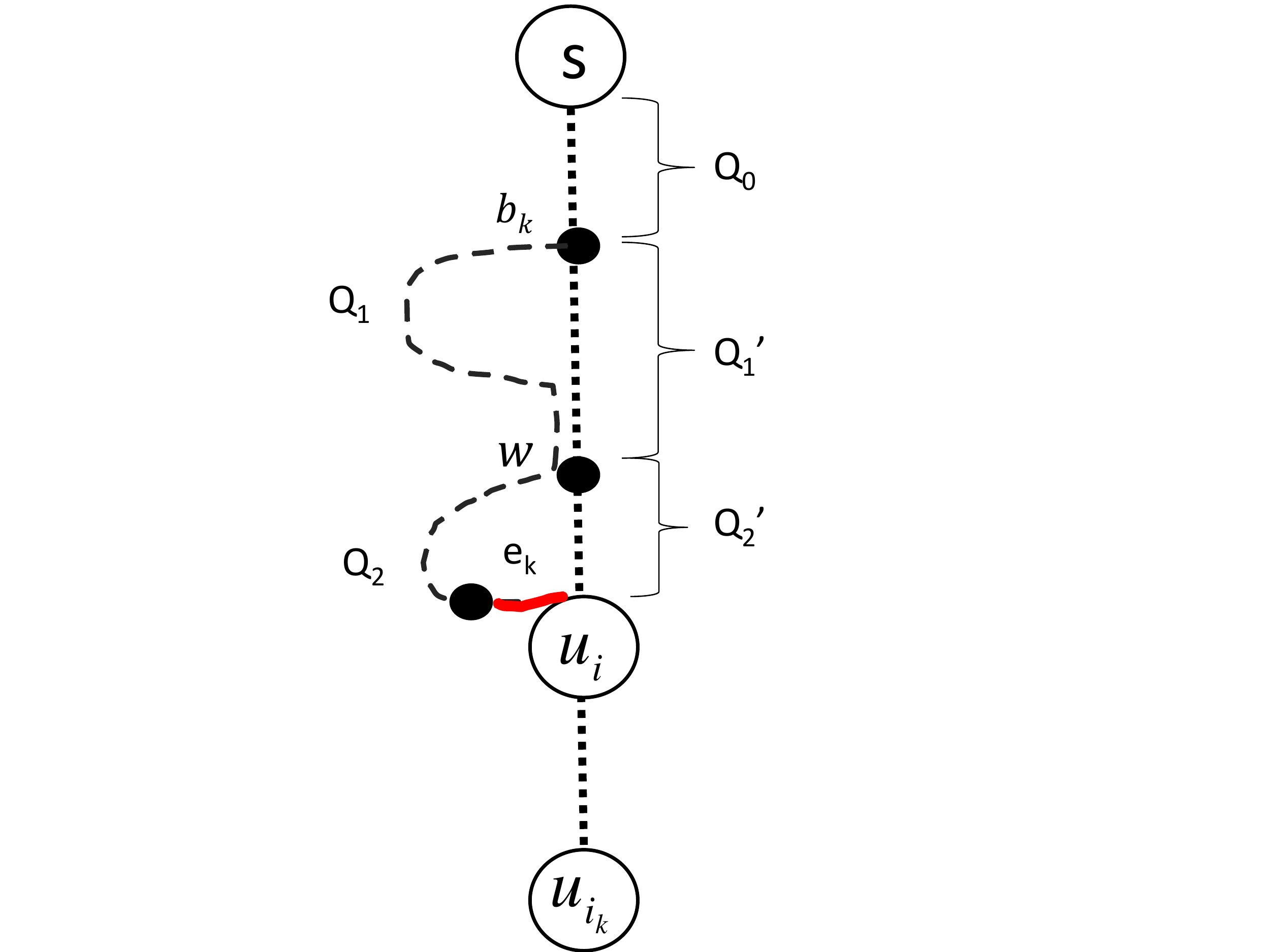}
\caption{The dotted straight line is $\pi(s,u_i)$ and the dashed line
depicts $Q_{Suff}[k]=P_k[b_k, u_i]$. The divergence point is unique.
The red $e_k$ is the new edge of $u_i$. The faulty edge $e_{j_k}$ can occur
in either $Q'_1$ or $Q'_2$.
\label{fig:figunique}}
\end{center}
\end{figure}

Consider the 4 truncated $s-u_i$ paths $P_k=P^*_{i_k, j_k}[s, u_i]$ for $k =\{1, \ldots, 4\}$. Note that since $\LastE(P_k)=e_k$ is the only new edge in $P_k$, i.e., $P_k \setminus \{\LastE(P_k)\}$ has no new edges, or,
\begin{equation}
\label{eq:newedgelast}
P_k \setminus \{\LastE(P_k)\} \subseteq T_0 \setminus \{e_{j_k}\}.
\end{equation}
Let $b_k$ be the first divergence point of $P_k$ from $\pi(s, u_i)$, namely, the last vertex on that $\pi(s, u_i)$ for which $\pi(s, b_k)=P_k[s,b_k]$.
Let $Q_{Pref}[k]=\pi[s, b_k]=P_k[s, b_k]$ (where the equality is by the definition of $b_k$) be the maximal common prefix of the paths $P_k$ and $\pi(s, u_{i_k})$, for $k \in \{1, \ldots, 4\}$. When $k$ is clear from the context, we may omit it and simply write $Q_{Pref}$. Let $Q_{Suff}[k]=P_k[b_k,u_i]$.
We now show that $b_k$ is the \emph{only} divergence point of $P_k$ and $\pi(s, u_i)$, or in other words, the paths meet again only at $u_i$. Formally, we show the following.
\begin{claim}
\label{cl:uniqe_div}
$\left( V(Q_{Suff}[k]) \cap V(\pi(s,u_i)) \right)\setminus \{b_k, u_i\}=\emptyset$ for $k \in \{1, \ldots, 4\}$.
\end{claim}
\Proof
Assume, towards contradiction, that the paths intersect again at some vertex
$$w \in \left( V(Q_{Suff}[k]) \cap V(\pi(s,u_i))\right) \setminus \{b_k, u_i\}.$$
Recall that $u_{i_k}$ was considered in round $i_k$ and let $j_k$ be the iteration in this round in which the edge $j_{k}$ was considered.
Let $Q_0=Q_{Pref}[k], Q_1=P_{k}[b_k, w], Q'_1=\pi(b_k,w), Q_2=P_{k}[w,u_i], Q'_2=\pi(w,u_i)$. For illustration, see Fig. \ref{fig:figunique}.
We distinguish between two cases concerning the faulty edge $e_{j_k}$: (C1) $e_{j_k} \in Q'_1$ or (C2) $e_{j_k} \in Q'_2$.

In case (C1), $Q'_2 \subseteq G \setminus \{e_{j_k}\}$, and as
it is part of the BFS tree, it holds that $|Q'_2|\leq |Q_2|$. Since $Q'_2$ is free of new edges but $\LastE(Q_2)=e_k$ is new, it holds that $\Cost_{i_k}(Q'_2)<\Cost_{i_k}(Q_2)$, in contradiction to the fact that $P^*_{i_k,j_k} \in SP(s, u_{i_k}, G \setminus \{e_{j_k}\}, W_{i_k})$.

In case (C2), $Q'_1 \subseteq G \setminus \{e_{j_k}\}$, and as it is part of the BFS tree, it holds that $|Q'_1|\leq |Q_1|$. Since $T_0$ contains a single $b_k-w$ path corresponding to $Q'_1$, it must hold that $Q_1 \nsubseteq T_0$ ($Q_1$ has at least one new edge) and therefore  $\Cost_{i_k}(Q'_1)<\Cost_{i_k}(Q_1)$, in contradiction again to the fact that $P^*_{i_k,j_k} \in SP(s, u_{i_k}, G \setminus \{e_{j_k}\}, W_{i_k})$.
\QED

%
It follows from Cl. \ref{cl:uniqe_div}, that $E(P_k) \cap E(\pi(s, u_i))=Q_{Pref}[k]$.
We now focus on the edge-set intersections
$$\cI_{k,\ell} ~=~ E(P_k) \cap E(\pi(s, u_{i_{\ell}}))$$
and establish the following auxiliary claim, showing that the same holds also
for the complete path $\pi(s, u_{i_\ell})$, for the $\ell$ values needed later.
\begin{claim}
\label{cl:pathinter}
$\cI_{k,\ell} = Q_{Pref}[k]$ \\
(a) for every $\ell \in \{k, \ldots, 4\}$, and\\
(b) for every $\ell <k$ such that $\depth(b_k) \leq \depth(b_{\ell})$.
\end{claim}
\Proof
Recall that $P_k=Q_{Pref}[k] \circ Q_{Suff}[k]$ and let $e_{j_k}=(x_k, y_k)$ and
$e_{j_\ell}=(x_\ell, y_\ell)$. We prove parts (a) and (b) in two steps.
We first show that $Q_{Pref}[k] \subseteq \pi(s, u_{i_{\ell}})$ and then show that
$Q_{Suff}[k]$ and $\pi(s, u_{i_{\ell}})$ are edge disjoint for $\ell$ satisfying
(a) or (b).
We begin by showing that $Q_{Pref}[k]=\pi(s, b_k) \subseteq \pi(s, u_{i_\ell})$.
Let $\ell \in \{k, \ldots, 4\}$. Since $e_{j_\ell} \in \pi(s, u_{i_\ell})$, by the ordering of the edges $e_{j_k}$, it holds that also $e_{j_k}=(x_k,y_k) \in \pi(s, u_{i_\ell})$. Since $P_k \subseteq G \setminus \{e_{j_k}\}$, the divergence point $b_k$ of $P_k$ and $\pi(s, u_i)$ occurred above $y_k$, hence $\pi(s, b_k) \subseteq \pi(s, u_{i_\ell})$. Next, let $\ell <k$ be such that $\depth(b_k) \leq \depth(b_{\ell})$. By part (a), $\pi(s, b_\ell) \subseteq \pi(s, u_{i_\ell})$. Since the divergence point $b_k$ occurred not after $b_{\ell}$ (and both $b_k$ and $b_\ell$ are in $\pi(s, u_i)$), it holds that also $\pi(s, b_k) \subseteq \pi(s, u_{i_\ell})$.
\par Next, we consider $Q_{Suff}[k]$ and show that it is edge disjoint from $\pi(s, u_{i_\ell})$ for $\ell \in \{k, \ldots, 4\}$. By the above argumentation, $\pi(s, u_{i_\ell})=\pi(s, y_k) \circ \pi(y_k, u_{i_\ell})$.  By Cl. \ref{cl:uniqe_div}, the paths $Q_{Suff}[k]$ and $\pi(s, u_i)$ are edge disjoint and hence the two paths $\pi(s, y_k)$ and $Q_{Suff}[k]$ are edge disjoint. It remains to show that $\pi(y_k, u_{i_\ell})$ and $Q_{Suff}[k]$ are edge disjoint. Since by Eq. (\ref{eq:newedgelast}) the path $P'=P_k \setminus \{\LastE(P_k)\}$ exists in $T_0 \setminus \{e_{j_k}\}$,
it holds that $V(P')\cap \Sensitive(e_{j_k})=\emptyset.$
However, $V(\pi(y_k, u_{i_{\ell}})) \subseteq \Sensitive(e_{j_k})$.
Hence $P'$ does not intersect with $\pi(s, u_{i_\ell})$. Finally, let
$\ell <k$ be as in (b), i.e., such that $\depth(b_k) \leq \depth(b_{\ell})$.
By Cl. \ref{cl:uniqe_div}, $Q_{Suff}[k]$ and $\pi(s, u_i)$ are edge disjoint and hence $\pi(s, y_{\ell}) \subseteq \pi(s, u_i)$ and $Q_{Suff}[k]$ are edge disjoint. It remains to show that $\pi(y_{\ell}, u_{i_{\ell}})$ and $Q_{Suff}[k]$ are edge disjoint. Since $b_k$ diverged from $\pi(s, u_i)$ not after $b_{\ell}$, it holds that $b_k$ is above $y_{\ell}$ hence $P' \subseteq T_0 \setminus \{e_{j_\ell}\}$.
Therefore $V(P')\cap \Sensitive(e_{j_\ell})=\emptyset$, but $V(\pi(y_{\ell}, u_{i_{\ell}})) \subseteq \Sensitive(e_{j_\ell})$, hence $\pi(y_{\ell}, u_{i_{\ell}})$ and $Q_{Suff}[k]$ are edge disjoint as required.
The claim follows.
\QED

\begin{figure}[htbp]
\begin{center}
\includegraphics[width=3in]{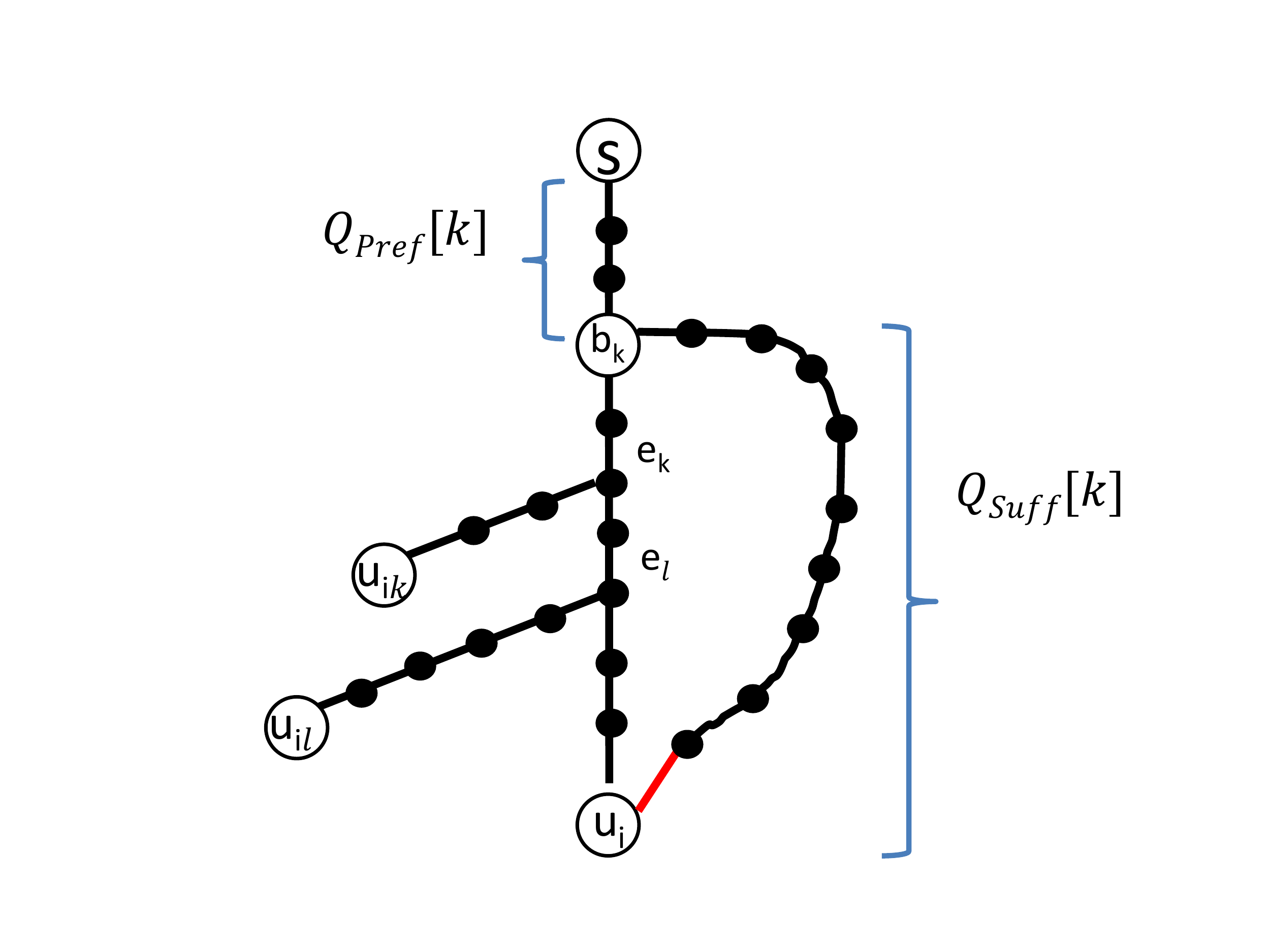}
~~
\includegraphics[width=3in]{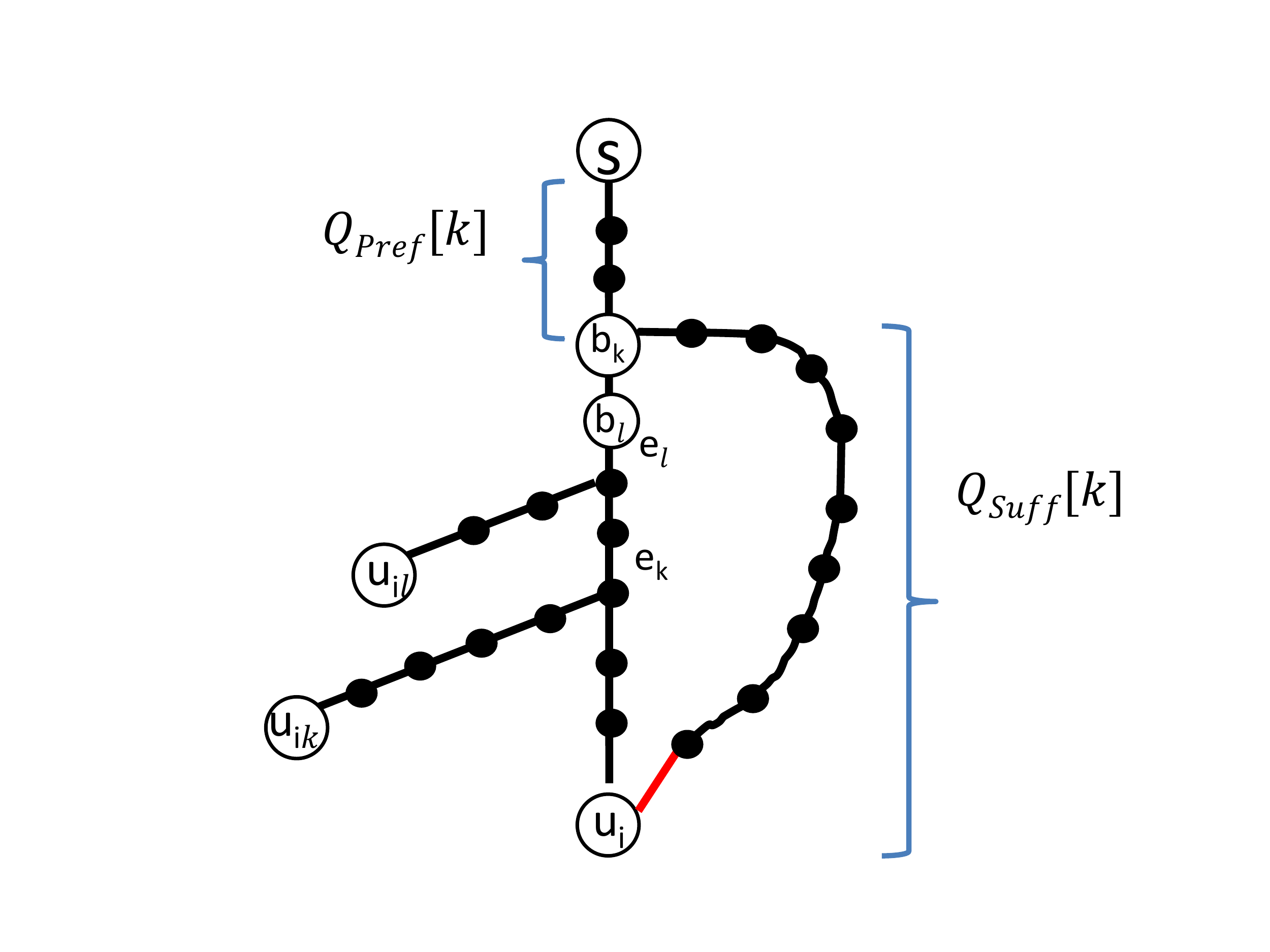}
\\
\small (a) \hbox{\hskip 3in} \small (b)
\end{center}
\caption{Illustration for Cl. \ref{cl:pathinter}. Black edges represent
the original BFS tree $T_0$. The red edge is a new edge.
\label{figure:approxcases}}
\end{figure}

\begin{claim}
\label{cl:depth}
$\depth(b_k, T_0)<\depth(b_{k+1}, T_0)$ for every $k \in \{1,2,3\}$.
\end{claim}
\Proof
Towards contradiction, assume that $\depth(b_k, T_0) \geq \depth(b_{k+1}, T_0)$ for some $k \in \{1, 2,3\}$.
We first claim that in this case both $P_{k+1}, P_{k} \subseteq G \setminus \{e_{j_k},e_{j_{k+1}}\}$.  Recall that $P_{k}=Q_{Pref}[k] \circ Q_{Suff}[k]$ and $P_{k+1}=Q_{Pref}[k+1] \circ Q_{Suff}[k+1]$.
Note that by Cl. \ref{cl:uniqe_div}, the paths $Q_{Suff}[k], Q_{Suff}[k+1]$ are edge disjoint with $\pi(s, u_i)$ and therefore $Q_{Suff}[k], Q_{Suff}[k+1] \subseteq G \setminus \{e_{j_k},e_{j_{k+1}}\}$. In addition, by the fact that $P_k$ diverged from $\pi(s,u_i)$ before the faulty edge $e_{j_k}$ (and by ordering also before $e_{j_{k+1}}$) it holds that $Q_{Pref}[k] \subseteq G \setminus \{e_{j_k},e_{j_{k+1}}\}$. Since $\depth(b_k, T_0) \geq \depth(b_{k+1}, T_0)$, it holds that $P_{k+1}$ diverged from $\pi(s, u_i)$ not after $b_k$, hence $Q_{Suff}[k], Q_{Suff}[k+1] \subseteq G \setminus \{e_{j_k},e_{j_{k+1}}\}$. Overall, we get that $P_{k+1}, P_{k} \subseteq G \setminus \{e_{j_k},e_{j_{k+1}}\}$.

We thus have two alternative replacement $s-u_{i_k}$ (resp., $s-u_{i_{k+1}}$) paths given by $\widetilde{P}_k=P_{k+1} \circ P^*_{i_k,j_k}[u_i, u_{i_k}]$ and
$\widetilde{P}_{k+1}=P_{k} \circ P^*_{i_{k+1},j_{k+1}}[u_i, u_{i_{k+1}}]$ respectively.
We now derive a contradiction by analyzing the costs of $P^*_{i_k,j_k}$
and $P^*_{i_{k+1},j_{k+1}}$ and showing that all costs components
(see Eq. (\ref{eq:well_multis_onef})) are equal except the last.
By the optimality of $P^*_{i_k,j_k}$ and $P^*_{i_{k+1},j_{k+1}}$ in round $i_{k}$ and $i_{k+1}$ respectively, i.e., by the fact that $P^*_{i_k,j_k} \in SP(s, u_{i_{k}}, G \setminus \{e_{j_k}\}, W_{i_k})$ and $P^*_{i_{k+1},j_{k+1}} \in SP(s, u_{i_{k+1}}, G \setminus \{e_{j_{k+1}}\}, W_{i_{k+1}})$,
it follows that $|P_k|=|P_{k+1}|$.
In addition, since $P_k$ is a subpath of $P^*_{i_k,j_k}$ and its last edge is the first new edge of $P^*_{i_k,j_k}$ (i.e., $\LastE(P_k)=\FirstEN(P^*_{i_k,j_k})$), it follows that $|\New(P_k)|=1$. By a similar argument, since $\LastE(P_{k+1})=\FirstEN(P^*_{i_{k+1},j_{k+1}})$, we also have that $|\New(P_{k+1})|=1$.
By the optimality of
$P^*_{i_k,j_k} \in SP(s, u_{i_k}, G \setminus \{e_{j_k}\},W_{i_k})$ according to
weight assignments $W_{i_k}$ (see Fact \ref{obs:weigh:prop}(c)) we get that
\begin{equation}
\label{eq:wineq}
|\cI_{k,k}| \leq |\cI_{k+1,k}|~.
\end{equation}
In the same manner, by the optimality of
$P^*_{i_{k+1},j_{k+1}} \in SP(s, u_{i_{k+1}}, G \setminus \{e_{j_{k+1}}, W_{i_{k+1}}\})$
according to weight assignments $W_{i_{k+1}}$, we get that
\begin{equation}
\label{eq:wineq2}
|\cI_{k+1,k+1}| \leq |\cI_{k,k+1}|~.
\end{equation}
Applying Cl. \ref{cl:pathinter}(a) with $\ell=k$ we have that
\begin{equation}
\label{eq:k_depth_a}
|\cI_{k,k}| = |Q_{Pref}[k]| = \depth(b_{k})~.
\end{equation}
Applying Cl. \ref{cl:pathinter}(a) with $\ell=k+1$ we have that
\begin{equation}
\label{eq:k_depth_b}
|\cI_{k,k+1}| = |Q_{Pref}[k]| = \depth(b_{k})
\end{equation}
and
\begin{equation}
\label{eq:k_depth2_a}
|\cI_{k+1,k+1}| = |Q_{Pref}[k+1]| = \depth(b_{k+1})~.
\end{equation}
By. Cl. \ref{cl:pathinter}(b), we also have that
\begin{equation}
\label{eq:k_depth2_b}
|\cI_{k+1,k}| = |Q_{Pref}[k+1]| = \depth(b_{k+1})~.
\end{equation}
Combining Eq. (\ref{eq:wineq}) with Eq. (\ref{eq:k_depth_a}) and
(\ref{eq:k_depth2_b}), we get that $\depth(b_k)\leq \depth(b_{k+1})$.
Combining Eq. (\ref{eq:wineq2}) with Eq. (\ref{eq:k_depth_b}) and
(\ref{eq:k_depth2_a}), we get the opposite inequality,
$\depth(b_{k+1})\leq \depth(b_{k})$. It follows that
$\depth(b_{k})=\depth(b_{k+1})$, hence
inequalities (\ref{eq:wineq}) and (\ref{eq:wineq2}) are in fact equalities.

As we have shown that the paths $P^*_{i_k, j_k}$ and $P^*_{i_{k+1}, j_{k+1}}$ have the same length, the same number of new edges and the same number of joint edges with the shortest-path, by Fact \ref{obs:weigh:prop}(d) their relative costs are determined by $\widetilde{W}_k=\sum_{e_j \in P_k} \Weight_j$ and $\widetilde{W}_{k+1}=\sum_{e_j \in P_{k+1}} \Weight_j$.
Hence, by the optimality of $P^*_{i_k,j_k}$ under $\Cost_{i_k}$ it follows that $\widetilde{W}_k <\widetilde{W}_{k+1}$, and by the optimality of $P^*_{i_{k+1}, j_{k+1}}$ under $\Cost_{i_{k+1}}$ we get that $\widetilde{W}_{k+1} <\widetilde{W}_{k}$, contradiction.
\QED
Claim \ref{cl:depth} implies that the vertices $b_1, b_2, b_3, b_4$ are distinct, and moreover, they appear in this order on $\pi(s,u_i)$. In addition, note that for every $k \in \{1,2,3\}$, $b_k \in P_{k+1}$ (since $b_{k+1}$ is below $b_k$ on $\pi(s,u_i)$).
\begin{claim}
\label{cl:length}
$|P_k[b_k,u_i]|>|P_{k+1}[b_{k},u_i]|$ for every $k \in \{1,2,3\}$.
\end{claim}
\Proof
By the uniqueness of the divergence point $b_k$ of $\pi(s,u_i)$ and $P_k$ (Cl. \ref{cl:uniqe_div}) and by Cl. \ref{cl:depth}, $P_k \subseteq G \setminus \{e_{j_k}, \ldots, e_{j_4}\}$ for every $k \in \{1, 2, 3, 4\}$.
Since $P_k$ is a replacement path in $G \setminus \{e_{j_{k'}}\}$ for $k' >k$,
but $P_k$ was nevertheless not chosen as part of the replacement path $P^*_{i_{k+1}, j_{k+1}}$, it follows that $\Cost_{i_{k+1}}(P_{k+1})<\Cost_{i_{k+1}}(P_{k})$.
Let us now analyze which cost component accounts for this difference.
By Cl. \ref{cl:pathinter}(a),
$|\cI_{k+1,k+1}| = \depth(b_{k+1})$
and
$|\cI_{k,k+1}| = \depth(b_{k})$.
Hence, since due to Cl. \ref{cl:depth}, $\depth(b_{k+1})>\depth(b_{k})$
(and thus $|\cI_{k+1,k+1}| > |\cI_{k,k+1}|$)
and $|\New(P_k)|=|\New(P_{k+1})|=1$, it follows by the optimality of $P^*_{i_{k+1}, j_{k+1}} \in SP(s, u_{i_{k+1}}, G \setminus \{e_{j_{k+1}}\}, W_{i_{k+1}})$ for $i_{k+1}$ (see Fact \ref{obs:weigh:prop}(c)) that $|P_{k+1}|<|P_{k}|$.
As $P_{k}[s, b_k]=P_{k+1}[s, b_k]=\pi(s, b_k)=Q_{Pref}[k]$, the claim follows.
\QED
Let $d_0=\depth(u_i)$ and let $d_k=\depth(v_k)$ for $k=\{1, \ldots, 4\}$.
Combining Cl. \ref{cl:depth} and \ref{cl:length}, we get that
$d_1 > d_2 > d_3 >d_4$, and as these are integers, necessarily $d_1-d_4 \geq 3$. Now consider the edges $e_k=(v_k,u_i)=\LastE(P_k)$ for $k=\{1, \ldots, 4\}$. The existence of these edges in $G$, implies that $d_k \in \{d_0-1, d_0, d_0+1\}$ for every $k=\{1, \ldots, 4\}$, implying that $d_1-d_4 \leq 2$, contradiction.
Lemma \ref{lem:bfsspanner_sizebound} follows.
\QED

\subsection{Multiple edge faults}

\begin{figure}[htbp]
\begin{center}
\includegraphics[scale=0.3]{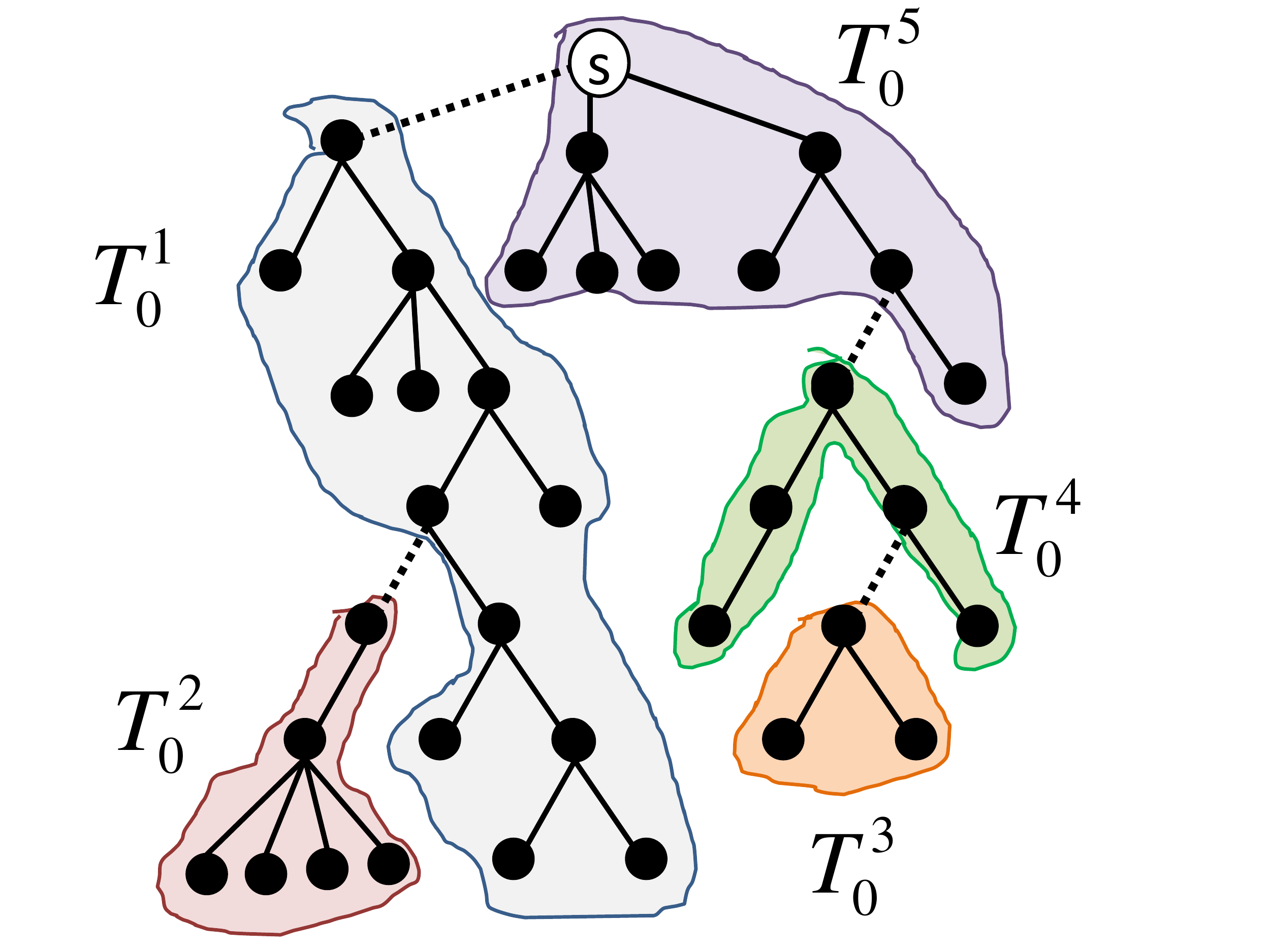}
\caption{Upon the failure of the BFS edges $F$ (dotted lines), the BFS tree
$T_0$ is decomposed into $c=|F|+1\leq f+1$ components, denoted
$T_0^1, \ldots, T_0^c$.
\label{fig:multibfsbrakes}}
\end{center}
\end{figure}

In this section, we consider the case of $f$ edge failures for constant
$f\geq 1$, and establish the following.
\begin{theorem}
\label{thm:multbfsmultf}
There exists a poly-time algorithm that for every $n$-vertex graph constructs
\begin{description}
\item{(1)}
a $(3(f+1), (f+1) \log n)$ \FTSPANNERBFS\ structure with $O(f n)$ edges and
\item{(2)}
a $(3(f+1)+1,0)$ \FTSPANNERBFS\ structure with
$O(f n+ n^{1+1/k}+n\cdot ((f+1) \cdot (2k-1))^{f+1})$ edges,
\end{description}
overcoming up to $f$ edge faults, for every $k \geq 3$.
\end{theorem}
For an edge set $F \subseteq E$, let $P^*_{F}(u_i) \in SP(s, u_i, G \setminus F)$ be the $s-u_i$ replacement path upon the failure of $F$ in $G$. To avoid complications due to shortest-paths of the same length, we assume all shortest-path are computed with a weight assignment $W$ that guarantees the uniqueness of the shortest-paths.
\paragraph{Algorithm Description.}
The algorithm consists of three phases.
The first phase constructs a (possibly dense) $f$-edge \FTBFS\ structure $T_1$ with respect to $s$ by using Alg. $\FBFS(s, G, f)$ to be defined later. 
The second phase constructs an $f$-edge $(3(f+1), 0)$ \FTSPANNERBFS\ structure $T_2 \subseteq T_1$ by carefully sparsifying the edges of $T_1$.
However, $T_2$ might still be dense.
Finally, the last phase obtains a \emph{sparse}
$(3(f+1),(f+1) \cdot \log n)$ \FTSPANNERBFS\ structure $H \subseteq T_2$
where $|H|=O(n)$.
We rely on the following fact.
\begin{lemma}
\label{fc:pairwise_ftspanner}\cite{CLPR09-span}
There exists an algorithm $\MSPAANER(G, \alpha, f)$ that given an $n$-vertex graph $G$ constructs an $f$ \emph{edge} fault tolerant $(\alpha,0)$ spanner $G' \subseteq G$ such that $|G'| \leq O(f \cdot n^{1+1/\alpha})$.
\end{lemma}
\begin{description}
\label{desc:upmultif}
\item
\textbf{Algorithm} $\APPBFS(s, G, f)$ {\bf - overview}
\item{(1)}
Invoke Alg. $\FBFS(s, G, f)$ to generate an $f$-edge \FTBFS\ structure $T_1$
with respect to $s$.

\item{(2)}
Sparsify the new edges of $T_1$ to obtain an $(3(f+1), 0)$ \FTSPANNERBFS\
structure $T_2 \subseteq T_1$.
\item{(3)}
Set $T'\gets \MSPAANER(T_2 \setminus T_0, \log n, f)$.
\item{(4)}
$H= T_0 \cup T'$.
\end{description}
Algorithm $\FBFS(s, G, f)$ of phase (1) operates in a ``brute-force" manner. For every $u_i \in V$, and every subset $F \subseteq E$, $|F|\leq f$, it constructs an $s-u_i$ replacement path $P^*_{F}(u_i)\in SP(s, u_i, G \setminus F,W)$ of minimal length. The $f$-edge \FTBFS\ structure $T_1$ is then given by $T_1=\{P^*_{i,F} \in SP(s, u_i, G \setminus F) ~\mid~  F \subseteq E, |F| \leq f, u_i \in V\}$.
In phase (2), each of these replacement paths $P=P^*_{F}(u_i)$ is considered and at most $f+1$ of the set of new edges $\New(P)$ are taken into $T_2$ at the expense of introducing a stretch. To choose these edges from $\New(P)$ for a given replacement path $P=P^*_{F}(u_i)$, the algorithm labels the vertices $V(P)$ according to their sensitivity to the set of failed edges $F$, where vertices are given the same label iff they appear in the same connected tree in the surviving forest $T_0 \setminus F$.
Since vertices $u', u''$ of the same label remain connected in $T_0 \setminus F$, the algorithm exploits their path in $T_0 \setminus F$ as a replacement to the path $P[u',u'']$ used by the optimal replacement path. The benefit of these bypasses is that they use only edges of the original BFS, allowing us to save the new edges that occur on $P[u',u'']$.
The potential drawback is that these bypasses might be longer than their counterparts in $P$. In the analysis we argue that the stretch introduced by these replacements is bounded by $3(f+1)$. Hence phase (2) turns the \emph{exact} FT-BFS structure $T_1$ to an \emph{approximate} FT-BFS structure $T_2$ with bounded multiplicative stretch.
We now describe formally the construction of $T_2$, beginning with
the labeling scheme $\LAB_{F}: V \to \mathcal{C}(T_0 \setminus F)$.
Let $\mathcal{C}(T_0 \setminus F)=\{T_0^1, \ldots, T_0^{\ell}\}$ be the set of connected components (subtrees) of the forest $T_0 \setminus F$. For pictorial illustration, see Fig. \ref{fig:multibfsbrakes}.
Then the label $\LAB_{F}(u')$ of every vertex $u' \in P$ is set to
$\ell' \in \{1, \ldots, \ell\}$ iff $u' \in T_0^{\ell'}$.
The procedure for selecting at most $f+1$ new edges of $P$ using the labels
$\LAB_{F}(u')$, $u' \in P$ is as follows.
Let $\New(P)=\{(u_1,v_1), \ldots, (u_k,v_{k})\}$ be the sorted set of new edges $E(P) \setminus E(T_0)$ according to their order of appearance on the replacement path $P$ (from $s$).
Pair the vertices $v_i$ by matching $v_i$ with the farthest $v_{i'}$ for $i' \in \{i, \ldots, k\}$ of the same label $\LAB_{F}(v_i)=\LAB_{F}(v_{i'})$, setting $M(v_i)=v_{i'}$ and $\widehat{m}(v_i)=i'$. Initialize $i=1$ and $T_2=T_0$. Now, repeatedly (until $i \geq k$) add the new edge $e_i=(u_i, v_i)$ to $T_2$ and set $i=\widehat{m}(v_i)+1$.

\begin{figure}[htb!]
\begin{center}
\includegraphics[scale=0.3]{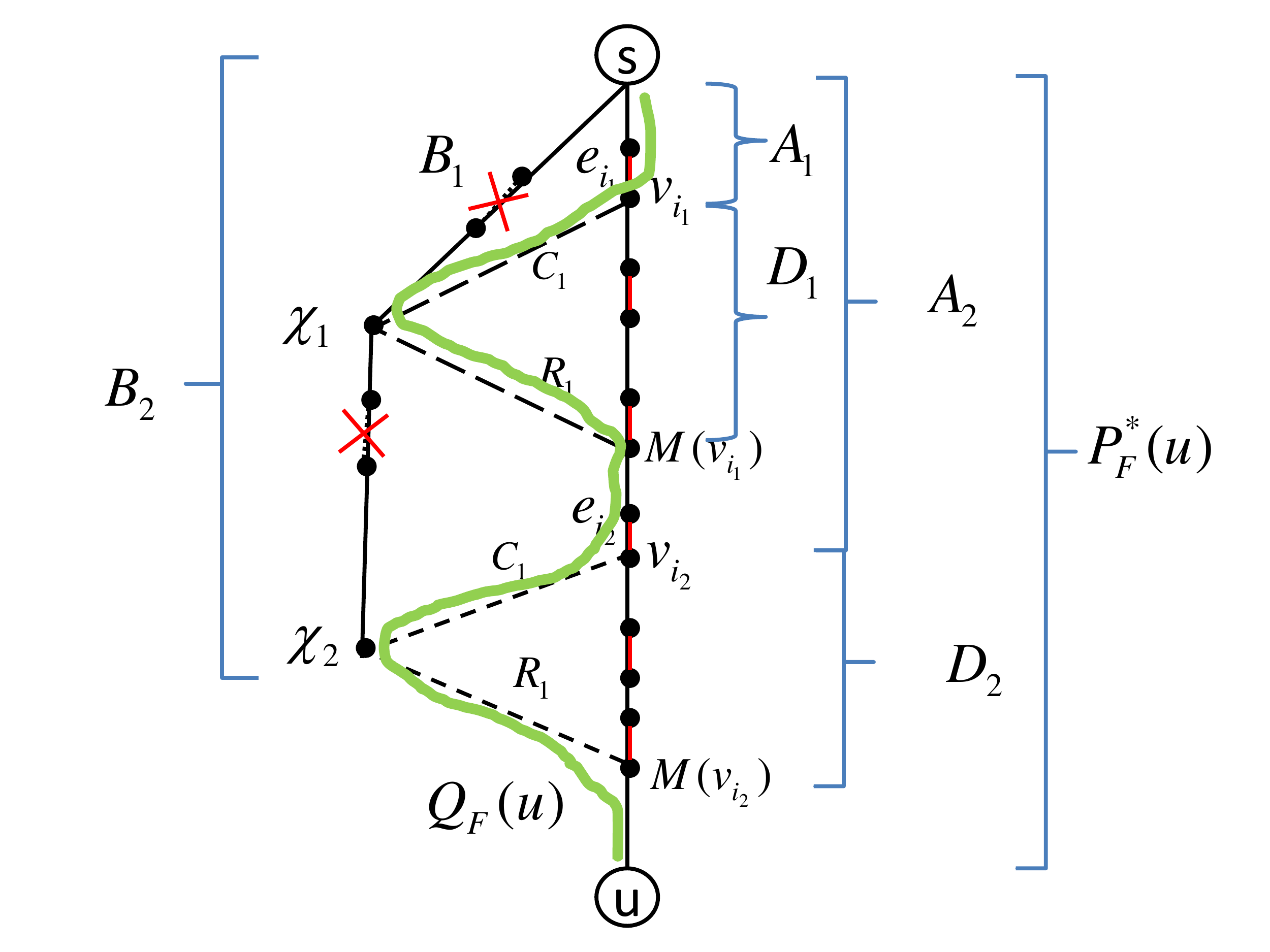}
\caption{Illustration of the replacement paths used in $T_2$. Black lines correspond to BFS edges. Shown is the $P^*_{F}(u)$ replacement path where new edges $\New(P^*_{F}(u))$ are marked in red. The dashed lines correspond to the BFS path used to bypass  segments in $P^*_{F}$. The replacement $s-u$ path $Q_f(u)$ with at most $f+1$ new edges is marked in green. \label{fig:multifbypass}}
\end{center}
\end{figure}

Finally, we describe phase (3) of the algorithm. Given the $(3(f+1), 0)$ \FTSPANNERBFS\ structure $T_2$, a subgraph $(3(f+1), (f+1)\log n))$ \FTSPANNERBFS\ structure $T_3$ of $O(n)$ edges is constructed as follows. Let $G'=T_2 \setminus T_0$ be the subgraph obtained by removing all original BFS edges of $T_0$ from $T_2$. Let $T' \gets \MSPAANER(G', \log n,f)$ be an $f$-edge fault tolerant $(\log n,0)$ spanner for $G'$. The resulting structure is $H=T_0 \cup T'.$
\paragraph{Analysis.}
We first provide an auxiliary claim regarding the labeling scheme $\LAB_{F}(u_i)$, $F \subseteq E$.
\begin{observation}
\label{obs:pathinfgraph}
\begin{description}
\item{(1)}
$\pi(u_i, u_{i'}) \subseteq T_0 \setminus F$ for every pair of vertices
$u_i, u_{i'}$ such that $\LAB_{F}(u_i)=\LAB_{F}(u_{i'})$.
\item{(2)}
$|\bigcup \{\LAB_{F}(u') ~\mid~ u' \in V\}| =
|\mathcal{C}(T_0 \setminus F)|\leq |F|+1$.
\end{description}
\end{observation}
\Proof
Part (1) follows by definition, since two vertices are assigned the same label for a given edge fault $F$ iff they belong to the same tree in the forest $T_0 \setminus F$. Part (2) is proven by induction. Let $e_1, \ldots, e_{\ell} \in F \cap E(T_0)$ be the faulty edges in $T_0$. We claim that $|\mathcal{C}(T_0 \setminus F)|= \ell+1$.  Since $\ell \leq |F|$, this would establish the observation. For the base of the induction, $\ell=1$, note that every faulty edge $e=(x,y)$ disconnects the tree $T_0$ into two nonempty components, one containing $x$ and one containing $y$, hence $|\mathcal{C}(T_0 \setminus \{e\})|= 2$. Assume this holds for every $\ell' \leq \ell$ and consider $\ell'+1$. By the induction assumption, the number of components in $T'=T_0 \setminus \{e_1, \ldots, e_{\ell'}\}$ is $|\mathcal{C}(T')|=\ell'+1$.
Let $T'' \in \mathcal{C}(T')$ be the connected tree in $T'$ that contains $e_{\ell'+1}$. By the definition of $\mathcal{C}(T')$, such $T''$ exists. Then by the induction base, $\mathcal{C}(T'' \setminus \{e_{\ell'+1}\})=2$, since $T''$ is broken into two components upon the removal of the edge $e_{\ell'+1}$. We get that $\mathcal{C}(T \setminus \{e_1, \ldots, e_{\ell'+1}\})=|\mathcal{C}(T')|+1=\ell'+2$. The observation follows.
\QED
Fix a vertex $u$, edge faults $F \subseteq E$, $|F|\leq f$, and a replacement path $P=P^*_{F}(u)$. Let $\New(P)=\{e_1, \ldots, e_{\ell}\}$, where $e_i=(u_i, v_i)$ for $i \in \{1, \ldots, \ell\}$ be the set of new edges in order of appearance on $P$ (from $s$) and let $\New^{+}(P)=\{e_{i_1}, \ldots, e_{i_{\ell'}}\} \subseteq \New(P)$ be the corresponding ordered set taken into $T_2$. Let $\New^{-}(P)=\New(P) \setminus \New^{+}(P)$ the set of new edges not included in $T_2$. The following observation is immediate by the structure of the algorithm.
\begin{observation}
\label{obs:multifnewedge_prop}
\begin{description}
\item{(1)} $e_{i_1}=e_1$.
\item{(2)} $M(v_{i_y}) \in V(P[v_{i_y}, v_{i_{y'}}])\setminus \{v_{i_{y'}}\}$ for every $x'>x$.
\item{(3)} $P[M(v_{i_{y-1}}), v_{i_{y}}] \subseteq T_2 \setminus F$.
\item{(4)} $\LAB_{F}(v_{i_y})\neq \LAB_{f}(v_{i_{y'}})$
for every $y, y' \in \{1, \ldots, \ell'\}$.
\end{description}
\end{observation}
\Proof
Parts (1) and (2) follow by the description of the algorithm.
To see (3), note that $e_{i_y}$ is the first  new edge that appears after $M(v_{i_{y-1}})$, i.e., $e_{i_y}=\FirstEN(P[M(v_{i_{y-1}}), u])$. Hence,
$\New(P[M(v_{i_{y-1}}), v_{i_{y}}])=\{e_{i_y}\}$.
In addition, since $P[M(v_{i_{y-1}}), v_{i_{y}}]$ appeared on the replacement path $P \subseteq T_1 \setminus F$, it holds that $E(P[M(v_{i_{y-1}}), v_{i_{y}}]) \cap F=\emptyset$. Since $e_{i_y}$ (the only new edge of $P[M(v_{i_{y-1}}), u]$ is taken to $T_2$, (3) follows.
Finally, consider Part (4). Assume towards contradiction that there exists some $x < x'$ such that $\LAB_{F}(v_{i_y})=\LAB_{f}(v_{i_{y'}})$.
By Part (2), $M(v_{i_y}) \in V(P[v_{i_y}, v_{i_{y'}}]) \setminus \{v_{i_{y'}}\}$. Since  $\LAB_{F}(v_{i_y})=\LAB_{f}(v_{i_{y'}})$ and $v_{i_{y'}}$ appears \emph{after} $M(v_{i_y})$, we get a contradiction to the selection of $M(v_{i_y})$. The observation follows.
\QED
We proceed by showing that the multiplicative stretch introduced by excluding $\New^{-}(P)$ from the intermediate structure $T_2$ is at most $3(f+1)$.
For $P^*_{F}(u)$, define the corresponding $s-u$ replacement path $Q_{f}(u)$. Let $Q_1=P[s, v_{i_1}]\circ \pi(v_{i_1}, M(v_{i_1}))$ and $Q_y=P[M(v_{i_{y-1}}), v_{i_y}]\circ \pi(v_{i_y}, M(v_{i_{y+1}}))$ for every $y \in \{2, \ldots, \ell'-1\}$. Then
$$Q_{f}(u)=Q_1 \circ Q_2 \ldots \circ Q_{\ell'} \circ P[M(v_{i_{\ell'}}),u].$$
For pictorial illustration, see Fig. \ref{fig:multifbypass}.

We state that $Q_{f}(u)$ is an $s-u$  replacement path in $T_2 \setminus F$ and it is longer than the optimal replacement path $P^*_{F}(u)$ by a factor of at most $3(f+1)$.
%
\begin{lemma}
\label{lemma:multspt2}
\begin{description}
\item{(1)}
$Q_{F}(u) \subseteq T_2 \setminus F$.
\item{(2)} $|Q_{F}(u)| \leq 3(f+1) \cdot |P^*_{F}(u)|$.
\end{description}
\end{lemma}
\Proof
Let $Q^*=Q_{f}(u)$. We begin with (1). Note that the missing new edges $\New(P)^{-}$ appear on $P$ between some two vertices of the same label. Specifically, $\New(P)^{-} \subseteq \bigcup_{y=1}^{\ell'} P[v_{i_y}, M(v_{i_{y}})]$. Since $\LAB_{F}(v_{i_y})=\LAB_{F}(M(v_{i_{y}}))$,
by Obs. \ref{obs:pathinfgraph}, it holds that $\pi(v_{i_y}, M(v_{i_{y}})) \subseteq T_0 \setminus F$ and hence also $\pi(v_{i_y}, M(v_{i_{y}})) \subseteq T_2 \setminus F$. Formally, by Obs. \ref{obs:multifnewedge_prop}(1)
$e_{i_1}=e_1=\FirstEN(P)$ hence $Q_1 \subseteq T_0\setminus F$. By Obs. \ref{obs:multifnewedge_prop}(3) the subpath $P[M(v_{i_{y-1}}), v_{i_y}] \in T_2 \setminus F$ for every $y \in \{2, \ldots, \ell'-1\}$. Finally, by Cl. \ref{obs:pathinfgraph}, $\pi(v_{i_y}, M(v_{i_{y+1}})) \subseteq T_0 \setminus F$ for every $y \in \{1, \ldots, \ell'-1\}$, since $\LAB_F(v_{i_y})=\LAB_F(M(v_{i_{y+1}}))$.
It follows that $Q^* \subseteq T_2 \setminus F$.
\par We now turn to (2) and show that $|Q^*| \leq 3(f+1)|P|$.
For every $y \in \{1, \ldots, \ell'\}$, we define the following in order to be able to bound the length of the bypass. Let $\chi_y=\LCA(v_{i_y}, M(v_{i_y}))$, $A_y=P[s, v_{i_y}], B_y=\pi(s, \chi_y), C_y=\pi(\chi_y, v_{i_y}), D_y=P[v_{i_y}, M(v_{i_y})]$ and $R_y=\pi(\chi_y, M(v_{i_y}))$.
Let $\widetilde{P}_y=A_y \circ C_y \circ R_y$ and $P_y=P[s,M(v_{i_y})]$.
\begin{claim}
\label{cl:aux_multif3}
$|\widetilde{P}_y|\leq 3 \cdot |P_y|$ for every $y \in \{1, \ldots, \ell'\}$.
\end{claim}
\Proof
Note that since $B_y \circ C_y=\pi(s, v_{i_y})$, the alternative $s-v_{i_y}$ path
$A_y$ satisfies $|A_y|\geq |B_y|+|C_y|$. In addition, since $C_y \circ D_y$ is an alternative $\chi_y-M(v_{i_y})$ path and $R_y=\pi(\chi_y, M(v_{i_y}))$ is the $\chi_y-M(v_{i_y})$ shortest-path, it holds that $|R_y| \leq |C_y|+|D_y|$. Hence, $|\widetilde{P}_y|\leq |A_y|+2|C_y|+|D_y|\leq 3|A_y|+|D_y|\leq 3|P_y|$.
\QED
We now claim that the replacement $s-M(v_{i_y})$ path
$$\widetilde{Q}_y=Q_1 \circ Q_2 \ldots \circ Q_y \subseteq T_2 \setminus F$$
satisfies $|\widetilde{Q}_y| \leq 3y|P_y|$. This is proved by induction on $y$. For the base of the induction consider $y=1$. In this case $\widetilde{Q}_1=\widetilde{P}_1$. Hence, the claim follows by Cl. \ref{cl:aux_multif3} and $|\widetilde{Q}_1|\leq 3 \cdot |P_1|$.
Assume the claim holds up to $y-1$ and consider $y$.
\begin{eqnarray*}
|\widetilde{Q}_{y}|&=& |\widetilde{Q}_{y-1}|+|Q_y|
\leq
3(y-1)|P_{y-1}|+|P[M(v_{i_{y-1}}), v_{i_y}]|+|C_y \circ R_y|
\\ & \leq & 3(y-1)|P[s, v_{i_y}]|+|C_y \circ R_y|
\\&=&
3(y-1)|A_{y}|+|C_y \circ R_y|
\leq
3(y-1)|A_{y}|+3|A_y \circ D_y|
\\&\leq& 3y \cdot |P_y|~,
\end{eqnarray*}
where the first inequality follows by the induction assumption, as $|\widetilde{Q}_{y-1}| \leq 3(y-1) |P_{y-1}|$, and the third inequality follows by Cl. \ref{cl:aux_multif3}.
Since by Obs. \ref{obs:multifnewedge_prop}(4) and Obs. \ref{obs:pathinfgraph}(2), $\ell' \leq f+1$, we get that $|\widetilde{Q}_{\ell'}|\leq 3 \cdot (f+1) |P_{\ell'}|$ and
$|Q^*|=|\widetilde{Q}_{\ell'}|+|P[M(v_{i_{\ell'}}), u]|\leq 3(f+1)|P_{\ell'}|+|P[M(v_{i_{\ell'}}), u]|\leq 3(f+1)|P|$. The lemma follows.
\QED
We therefore have the following.
\begin{corollary}
\label{cor:T_2_PROP}
For every $u$ and $F \subseteq E$, there exists a replacement path
$Q_{F}(u) \in T_2 \setminus F$ such that
\begin{description}
\item{(1)}
$|\New(Q_{F}(u))|\leq f+1$ and
\item{(2)}
$|Q_{F}(u)|\leq 3(f+1) \cdot P^*_{F}(u)$.
\end{description}
Hence $T_2$ is an $f$-edge $(3(f+1),0)$ \FTSPANNERBFS\ structure.
\end{corollary}
Finally, we prove the correctness of the last phase and show the following.
\begin{claim}
\label{cl:multspt3}
\begin{description}
\item{(1)}
$H$ is an $f$-edge $(3(f+1), (f+1)\log n)$ \FTSPANNERBFS\ structure
with respect to $s$.
\item{(2)}
$|H|=O(f n)$.
\end{description}
\end{claim}
\Proof
Let $P=P^*_{F}(u)$ be the optimal replacement path and let $Q=Q_{F}(u)$ be the corresponding replacement path in $T_2$ obtained by using at most $f+1$ bypasses between vertices of the same label on $P$. Then, by Cor. \ref{cor:T_2_PROP}, $|\New(Q)|\leq f+1$ and $|Q|\leq 3(f+1) \cdot P$.
Let $G'=T_2 \setminus T_0$ and $G''=\MSPAANER(G', \log n, f)$ be the
$f$-edge FT $(\log n,0)$ spanner for $G'$ (see Fact \ref{fc:pairwise_ftspanner}).
Let $E'=\New(Q) \setminus H$ be the set of new edges in $Q$ that are missing in the final $H$. Hence, $E' \subseteq G' \setminus G''$. Since $E' \subseteq \New^+(Q)$, it holds that $|E'| \leq |\New^+(Q)|\leq f+1$. We now claim that for every missing edge $e=(x,y) \in E'$ there exists an $x-y$ path in the surviving structure $G'' \setminus F$ of length at most $\log n$. By the fact that $G''$ is an $f$-edge $(\log n,0)$ spanner for $G'$, it holds that $\dist(x, y, G''\setminus F) \leq \log n \cdot \dist(x, y, G' \setminus F)=\log n$, where the last equality follows by the fact that
$e=(x,y)$ appears on the replacement path $Q$, hence $e \notin F$, so $\dist(x, y, G' \setminus F)=1$.
We therefore have that $Q$ contains at most $f+1$ missing edges, and for each there exists a path in $G"$ of length at most $\log n$. Overall, $\dist(s, u, G \setminus F) \leq |E(Q) \setminus E'|+(\log n )\cdot |E'| \leq 3(f+1)|P|+(f+1)\log n$. Part (1) is established.
We now consider part (2). By Fact \ref{fc:pairwise_ftspanner},
$|G''|=O(n)$, hence $H=T_0 \cup G''$ has $O(f n)$ edges as well.
The claim follows.
\QED
This completes the proof of Thm. \ref{thm:multbfsmultf}(1). We now consider part (2) of the theorem asserting that it is possible to get rid of the additive factor, albeit at the expense of considerably increasing the size of the \FTSPANNERBFS\ structure.
\Proof[Thm. \ref{thm:multbfsmultf}(2)]
For a given $k \geq 3$, we first construct the collection of all replacement paths $P^*_{F}(u_i) \in SP(s, u_i, G \setminus F,W)$ for every $F \subseteq E$ and every $u_i \in V$. Let $\ell=(f+1)\cdot(2k-1)$.
The $(3(f+1)+1,0)$ \FTSPANNERBFS\ structure $H$ is constructed in two steps. A replacement path $P^*_{F}(u)$ is \emph{short} iff $|P^*_{F}(u)|\leq \ell$.
In the first step, a subgraph $H_1 \subseteq G$ is constructed containing the set of all short replacement paths, i.e., $H_1=\{\LastE(P^*_{F}(u_i))~\mid~ |P^*_{F}(u_i)|\leq \ell\}$.
In the second step, an $(3(f+1),\ell)$ \FTSPANNERBFS\ structure $H_2$ is constructed by employing the algorithm of Thm. \ref{thm:multbfsmultf}(1) with one minor modification;  in step (3) of the algorithm we construct an $(2k-1,0)$ spanner instead of $(\log n,0)$ spanner, i.e., step (3) is given by $T'\gets \MSPAANER(T_2 \setminus T_0, 2k-1, f)$ for the given $k$ (while $2k-1=\log n$ in the original algorithm). By Fact \ref{fc:pairwise_ftspanner} and the proof of Cl. \ref{cl:multspt3}, it holds that $|E(H_2)|=O(f \cdot n +n^{1+1/k})$. We next bound the size of $H_1$.
\begin{claim}
\label{cl:h1_size}
$|H_1| \leq n \cdot ((f+1)\cdot(2k-1))^{f+1}$.
\end{claim}
\Proof
Let $\mathcal{P}_i=\{P^*_F(u_i), F \subseteq E\}$ be the collection of $s-u_i$ replacement paths.
Define $\mathcal{P}_i^{0}=\{\pi(s,u_i)\}$ and $\mathcal{P}_i^{f'}=\{P^*_{F}(u_i) ~\mid~ |F|=f'\}$ for
$f' \geq 1$ as the collection of $s-u_i$ replacement paths supporting a sequence of $f'$ edge faults. Hence,
$\mathcal{P}_i=\bigcup_{f'=1}^f \mathcal{P}_i^{f'}$.

We prove that for every $f' \in \{1, \ldots,f\}$ if holds that $|\mathcal{P}_i^{f'}|\leq \ell \cdot |\mathcal{P}_i^{f'-1}|$. To see this, observe that
every replacement path $P^*_{F}(u_i) \in \mathcal{P}_i^{f'} \setminus \mathcal{P}_i^{f'-1}$ protects an edge failure $e$ in some $P^*_{F'}(u_i)\in \mathcal{P}_i^{f'-1}$.
Hence, $F=F' \cup \{e\}$. Since $\mathcal{P}_i^{f'-1}$ contains only short replacement paths, it holds that each short $s-u_i$ replacement path $P^*_{F'}(u_i)\in \mathcal{P}_i^{f'-1}$ has at most $\ell$ replacement paths of the form $P^*_{F' \cup \{e\}}(u_i)$ in $\mathcal{P}_i^{f'}$ that protect against the failure of $e \in P^*_{F'}(u_i)$. Concluding that $|\mathcal{P}_i^{f'}|\leq \ell \cdot |\mathcal{P}_i^{f'-1}|$.
Overall, $|\New_i|\leq |\bigcup_{f'=1}^f \mathcal{P}_i^{f'}| \leq \ell^{f+1}$ and $|H_1| \leq n \cdot \ell^{f+1}$. The claim follows.
\QED

Finally, we show that $H=H_1 \cup H_2$ is an $(3(f+1)+1,0)$ \FTSPANNERBFS\ structure. Consider a vertex edge-set pair $(i,F)$ corresponding to a vertex $u_i \in V$ and edge set $F \subseteq E$. There are two cases. Case (a) is where the replacement path $P^*_{F}(u_i)$ is short, i.e., $|P^*_{F}(u_i)| \leq \ell$. By the same argumentation as in Lemma \ref{lemma:correctness} it holds that $\dist(s, u_i, H_1\setminus F)=\dist(s, u_i, G\setminus F)$ (since $H_1$ contains the last edges of these replacement paths, which was shown to be sufficient).
The complementary case is where the replacement path is long, $|P^*_{F}(u_i))|>\ell$.
By the proof of Thm. \ref{thm:multbfsmultf}(1), it holds that $\dist(s, u_i, H_2\setminus F) \leq (3(f+1))\dist(s, u_i, G\setminus F)+\ell \leq (3(f+1)+1)\dist(s, u_i, G\setminus F)$. The claim follows.
\QED

\section{Lower Bounds for Additive FT-ABFS Structures}
\label{sec:lowerbound_add}
In this section we provide lower bound constructions for $(1, \beta)$
\FTSPANNERBFS\ structures for various values of the additive stretch $\beta$.
The starting point for these constructions is the lower bound construction for the exact \FTBFS\ structures in \cite{PPFTBFS13}. In this construction, the bulk of the edges was due to a complete bipartite graph $B=(L,R)$ where $|L|=\Theta(n)$ and $|R|=\Theta(\sqrt{n})$. When relaxing to additive stretch $2$, it is no longer required to include $B$ entirely in the spanner; in fact, one can show that it suffices to include in the $(1,2)$ \FTSPANNERBFS\ structure only a subgraph $B' \subseteq B$ of $O(n)$ edges. This subgraph $B'$ is obtained by connecting an arbitrary vertex $\ell^* \in L$ to every vertex $r \in R$ and connecting an arbitrary vertex $r^* \in R$ to every vertex $\ell \in L$. Thus our goal is to replace $B$ with a dense subgraph of high girth. Constructions of this type are known, but our requirement is in fact more stringent. Note that an essential characteristic of the construction of \cite{PPFTBFS13} is that the $R$ layer consists of at most $c \cdot \sqrt{n}$ vertices, for some constant $c>0$. A larger layer could not be supported since the upper bound analysis of \cite{PPFTBFS13} implies that every vertex in $L$ requires at most $O(\sqrt{n})$ edges in every \FTBFS\ structure. Since the known lower bound on the number of edges in a $\Theta(n)\times \Theta(\sqrt{n})$ bipartite graph with high girth (greater than 4) is $O(n)$, such graphs are not good candidates for replacing the bipartite graph $B$ in the construction, as using them results in graphs of $O(n)$ rather than $O(n^{1+\epsilon(\beta)})$ edges. Instead, our constructions replace the complete \emph{unbalanced} bipartite graph $B$ by a \emph{multiple} copies of \emph{balanced} $\Theta(\sqrt{n}) \times \Theta(\sqrt{n})$ bipartite graphs with high girth. Hence, the lower bound constructions of $(1, \beta)$ \FTSPANNERBFS\ structure rely on known construction of balance bipartite graphs with high girth.  The desired construction is achieved by carefully inserting multiple copies of these graphs.

\paragraph{Additive Stretch $\beta=1$.}
\label{sec:add1}
By a proof very similar to that of the exact case (see \cite{PPFTBFS13})
one can show the following (details are omitted).
\begin{theorem}
\label{thm:lowerbound_edgeonef}
There exists an $n$-vertex graph $G(V, E)$ and a source $s \in V$ such that any $(1,1)$ \FTSPANNERBFS\ structure rooted at $s$ has $\Omega(n^{3/2})$ edges.
\end{theorem}

\paragraph{Additive stretch $3 \leq \beta \leq O(\log n)$.}
The \emph{girth} of a graph $G$ is the minimum number of edges on any cycle in $G$. Let $B(n, g)=(L, R,E)$ be a bipartite graph, where $|L|=|R|=n$, the girth is at least $g$ and the number of edges is $|E|=\Omega(n^{m(g)})$ for some function $m(g)>1$. Removing any edge $(u,v)$ in such a graph $B(n,g)$ increases the distance between its endpoints $u$ and $v$ from $1$ to $g-1$, which implies that any (strict) subgraph of $B(n,g)$ has additive stretch $\beta \geq g-2$.
\par
\begin{figure}[htbp]
\begin{center}
\includegraphics[scale=0.4]{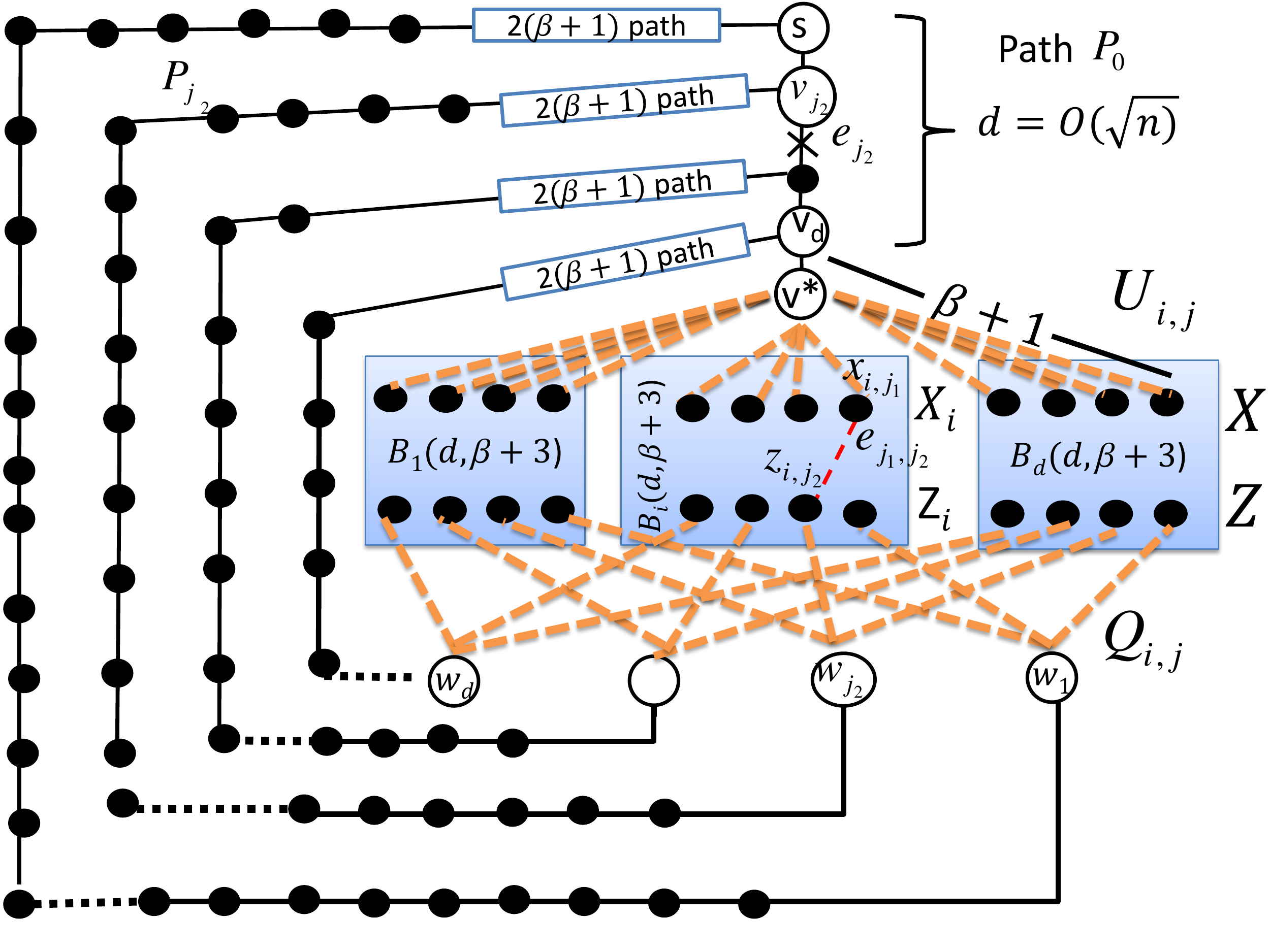}
\caption{Lower bound construction for \FTSPANNERBFS\ structures.
The original BFS tree consists of the non-dashed edges.
The dotted edges are necessary to make it an \FTSPANNERBFS\ structure.
The orange dashed lines correspond to $\beta+1$ length path. The edges of the bipartite graphs $B_{i}(d, \beta+3)$ are required in every $(1,\beta)$ \FTSPANNERBFS\ structure with respect to $s$. For example the red edge $e_{j_1, j_2}$ is necessary upon the fault of $e_{j_2} \in P_0$. \label{fig:lowerbound1f}}
\end{center}
\end{figure}

 In what follows, we embed copies of $B(d, g)$ graphs for some parameters $d$ and $g$ in order to achieve a lower bound for the case of a $(1,\beta)$ \FTSPANNERBFS\ structure with \emph{edge faults}. We show the following.
\begin{theorem}
\label{thm:lowerbound_additive}
For every integer $n$ and constant integer $\beta$, there exists an $n$-vertex graph $G(n, \beta)=(V, E)$ and a source $s \in V$ such that any $(1,\beta)$ \FTSPANNERBFS\ structure with respect to $s$, $H$, has:
\begin{description}
\item{(1)}
$|E(H)|=\Omega(n^{5/4})$, if $\beta \leq 3$.
\item{(2)}
$|E(H)|=\Omega(n^{1+1/2(\beta+3)})$, if $4 \leq \beta \leq O(\log n)$.
\item{(3)}
$|E(H)|=\Omega(n^{7/6})$, if $\beta \leq 5$.
\item{(4)}
$|E(H)|=\Omega(n^{11/10})$, if $\beta \leq 9$.
\end{description}
\end{theorem}
\Proof
Given integers $n$ and $\beta$, the graph $G(n, \beta)=G(V,E)$ consists of the following main components (illustrated in Fig. \ref{fig:lowerbound1f}):
\begin{description}
\item{(1)}
A path $P_0=[s=v_1, \ldots, v_{d+1}=v^*]$ of length $d$, for an integer parameter $d$ fixed later. Our focus in the analysis is on what happens when some edge on this path fails.
\item{(2)}
A set $X=\bigcup_{i=1}^d X_i$ of $d^2$ vertices organized in $d$ sets $X_1, \ldots, X_d$ each of size $d$, where $X_i=\{x_{i,1},\ldots,x_{i,d}\}$ for every $i \in \{1, \ldots, d\}$.
\item{(3)}
A set of $d^2$ vertex disjoint paths of length $\beta+1$ each, $U_{i,j}=[v^*=u_1^i, \ldots, u_{\beta+2}^i=x_{i,j}]$, for every $i,j \in \{1, \ldots, d\}$ connecting the vertex $v^* \in P_0$ to the vertices of $X$.
\item{(4)}
A set $Z=\bigcup_{i=1}^d Z_i$ of $d^2$ vertices organized in $d$ sets $Z_1, \ldots, Z_d$ each of size $d$, where $Z_i=\{z_{i,1},\ldots,z_{i,d}\}$ for every $i \in \{1, \ldots, d\}$.
\item{(5)}
A vertex set $W=\{w_1, \ldots, w_d\}$, where vertex $w_j$ is connected by $d$ vertex disjoint paths $Q_{i,j}$ of length-$\beta+1$ to the $d$ vertices $\{z_{i, j} \in Z_{i} \mid i=1, \ldots, d\}$. Altogether there exist $d^2$ vertex disjoint paths $Q_{i,j}=[w_j=q_1^j, \ldots, q_{\beta+2}^j=z_{i,j}]$, one for every $i,j \in \{1, \ldots, d\}$.
\item{(6)}
A collection of $d$ vertex disjoint paths of decreasing length, $P_1, \ldots, P_d$,
where for $j \in \{1,\ldots,d\}$, $P_j=[v_j=p^j_1, \ldots, p^j_{\ell_j}=w_j]$ connects $v_j$ with $w_j$ and its length is
$\ell_j=|P_j|=d+4+(\beta+1)\cdot (d-j+1).$
\item{(7)}
$d$ copies of the bipartite graph $B(d, \beta+3)$, where
$B_i(d, \beta+3)=(X_i, Z_i, E_i)$ connects $X_i$ to $Z_i$ for every
$i \in \{1, \ldots, d\}$. This graphs contribute most of the edges in $G$
(with all the other components containing only $O(n)$ edges).
\item{(8)}
A path $R=[s=r_1, \ldots, r_{d'+1}]$ for $d'\geq 0$ to be defined later, added in order to complete the number of vertices in $G$ to exactly $n$. (This path has no special role in the construction.)
\end{description}
Overall, the vertex set of $G(n, \beta)$ is
\begin{eqnarray*}
V&=&X \cup Z \cup V(P_0) \cup \left( \bigcup_{i=1}^d V(P_i) \right) \cup
\bigcup_{i=1}^d \bigcup_{j=1}^d\left(V(Q_{i,j}) \cup V(U_{i,j})\right) \cup V(R).
\end{eqnarray*}
and its edge set is
\begin{eqnarray*}
E&=&\bigcup_{i=1}^d E(B_i(d,\beta+3)) \cup E(P_0) \cup E(R)\cup \bigcup_{i=1}^d E(P_i) \cup \bigcup_{i,j} (E(Q_{i,j}) \cup E(U_{i,j})).
\end{eqnarray*}
Set $d=\lfloor \sqrt{n/(14\beta)} \rfloor$,
and let $d'=n-|V \setminus (R \setminus \{s\})|$.
We now turn to prove the correctness of the construction and establish Thm. \ref{thm:lowerbound_additive}.
Note that without the path $R$, the rest of $G(n, \beta)$ contains fewer than $n$ vertices, as $|V(P_0)|=d$, $|X \cup Z|=2d^2$, $|\bigcup_{i=1}^d V(P_i)|=\sum_{j=1}^d \left(d+4+(\beta+1)(d-j+1)\right) \leq 10\beta \cdot d^2$, $|\bigcup_{i,j}V(U_{i})| = d^2 \cdot (\beta+2)$ and $|\bigcup_{i,j} V(Q_{i,j})| = d^2 \cdot (\beta+2)$.
Hence overall, $|V \setminus \left(R \setminus \{s\}\right)| \leq 14 \beta\cdot d^2 \leq n$. Fix $d'=|R|$ to complete the size of $G(n, \beta)$ to $n$, i.e., set $d'=n-|V \setminus (R \setminus \{s\})|$.
\begin{observation}
\label{cl:numbervertices_lbaddtive}
$|E|=\Omega(n^{1/2+m(\beta+3)/2})$.
\end{observation}
\Proof
The bulk of the edges is due to the bipartite graphs $B_{i}(d, \beta+3)$. Since there are $d$ such graphs, each with $\Omega(d^{m(\beta+3)})$ edges, we get that $|E| \geq \sum_{i} E(B_{i}(d, \beta+3))= d \cdot \Omega(d^{m(\beta+3)})=\Omega(n^{1/2+m(\beta+3)/2})$. The Claim follows.
\QED
%
We now show the following.
\begin{claim}
\label{cl:lb_alledges}
Every $(1,\beta)$ \FTSPANNERBFS\ structure $H$ for $G$ with respect to $s$, must contain all the edges of $B_i(d,\beta+3)$ for every $i \in \{1, \ldots, d\}$.
\end{claim}
\Proof
Let $G=G(n,\beta)$. Assume, towards contradiction, that there exists some $i \in \{1, \ldots, d\}$ and an $(1,\beta)$ \FTSPANNERBFS\ structure $H$ for $G$ such that $B_{i}(d, \beta+3) \nsubseteq H$. Let $e_{j}=(v_{j},v_{j+1})$ for every $j \in \{1, \ldots, d\}$. Let $e_{j_1, j_2}=(x_{i,j_1}, z_{i,j_2})$, where $x_{i,j_1}\in X_i$ and $z_{i,j_2}\in Z_i$, be a missing edge in $H$.
Note that for $i',j' \in \{1, \ldots, d\}$, the shortest $s-z_{i',j'}$ path in $G \setminus \{e_{j'}\}$ is $P_{i',j'}= P_0[s=v_1,v_{j'}] \circ P_{j'} \circ Q_{i',j'}$ .
\par Upon the failure of the edge $e_{j_2}=(v_{j_2},v_{j_2+1}) \in P_0$, the unique shortest $s-x_{i,j_1}$ path in $G \setminus \{e_{j_2}\}$ is $P^*=P_{i,j_2} \circ e_{j_1, j_2}$.  Since $e_{j_1, j_2} \notin H$ and therefore also $P^* \nsubseteq H$, the $s-x_{i, j_1}$ distance in $H \setminus \{e_{j_2}\}$ is strictly larger than that in $G \setminus \{e_{j_2}\}$. Let $P' \in SP(s, x_{i,j_1}, H\setminus \{e_{j_2}\})$ be the $s-x_{i,j_1}$ shortest-path in $H \setminus \{e_{j_2}\}$. By construction, $P'$ must traverse the $Z$ vertices (as the path $P_0$ is disconnected).

Let $z_{i',j'}$ be the first $Z$ vertex occurring on $P'$.
There are three cases to consider. \\
Case (C1) $j' > j_2$ and $i' \in \{1, \ldots, d\}$. Since the faulty edge disconnects the shortest-path of $G$ between $s$ and $v_{j'}$ for every $j' > j_2$, the shortest-path between $s$ and $z_{i',j'}$ in $G \setminus \{e_{j_2}\}$ must visit another $Z$ vertex, in contradiction to the fact that $z_{i',j'}$ is the first $Z$ vertex on $P'$. \\Case (C2) $j' < j_2$ and $i' \in \{1, \ldots, d\}$. In this case, the shortest-path between $s$ and $z_{i',j'}$ in $G \setminus \{e_{j'}\}$ is $P_{i',j'}$. Since $|P_{j'}|\geq |P_{j_2}|+\beta+1$ for every $j'<j_2$, it holds that $|P'|>|P^*|+\beta+1$, in contradiction to the fact that $H$ is a $(1, \beta)$ \FTSPANNERBFS\ structure. \\
Case (C3) $j'=j_2$. This case is further divided into two cases.\\
Subcase(C3A): $i \neq i'$. The shortest-path $P'$ must be of the form $P'=P_{i',j_2} \circ Q'$ where $Q' \in SP(z_{i',j_2}, x_{i,j_1}, H \setminus  \{e_{j_2}\})$. Since any path connecting vertex $w \in B_{i}(d, \beta+3)$ and $w' \in B_{i'}(d, \beta+3)$ is of length at least $\beta+1$, it holds that $|P'| \geq |P^*|+\beta+1$, and we end with contradiction again. \\
Subcase (C3B): $i = i'$. Since the edge $e_{j_1, j_2} \notin T'$ is missing, $z_{i, j_2}$ is the first $Z$ vertex on $P'$, and the distance between $x_{i,j} \in B_{i}(d, \beta+3)$ and $w' \in B_{i'}(d, \beta+3)$ for $i \neq i'$, is of length at least $\beta+1$, it follows that $x_{i, j_2}$ is connected to $z_{i,j_2}$ via vertices in $B_{i}(d, \beta+3)$, i.e., $P'=P_{i, j_2} \circ Q'$ where $Q'$ is an $z_{i, j_2}-x_{i, j_1}$ shortest path in $H$. Since the girth of $B_{i}(d, \beta+3)$ is $g \geq \beta+3$, it holds that $\dist(x_{i, j_1}, z_{i, j_2}, B_{i}(d, \beta+3) \setminus \{e_{j_1, j_2}\}) \geq \beta+2$, hence $|Q'|\geq \beta+2$ and $|P'|\geq |P^*|+\beta+1$, contradiction.
\QED
\begin{corollary}
\label{cor:lb}
Every $(1,\beta)$ \FTSPANNERBFS\ structure $H$ for $G$ must contain at least $d \cdot d^{m(\beta+3)}$ edges.
\end{corollary}
To establish the theorem we now make use of the following fact concerning the existence of graphs with ``many'' edges and high girth.
\begin{fact}
\label{cl:bipartite_densegirth}
For every sufficiently large $n$ there exists a connected bipartite graph $B(n,g)=\{L, R, E\}$, $|L|=|R|=n$ such that
\begin{description}
\item{(1)}
$m(g)=3/2$ for $g=6$, Lemma 5.1.1 of \cite{Godsil:book},
\item{(2)}
$m(g)=4/3$ for $g\geq 8$ \cite{MKD05}
\item{(3)}
$m(g)=6/5$ for $g \geq 12$ \cite{MKD05} and
\item{(4)}
$m(g)=1+1/g$ for every $6 \leq g \leq 2\log n$, \cite{Smid06}.
\end{description}
\end{fact}
Thm. \ref{thm:lowerbound_additive} follows by Cor. \ref{cor:lb} and by applying Fact \ref{cl:bipartite_densegirth}.
\QED

\section{Upper Bound for Additive Stretch $4$}
\label{sec:upperbound_add}

In this section, we establish the following.

\begin{theorem}
\label{thm:add_ub}
There exists a poly-time algorithm that for every $n$-vertex  unweighted undirected graph $G$ and source $s$ constructs a $(1, 4)$ \FTSPANNERBFS\ structure $H$ with $O(n^{4/3})$ edges.
\end{theorem}
As usual, the starting point of our construction is the shortest path tree $T_0$, and this tree should be fortified against possible failure of any of its edges, by adding edges and augmenting $T_0$ into $H$.

\paragraph{Overview.}
The construction of $H$ consists of 3 main stages and revolves around
the following key observation.
Let $E' \subseteq E(G)$ be a subset of edges.
A replacement path $P^*_{i,j} \in SP(s, v_i, G \setminus \{e_j\})$ is ``missing-ending" with respect to  $E'$ if the last edge of $P^*_{i,j}$ does not belong to $E'$, i.e., $\LastE(P^*_{i,j}) \notin E'$. Let $\mathcal{P}^*=\{P^*_{i,j} \mid v_i \in V \mbox{~and~} e_j \in \pi(s, v_i)\}$ be the collection of all $s-v_i$ replacement paths for every $e_j \in \pi(s, v_i$) and consider its partition  $\mathcal{P}^*_{E'}=\mathcal{P}^{miss}_{E'} \cup \mathcal{P}^{+}_{E'}$ into missing-ending paths $\mathcal{P}^{miss}_{E'}=\{P^*_{i,j} ~\mid~ \LastE(P^*_{i,j}) \notin E'\}$ and non missing-ending paths $\mathcal{P}^{+}_{E'}=\{P^*_{i,j} ~\mid~ \LastE(P^*_{i,j}) \in E'\}$.
Clearly, a subgraph $H$ containing all replacement paths $\mathcal{P}^*$ is an exact \FTBFS\ structure, however by \cite{PPFTBFS13} such a subgraph might contain $\Omega(n^{3/2})$ edges. One of the key observations in our analysis is that it is sufficient to ``take care'' only of the missing ending path collection $\mathcal{P}^{miss}_{E'}$.
\begin{observation}
\label{obs:only_new_h}
Let $H \subseteq G$ be a subgraph containing $E'$ and satisfying
$\dist(s, v_i, H \setminus \{e_j\}) \leq
\dist(s, v_i, G \setminus \{e_j\})+\beta$ for every $v_i \in V$
and $e_j \in \pi(s, v_i)$ such that $P^*_{i,j} \in \mathcal{P}^{miss}_{E'}$.
Then $H$ is a $(1, \beta)$ \FTSPANNERBFS\ structure, i.e.,
$\dist(s, v_i, H \setminus \{e_j\}) \leq
\dist(s, v_i, G \setminus \{e_j\})+\beta$
also holds for non missing-ending paths $P^*_{i,j} \in \mathcal{P}^{+}_{E'}$.
\end{observation}
Observation \ref{obs:only_new_h} provides the basis for the general structure
of the algorithm.
Its proof is based on ideas resembling those appearing in the proof of Lemma \ref{lem:correct_add_up_clust}).

\begin{figure}[htb!]
\begin{center}
\begin{minipage}{3in}
\includegraphics[width=3in]{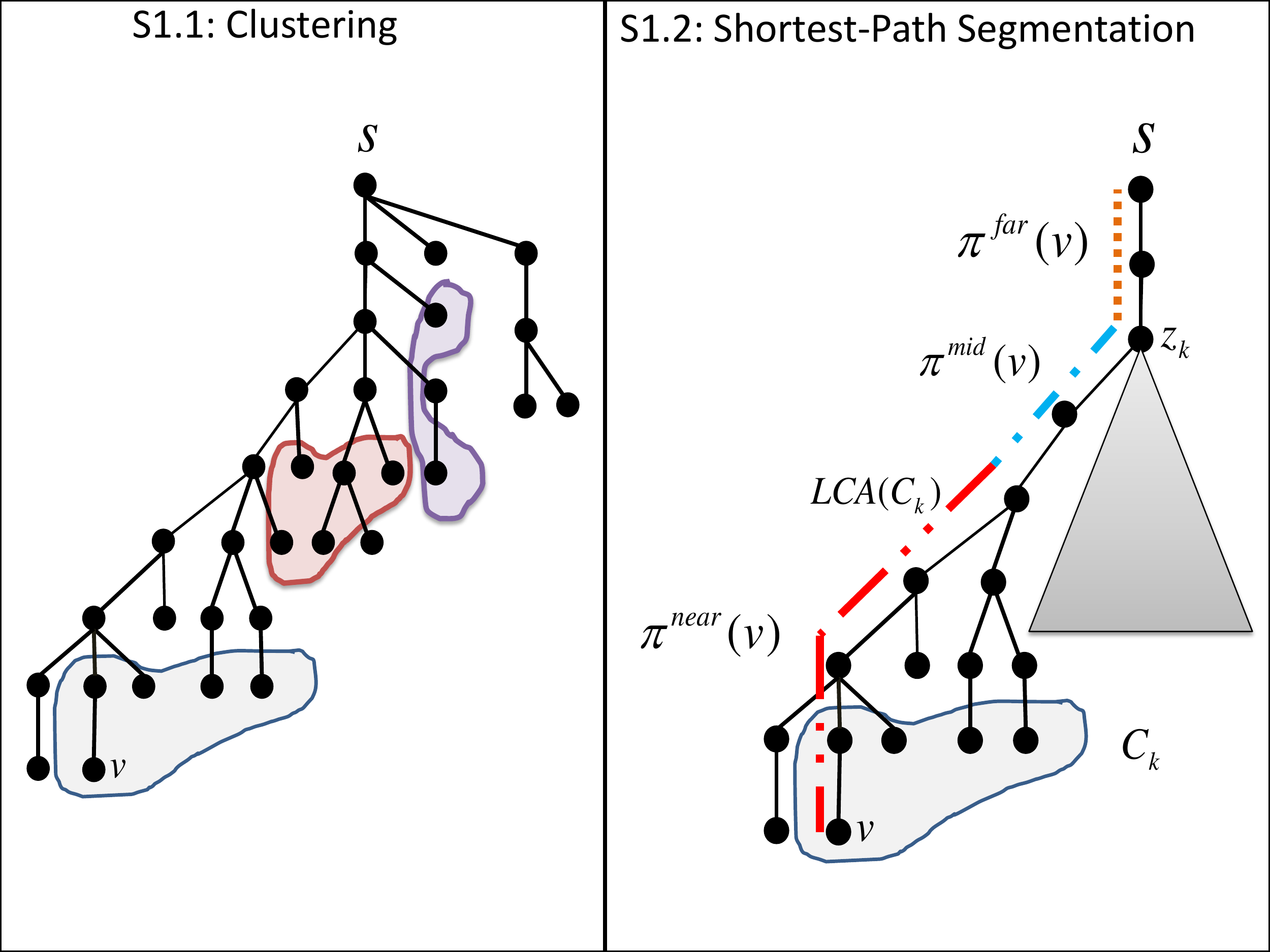}
\includegraphics[width=3in]{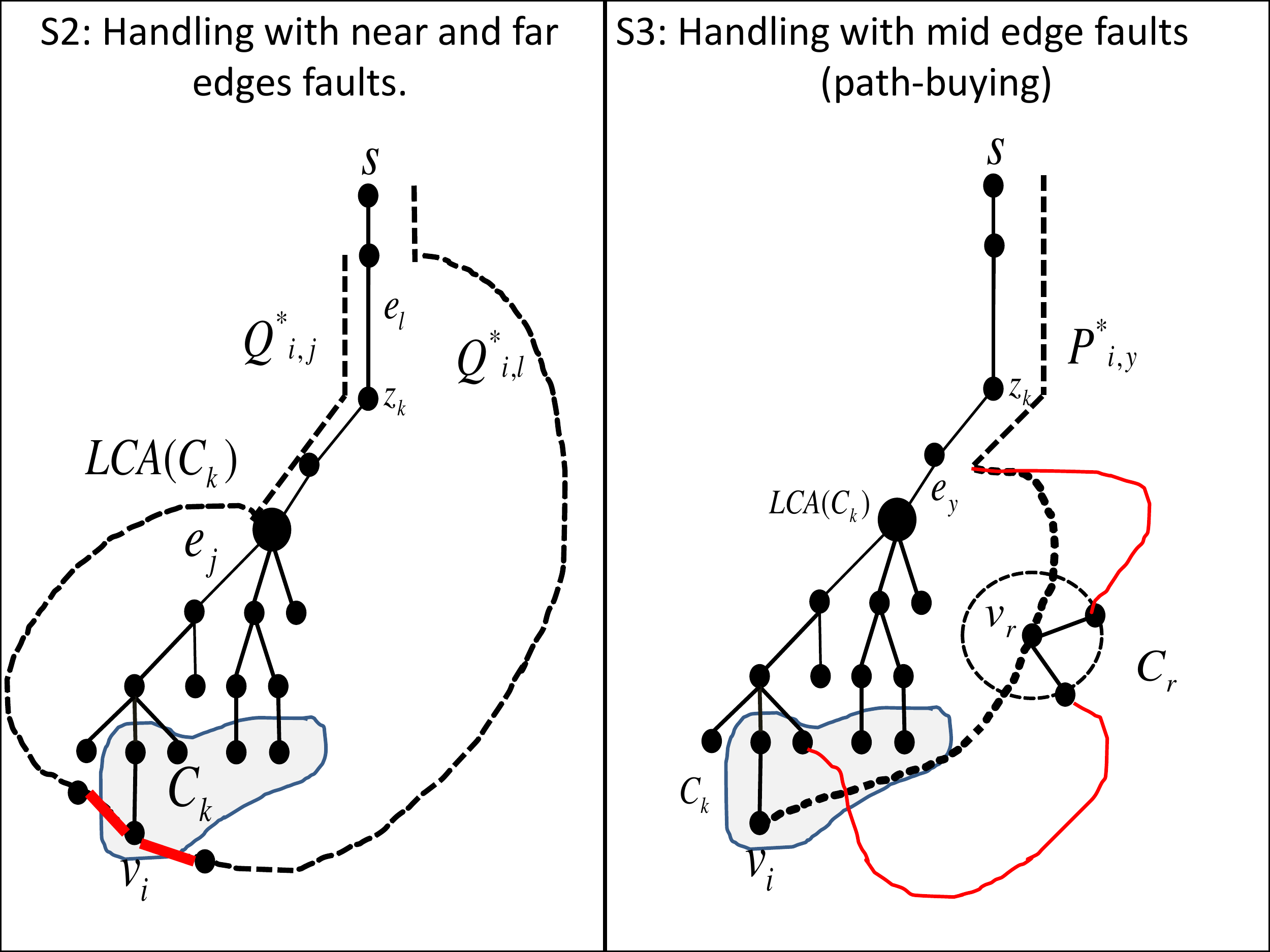}
\end{minipage}
\end{center}
\caption{
Schematic illustration of the main stages in $H$ construction. Stage S1.1: Clustering in the graph $G \setminus T_0$. Stage S1.2: Segmentation of the shortest-path $\pi(s, v_i)$ into three segments: near, mid and far. Stage S2: Handling with the near and far edge faults by adding the last edge of their corresponding replacement paths $Q^*_{i,j}$. Stage S3: Handling with mid edge faults by applying a path buying procedure on the detour segments $P^-_{i,j}$ of the corresponding replacement paths $P^*_{i,j}$. A candidate
detour is not added to $H$ if there exists an alternative "safe" and short path between some clusters representatives.
\label{figure:ubstages}}
\end{figure}

A vertex-edge pair $(i,j)$, representing a vertex $v_i$ and an edge
$e_j \in \pi(s, v_i)$, is \emph{satisfied} by a subgraph $H \subseteq G$
if there exists a replacement path $P \in SP(s, v_i,  G \setminus \{e_j\})$
whose last edge $\LastE(P)$ is in $H$, otherwise it is \emph{unsatisfied}.
For a subgraph $H \subseteq G$, let
\begin{align*}
\Pairs(H) = & \{(i,j) ~\mid~ e_j \in \pi(s, v_i) \mbox{~and~}
\\ &
\LastE(P) \notin H \mbox{~for every~} P \in
SP(s, v_i, G \setminus \{e_j\})\}
\end{align*}
be the collection of unsatisfied pairs in $H$.
Starting with the collection of all pairs
$\Pairs_0=\{(i,j) ~\mid~ e_j \in \pi(s, v_i) \}$
that are required to be satisfied in $H$, the algorithm consists of
three
stages, aiming towards increasing the set of satisfied pairs by adding a
suitable collection of edges to the constructed \FTSPANNERBFS\ structure $H$.
Specifically, in each stage $k$, the algorithm is given a ``partial"
\FTSPANNERBFS\ structure $\CurrSpanner_k$ and a list of pairs
$\Pairs_k\subseteq\Pairs_{k-1}$ that might not be satisfied yet
in $\CurrSpanner_k$. Essentially, the pairs $\Pairs_{k-1} \setminus \Pairs_k$
are satisfied in $\CurrSpanner_k$.
The algorithm then defines a subset of target pairs
$\Delta_k \subseteq \Pairs_k$ and a corresponding collection of edges
$\LEdgeDelta(\Delta_k)$ that aim to satisfy $\Delta_k$ in the final spanner.
At the end of this stage, the algorithm sets $\CurrSpanner_{k+1}=\CurrSpanner_k \cup \LEdgeDelta(\Delta_k)$ and the updated list of unsatisfied pairs is reduced to $\Pairs_{k+1}=\Pairs_{k} \setminus \Delta_k$.
To compute a sparse $\LEdgeDelta(\Delta_k)$,  the algorithm considers for every pair $(i,j) \in \Delta_k$, a \emph{specific} replacement path $\widetilde{P}_{i,j} \in SP(s, v_i, G \setminus \{e_j\})$. Only the last edge $\LastE(\widetilde{P}_{i,j})$ is added to $\LEdgeDelta(\Delta_k)$.
Hence, $\LEdgeDelta(\Delta_k)=\{\LastE(\widetilde{P}_{i,j})\mid~ (i,j) \in \Delta_k\}.$
Finally, in the last stage 3, the replacement paths of the yet unsatisfied
pairs in the current \FTSPANNERBFS\ structure $\CurrSpanner_3$ are considered to be added \emph{entirely} to $H$, by employing a modified path-buying procedure.
We now describe these stages in more detail.
For a schematic illustration of the scheme, see Fig. \ref{figure:ubstages}.

At the first stage S1 of Subsec. \ref{s1:cluster}, the algorithm clusters some of the vertices of $V(G)$, resulting in a clustered graph $G_C \subseteq G \setminus T_0$ which is shown to have $O(n^{4/3})$ edges and a set $\mathcal{C}$ of $O(n^{2/3})$ clusters.
For this we use the clustering algorithm of \cite{CGK13}, where not all the vertices of $V(G)$ are clustered but the only $G$ edges missing in $G_C$ are those incident to clustered vertices.
Hence, letting
$\Delta_1=\{(i,j) \mid v_i\mbox{~is not clustered and~}e_j \in \pi(s, v_i) \}$,
the edge set $\LEdgeDelta(\Delta_1)=G_C$ satisfies $\Delta_1$
in the current spanner $\CurrSpanner_1$.
It therefore follows that it remains to handle only pairs $(i',j)$ for clustered vertices $v_{i'}$.
At the end of stage S1 (Sec. \ref{append:sp_segment}), the algorithm uses the clustering to divide the shortest path
$\pi(s, v_i)$ of every clustered vertex $v_i$ into three consecutive segments
$$\pi(s, v_i)=\pi^{far}(v_i) \circ \pi^{mid}(v_i) \circ \pi^{near}(v_i),$$
where the breakpoints depend on the clustering.
Let $C(v_i) \in \mathcal{C}$ be the cluster of $v_i$ and $\LCA(C(v_i))$ be the least common ancestor in $T_0$ of the members of $C(v_i)$. Then, our segmentation satisfies that $\pi^{far}(v_i) \circ \pi^{mid}(v_i)=\pi(s, \LCA(C(v_i)))$ and that $|\pi^{mid}(v_i)| \leq \left \lceil n^{2/3} \right \rceil$, and the vertices of $\pi^{far}(v_i)$ are at distance at least $n^{2/3}$ from $v_i$.
\par Equipped with this segmentation, the algorithm now handles separately edge failures $e_j \in T_0$ in each of these three segments. Specifically,
stage S2 (in Sec. \ref{s2:nearfar}) deals with failures of
$e_j \in \pi^{far}(v_i) \cup \pi^{near}(v_i)$,
and stage S3 (in Sec. \ref{sec:pathbuyingproc}) deals with failures of $e_j \in \pi^{mid}(v_i) $ by employing a modified path-buying procedure.

Generally speaking, the edge faults in the far and near segments are handled by
adding the collection of all last edges of the corresponding replacement paths.
For a suitable construction of the replacement paths,
this last edge collection is shown to be small.
In contrast, the edge faults in the mid segments are handled by considering every replacement path $P^*_{i,j} \in SP(s, v_i, G \setminus \{e_j\})$ satisfying that $e_j \in \pi^{mid}(v_i) $ and adding it \emph{entirely} to $H$ if it satisfies some \emph{cost} to \emph{value} balance.  To efficiently handle the faults, either in the near and far sections or in the mid section, a (nontrivial) preprocessing step of replacement path construction is required.

Let $IC=\{ i \mid v_i \mbox{~is clustered}\}$.
In stage S2, the algorithm defines, for every $i \in IC$,
$\Delta^{near}_i ~=~ \{(i,j) ~\mid~ e_j \in \pi^{near}(v_i) \}$ ~~ and ~~
$\Delta^{far}_i ~=~ \{(i,j) ~\mid~ e_j \in \pi^{far}(v_i) \}$.
\\
The goal of this stage is to satisfy the pairs of
$\Delta^{near}=\bigcup_{i\in IC} \Delta^{near}_i$ ~~~and~~~
$\Delta^{far}=\bigcup_{i\in IC} \Delta^{far}_i$
in the constructed \FTSPANNERBFS\ structure $H$.
This stage consists of two substages. In Substage S2.1,
a collection of replacement path $Q^*_{i,j}$ is constructed. In Substage S2.2,
the algorithm creates a sparse
set $\LEdgeDelta(\Delta^{near})$ (resp., $\LEdgeDelta(\Delta^{far})$) containing the last edges of some replacement paths $Q^*_{i,j} \in SP(s, v_i, G \setminus \{e_j\})$ for every $(i,j) \in \Delta^{near}$ (resp., $(i,j) \in \Delta^{far}$).
Due to some nice properties of the $Q^*_{i,j}$ collection, the analysis shows that
$\left|\LEdgeDelta(\Delta^{near}) \cup \LEdgeDelta(\Delta^{far})\right|=O(n^{4/3})$.
In particular, the size of the set $\LEdgeDelta(\Delta^{near}_i)$, consisting of the new edges appearing as last edges on $Q^*_{i,j} \in SP(s, v_i, G \setminus \{e_j\})$ satisfying that $e_j \in \pi^{near}(v_i) $, is bounded by a constant (which will be shown to follow from Fact. \ref{fc:clustering} that diameter of the clusters is constant). In addition, the number of last edges $\LEdgeDelta(\Delta^{far}_i)$ appearing on paths $\widetilde{P}_{i,j} \in SP(s, v_i, G \setminus \{e_j\})$ that protect against failures $e_j$ in the far segment $\pi^{far}(v_i)$,  is bounded by $O(n^{1/3})$ for every clustered vertex $v_i$, and as there are $n$ vertices, in total there are $O(n^{4/3})$ edges in $\LEdgeDelta(\Delta^{far})$.
This holds since one can show that the detours of the replacement paths
for every clustered vertex $v_i$ are \emph{long}, {\em vertex disjoint}, and fully contained in the graph $G$.
Hence, the remaining pairs $(i,j)$ that should be handled in the subsequent steps such that $v_i$ is clustered and $e_j \in \pi^{mid}(v_i) $. The fact that the length of $\pi^{mid}(v_i)$ is bounded plays an important role in the subsequent steps.

The remaining set of replacement paths are handled in Stage S3. This stage consists of two substages as well. In Substage S3.1, a collection of replacement paths $\{P^*_{i,j}\}$ that satisfies
some key properties is constructed.  Then, in Substage S3.2, the algorithm employs a modified path-buying procedure, first developed by Baswana et al. \cite{BSADD10} and recently revisited by Cygan et al. \cite{CGK13}. This modified path-buying procedure heavily exploits the key properties of the paths $P^*_{i,j}$
constructed in Substage S3.1.

We now provide a detailed description of the algorithm and establish Theorem \ref{thm:add_ub}.

\subsection{S1.1: Clustering}
\label{s1:cluster}
The following fact is taken from \cite{CGK13}.
\begin{fact}[\cite{CGK13}]
\label{fc:clustering}
There is a poly-time algorithm $\CLUSTER(G, \gamma)$ that given a parameter
$\gamma \in [0,1]$ and a graph $G=(V,E)$ constructs a collection $\mathcal{C}$
of at most $n^{1-\gamma}$ vertex-disjoint clusters, each of size $n^{\gamma}$,
and a subgraph $G_{C}$ of $G$ with $O(n^{1+\gamma})$ edges,
such that
\begin{description}
\item{(1)}
for any missing edge $(u,v) \in E(G) \setminus E(G_C)$,
$u$ and $v$ belong to two different clusters, and
\item{(2)}
the diameter of each cluster (i.e., the maximum distance in $G_C$ between
any two vertices of the cluster) is at most $2$.
\end{description}
\end{fact}
Note that the clustering does not necessarily form a partition of $V$, i.e., the set $\widetilde{V}_{C}=V \setminus \bigcup_{C_i \in \mathcal{C}} C_i$ may be nonempty. However, in that case, all edges connecting vertices of $\widetilde{V}_{C}$ among themselves or to vertices of one of the clusters belong to $G_C$. Also, the clusters are not guaranteed to be connected.

Invoke the algorithm $\CLUSTER$ on $G \setminus T_0$ and $\gamma=1/3$,
getting the subgraph
$G_C=\CLUSTER(G \setminus T_0, \gamma)$ and
let $\mathcal{C}=\{C_1, \ldots, C_\kappa\}$ be the clusters of $G_C$. Define
$$\CurrSpanner_1=T_0 \cup G_C.$$
By Fact \ref{fc:clustering}, $|E(\CurrSpanner_1)|=O(n^{4/3})$ and
the number of clusters is $\kappa=O(n^{2/3})$.
For a clustered vertex $v' \not\in \widetilde{V}_{C}$, denote by
$C(v') \in \mathcal{C}$ its cluster in $G_C$.
For every cluster $C_\ell \in \mathcal{C}$, let $\LCA(C_\ell)$ be the least common ancestor of all $C_\ell$ vertices in $T_0$. Formally, $\LCA(C_{\ell})$ is the vertex $\widehat{v}$ of maximal depth satisfying that $\pi(s, \widehat{v}) \subseteq \pi(s, v')$ for every $v' \in C_{\ell}$.

\subsection{S1.2: Shortest path $\pi(s,v_i)$ segmentation}
\label{append:sp_segment}
For every clustered vertex $v \in C_k$, divide its shortest path $\pi(s, v)$
into 3 segments
$$\pi(s, v)=\pi^{far}(v) \circ \pi^{mid}(v) \circ \pi^{near}(v)$$
in the following manner.
Define $z_k \in \pi(s, \LCA(C_k))$
to be the upmost vertex on $\pi(s, \LCA(C_k))$ (closest to $s$) satisfying that $\dist(z_k, \LCA(C_k))\leq \lceil n^{2/3} \rceil$. Then, let $\pi^{far}(v)=\pi(s, z_k)$, $\pi^{mid}(v)=\pi(z_k, \LCA(C_k))$ and $\pi^{near}(v)=\pi(\LCA(C_k), v)$.
Note that for every two vertices $v, v' \in C_{\ell}$ in the same cluster it holds that $\pi^{mid}(v)=\pi^{mid}(v')$ and also $\pi^{far}(v)=\pi^{far}(v')\subseteq \pi(s, v)$.
For an illustration of the segmentation of $\pi(s, v)$, see part S1.2
of Fig.  \ref{figure:ubstages}; the vertex $\LCA(C_k)$ is used to draw the line between the near segment $\pi^{near}(v)$ and the rest of the $\pi(s, v)$ path. By the definition of $z_k \in \pi(s, \LCA(C_k))$, it holds that $|\pi^{mid}(v)| \leq \lceil n^{2/3} \rceil$.
\par Define
$\pi(C_k)=\pi(s, \LCA(C_k))$ to be the maximal shortest path segment shared by the members of cluster $C_k$.
\begin{observation}
\label{obs:path_pi_cluster}
\begin{description}
\item{(a)}
$\pi(C_k)=\pi^{far}(v) \circ \pi^{mid}(v)$ for every $v \in C_k$.
\item{(b)}
$\dist(v', v, G)\geq \lceil n^{2/3} \rceil$ for every $v' \in \pi^{far}(v)$.
\end{description}
\end{observation}

Note that $\pi^{far}(v)$ is the same for every $v \in C_k$ and also $\pi^{mid}(v)$ is the same for every $v \in C_k$, therefore the common shortest-path section $\pi(C_k)$ can be divided into two segments $\pi^{far}(C_k)$ and $\pi^{mid}(C_k)$ such that
\begin{eqnarray*}
\pi^{mid}(C_k) &=& \pi^{mid}(v) \mbox{~for every~} v \in C_k~, \mbox{~and~}
\pi^{far}(C_k) = \pi^{far}(v) \mbox{~for every~} v \in C_k~.
\end{eqnarray*}

\subsection{S2: Handling near and far edge faults}
\label{s2:nearfar}
To protect against the edge failures occurring in the \emph{near} and \emph{far}
shortest-path segments, two sparse collections of new edges are added to $H$,
namely, $E^{near}$ and $E^{far}$.

To guarantee that $E^{near}$ and $E^{far}$ are sufficiently sparse,
the replacement paths are required to be \emph{nice} as defined below.
For a collection of replacement paths
$\mathcal{Q}=\{Q^*_{i,j} \in SP(s, v_i, G \setminus \{e_j\})\}$,
let $b_{i,j}$ be the first divergence point of
$Q^*_{i,j} \in \mathcal{Q}$ from $\pi(s, v_i)$
(i.e., the first vertex on $Q^*_{i,j}$ such that
the vertex $v'$ appearing after it on $Q^*_{i,j}$ is not on $\pi(s, v_i)$).
For every  clustered vertex $v_i$, define
\begin{eqnarray*}
E^{near}_i &=&
\{\LastE(Q^*_{i,j}) \notin \CurrSpanner_1 \mid e_j \in \pi^{near}(v_i) \}
\mbox{~~and~~}
E^{far}_i =
\{\LastE(Q^*_{i,j}) \notin \CurrSpanner_1 \mid e_j \in \pi^{far}(v_i)\}~.
\end{eqnarray*}

\begin{Definition}
\label{def:nice}
A collection $\mathcal{Q}=\{Q^*_{i,j}\}$ of replacement paths is \emph{nice}
if it satisfies the following properties:
\item{(N1)}
$|E^{near}_i|\leq 5$,
\item{(N2)}
$|E^{far}_i| \leq \lceil n^{1/3} \rceil$.
\end{Definition}
\def\APPENDNICE{
\paragraph{S2.1: The nice collection $\mathcal{Q}$ of replacement paths.}
Our starting point for constructing the nice collection
$\mathcal{Q}=\{Q^*_{i,j} \in SP(s, v_i, G \setminus \{e_j\})\}$ is the graph
$\CurrSpanner_1=G_C \cup T_0$.
Note that by Fact \ref{fc:clustering}, the edges in $G \setminus \CurrSpanner_1$ are those incident to clustered vertices.
A path $Q^*_{i,j} \in SP(s, v_i, G \setminus \{e_j\})$ is \emph{cute} if its divergence point $b_{i,j}$ is unique (i.e., $Q^*_{i,j}[b_{i,j}, v_i]$ and $\pi(b_{i,j}, v_i)$ are edge disjoint).
The following algorithm constructs a cute collection of replacement paths which are also shown to be nice.
\paragraph{Algorithm \mbox{\tt Qcons} for $Q^*_{i,j}$ construction.}
For every vertex-edge pair $(i,j)$ such that $v_i \in V(G)$ and $e_j \in \pi(s, v_i)$, define $Q^*_{i,j} \in SP(s, v_{i}, G \setminus \{e_j\})$ in the following manner.
First, the replacement path is classified according to whether or not it \emph{must} be missing ending (ending with an edge that is not in $\CurrSpanner_1$).
To do that, the algorithm checks if there exists a replacement path which is not missing ending and sets $Q^*_{i,j}$ accordingly.
This is done as follows. Let $E'=E(v_i, G\setminus E(\CurrSpanner_1))$ be the new edges incident to $v_i$ in $G$.
\\
Case (a): $\dist(s, v_i, G \setminus \left(E' \cup \{e_j\} \right))=\dist(s, v_i, G \setminus \{e_j\})$.
In this case, there is a replacement path which is not missing-ending, and the algorithm takes one such path $Q^*_{i,j} \in SP(s, v_{i}, G \setminus \left(E' \cup \{e_j\}\right)))$.
\\
Case (b): $\dist(s, v_i, G \setminus \left(E' \cup \{e_j\} \right))>\dist(s, v_i, G \setminus \{e_j\})$.
In this case, the chosen replacement path must be missing-ending. The algorithm attempts to select a \emph{cute}
replacement whose divergence point is highest. Let
$\mathcal{U}=\{P \in SP(s, v_i, G \setminus \{e_{j}\}) ~\mid~ P \mbox{~is cute}\}$ be the collection of $s-v_i$ replacement paths which are cute. In the analysis section, we show that $\mathcal{U}$ is nonempty. For every cute path $P_{\ell} \in \mathcal{U}$ with unique divergence point $b_{\ell}$, let the cost of the path be the depth (distance from $s$ in $T_0$) of $b_{\ell}$, i.e., $\Cost(P)=\depth(b_{\ell})$. Let $Q^*_{i,j} \in \mathcal{U}$, be the cute path of minimum cost. I.e., $\Cost(Q^*_{i,j})=\min\{\Cost(P), P \in \mathcal{U}\}$.

\paragraph{Analysis of the $Q^*_{i,j}$ paths.}
In this section we prove the following.
\begin{lemma}
\label{q_proper}
$\mathcal{Q}=\{Q^*_{i,j}\}$ is nice.
\end{lemma}
We begin by proving correctness.
\begin{claim}
\label{cl:q_correct}
For every $i,j$,
\begin{description}
\item{(a)}
$Q^*_{i,j} \in SP(s, v_i, G \setminus \{e_j\})$, \\
\item{(b)}
if $\LastE(Q^*_{i,j})\notin T_0$ then
$\dist(s, v_i, G \setminus \left(E' \cup \{e_j\} \right)) >
\dist(s, v_i, G \setminus \{e_j\})$ for $E'=E(v_i,G)\setminus T_0$.
\end{description}
\end{claim}
\Proof
If the path $Q^*_{i,j}$ is chosen by Algorithm \mbox{\tt Qcons} according to case (a), then it is not a missing-ending and the lemma follows trivially. Note that Part (b) is also immediate by the construction.
It remains to prove Claim (a) for paths $Q^*_{i,j}$ chosen according to Case (b), i.e., missing-ending $Q^*_{i,j}$ paths. In fact, since Algorithm \mbox{\tt Qcons} chooses the cute $s-v_i$ replacement path whose divergence point is closest to $s$, it is sufficient to show that the set of cute paths $\mathcal{U} \subseteq SP(s, v_i, G \setminus \{e_j\})$ is nonempty.
To do so, we exhibit at least one such path $P'$ in this set. Let $P_1\in SP(s, v_i, G \setminus \{e_j\})$ be an arbitrary replacement path. $P_1$ is converted into a \emph{cute} path $P'\in \mathcal{U}$.
Let $w \in \left( \pi(s, v_i) \cap P_1 \right) \setminus \{v_i\}$ be the last mutual point of $P_1$ and $\pi(s, v_i)$ which is not $v_i$. Define $P'=\pi(s, w) \circ P_1[w, v_i]$.  Clearly, $|P'| =|P_1|$, so it remains to show that $P' \subseteq \left( G \setminus \{e_j\} \right)$ and in particular it is sufficient to show that $e_j \notin \pi(s, w)$.
Let $\widehat{Q}_1=\pi(s,w)$ and $\widehat{Q}_2=\pi(w, v_i)$. We now claim  that $e_j \in \widehat{Q}_2$.
Assume, towards contradiction that $e_j \notin \widehat{Q}_2$.
This implies that the path $Q'=P_2[s, w] \circ \widehat{Q}_2$ satisfies $|Q'|=|P_2|$ and $Q' \subseteq G \setminus \{e_j\}$, hence $Q' \in SP(s, v_i, G \setminus \{e_j\})$ and it is \emph{not} missing-ending,
in contradiction to the fact that $Q^*_{i,j}$ was chosen according to Case (b). Hence, $e_j \in \widehat{Q}_3$ and $P' \subseteq G \setminus \{e_j\}$ is cute.
\QED
We next provide several preliminary claims.
\begin{claim}
\label{cl:order}
If $Q^{*}_{i,k}$ and $Q^{*}_{i,k'}$ are the missing-ending replacement paths
chosen by Algorithm \mbox{\tt Qcons} and $|Q^{*}_{i,k}|<|Q^{*}_{i,k'}|$,
then $e_{k}$ is below $e_{k'}$ on $\pi(s, v_i)$.
\end{claim}
\Proof
If $e_k$ is above $e_{k'}$, then $Q^{*}_{i,k} \in G \setminus \{e_k'\}$ so it can serve as a replacement path for $v_i$ and $e_{k'}$ as well, in contradiction to Cl. \ref{cl:q_correct}, by which $Q^*_{i,k'}$ is an $s-v_i$ shortest path in $G \setminus \{e_k'\}$.
\QED
\begin{claim}
\label{cl:same}
If $Q^{*}_{i,k}$ and $Q^{*}_{i,k'}$ are the missing-ending replacement paths
chosen by Algorithm \mbox{\tt Qcons} and  $|Q^*_{i,k}|=|Q^*_{i,k'}|$,
then $Q^*_{i,k}=Q^*_{i,k'}$.
\end{claim}
\Proof
Without loss of generality, let $e_{k}$ be above $e_{k'}$ on $\pi(s, v_i)$ and let $b_k$ (resp., $b_{k'}$) be the unique divergence point of $\pi(s, v_i)$ and $Q^*_{i,k}$ (resp., $Q^*_{i,k'}$). By construction, there exists such unique divergence point.  Let $b_{\ell}$ for  $\ell\in \{k, k'\}$ be the divergence point that is closer to $s$. Since $b_{k}$ must appear on $\pi(s, v_i)$ above $e_{k}$, it holds that $b_{\ell}$ appears above $e_k$ as well and hence that $Q^*_{i,\ell} \in G \setminus \{e_k, e_{k'}\}$.
In addition, since $|Q^*_{i,\ell}|=|Q^*_{i,k}|=|Q^*_{i,k'}|$ it holds that $Q^*_{i,\ell} \in SP(s, v_i, G \setminus \{e_k, e_{k'}\})$. By Algorithm \mbox{\tt Qcons} definition, it holds that $b_{k'}=b_{k}$ and by the uniqueness of the shortest-path under $W$ it holds that $Q^*_{i,k}=Q^*_{i,k'}$.
\QED
We now establish the niceness of $\mathcal{Q}$. Let us first prove property (N1) of Def. \ref{def:nice}.
\begin{lemma}
\label{lem:close_bound}
$|E^{near}_i| \leq 5$.
\end{lemma}
\Proof
Assume towards contradiction that there exists some
$v_i$ such that $|E^{near}_i| \geq 6$.
Let us consider 6 specific edges $e'_1, \ldots, e'_6 \in E^{near}_i$. Recall that $v_i$ must be clustered and
let $Q^*_{i,1}, ...,Q^*_{i,5}, Q^*_{i,6}$ be replacement paths whose last edge is $e'_1, \ldots, e'_6$ respectively.
It then holds that $e_j \in \pi^{near}(v_i)=\pi(w, v_i)$ where $w=\LCA(C(v_i))$ is the least common ancestor of $v_i$'s cluster $C(v_i)$.
Since $\LastE(Q^*_{i,j_1}) \neq \LastE(Q^*_{i,j_2})$ for every $j_1, j_2 \in \{1, \ldots, 6\}$, by Cl.
\ref{cl:same}, the paths $Q^*_{i,j}$ are of distinct lengths for every $j \in \{1, \ldots, 6\}$. Without loss of generality, assume that
$|Q^*_{i,1}|< ...<|Q^*_{i,6}|$.
We therefore have that
\begin{equation}
\label{eq:c1}
|Q^*_{i,6}|>|Q^*_{i,1}|+4~.
\end{equation}
In addition by Cl. \ref{cl:order}, it holds that the edges $e_j$ are sorted in decreasing distance from $s$, i.e., $\dist(s, e_1)>\ldots> \dist(s, e_6)$. Since $e_1, \ldots, e_6 \in \pi(w, v_i)$ appear strictly below $w$, it holds that there exists at least one vertex $v' \in C(v_i)$ such that
$e_6=(x,y)\notin \pi(s, v')$ and hence also $e'_1, \ldots, e'_5 \notin \pi(s, v')$. See Fig. \ref{fig:closeclust} for an illustration. This holds since otherwise, if $e_6$ belongs to $\pi(s, v'')$ for every $v'' \in C(v_i)$ then the vertex $y$ is a common ancestor of the cluster vertices $C(v_i)$ and it is deeper then $w$, in contradiction to the fact that $w$ is the least common ancestor).
Hence by the cluster diameter property of Fact \ref{fc:clustering}, it holds that
\begin{eqnarray*}
|Q^*_{i,6}|&=& \dist(s, v_i, G \setminus \{e_6\})
\leq
\dist(s, v', G \setminus \{e_6\})+2=|\pi(s, v')|+2
\\&= &
\dist(s, v', G \setminus \{e_1\})+2
\leq
\dist(s, v_i, G \setminus \{e_1\})+4=
|Q^*_{i,1}|+4~,
\end{eqnarray*}
which is in contradiction with Eq. (\ref{eq:c1}). The lemma follows.
\QED

\begin{figure}[htbp]
\begin{center}
\includegraphics[scale=0.35]{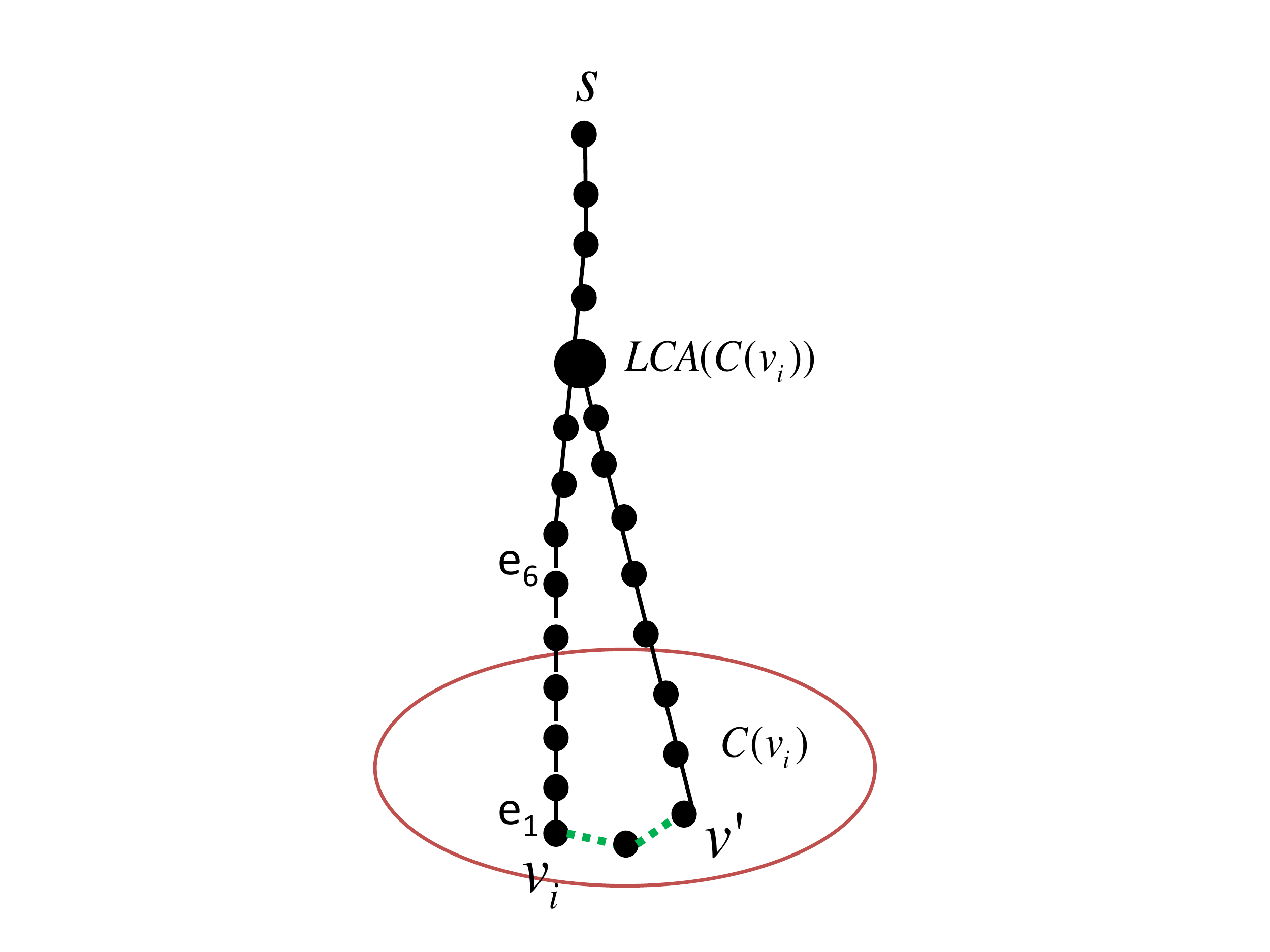}
\caption{ The set of $E^{near}_i$ must contain at most $6$ edges. Drawn are the edges $e_1, \ldots, e_6$ whose corresponding last edge of $Q^*_{i,1},...Q^*_{1,6}$ are in $E^{near}_i$. The edges of the cluster graph $G_C$ are dotted. The vertices $v_i$ and $v'$ are in the same cluster hence of distance 2 in $G_C$.  \label{fig:closeclust}}
\end{center}
\end{figure}


Finally, we turn to establish Property (N2) of Def. \ref{def:nice} and thus establish Lemma \ref{q_proper}.
For every path $P$, let $\Cost_{W}(P)=\sum_{e \in P} W(e)$.
\begin{claim}
\label{cl:far}
$|E^{far}_i|\leq \lceil n^{1/3} \rceil$.
\end{claim}
\Proof
We first claim that for every two missing-ending paths $Q^*_{i,j}$ and $Q^*_{i,j'}$ such that $\LastE(Q^*_{i,j}) \neq \LastE(Q^*_{i,j'}) \in E^{far}_i$, it holds that their detours $Q_1=Q^*_{i,j}[b_{i,j}, v_i]$ and $Q_2=Q^*_{i,j'}[b_{i,j'}, v_i]$ are vertex disjoint except for their common endpoint $v_i$. Assume towards contradiction that there exists some mutual point $w \in Q_1 \cap Q_2 \setminus \{v_i\}$ in the intersection.
Since $Q_\ell$ and $\pi(s, v_i)$ are edge disjoint for $\ell \in \{1,2\}$, we have that there are two alternative $w-v_i$ paths in $G\setminus \{e_j, e_{j'}\}$, namely $Q_1[w, v_i]$ and $Q_2[w, v_i]$.
By the optimality of $Q_1 \in SP(b_{i,j}, v_i, G \setminus \pi(s, v_i),W)$, we have that $\Cost_{W}(Q_1[w, v_i])<\Cost_{W}(Q_2[w, v_i])$. Similarly, by the optimality of $Q_2 \in SP(b_{i,j'}, v_i, G \setminus \pi(s, v_i),W)$, we have that $\Cost_{W}(Q_2[w, v_i])<\Cost_{W}(Q_1[w, v_i])$; contradiction. It follows that $Q_1$ and $Q_2$ are vertex disjoint.
We next claim that it also holds that $|Q^*_{i,j}[b_{i,j}, v_i]|\geq \lceil n^{2/3}\rceil$.
To see this, note that since $Q^*_{i,j}$ protects the fault of an edge $e_j \in \pi^{far}(v_i)$ and the unique divergence point $b_{i,j}$ must appear above $e_j$, it holds that $b_{i,j} \in \pi^{far}(v_i)$ as well. Hence, $|Q^*_{i,j}[b_{i,j}, v_i]| \geq \dist(b_{i,j}, v_i, G)>\lceil n^{2/3} \rceil$ where the last inequality follows from Obs. \ref{obs:path_pi_cluster}(b).
\par Assume towards contradiction that
$|E^{far}_i|> \lceil n^{1/3} \rceil$. Then there are $\lceil n^{1/3} \rceil$ vertex disjoint paths in $G$ each of length at least $\lceil n^{2/3} \rceil$. Overall the number of vertices in those paths is greater than $|V(G)|=n$, contradiction.
\QED
We conclude this section with the following immediate observation.
\begin{observation}
\label{obs:close_far}
For every clustered vertex $v_i\in V(G)$ and edge $e_j \in \pi(s, v_i)$, if $\LastE(Q^*_{i,j}) \notin E^{near}_i \cup E^{far}_i$ then $e_j \in \pi^{mid}(v_i)=\pi^{mid}(C(v_i))$.
\end{observation}
}
\APPENDNICE

\paragraph{S2.2: Creating $\CurrSpanner_2$.}
Having the nice collection of replacement paths $\mathcal{Q}$,
let $IH = \{ i \mid v_i \in V(G)\}$.
The current spanner of this step is given by
$$\CurrSpanner_2 = \CurrSpanner_1 \cup
\bigcup_{i\in IH} \left( E^{near}_i \cup E^{far}_i \right).$$
Note that by Definition \ref{def:nice}, $|\CurrSpanner_2|=O(n^{4/3})$.

\subsection{S3: Handling mid edge faults}
\label{sec:pathbuyingproc}
\def\APPENDPINTRO{
At this point, we handled faults of edges $e_j \in \pi(s, v_i)$ except those
occurring on the middle sections $\pi^{mid}(v_i)$. Unfortunately, those cannot
be handled by adding the last edges of all missing-ending paths $Q^*_{i,j}$,
as there may be too many such missing edges.
Instead, we would like to ``aggregate" the remaining problems,
and handle them by adding only \emph{some} of the missing edges relying on
the properties of the clustering to provide approximately shortest replacement
paths whenever the optimal path was not included.
One property that could have helped this aggregation process is
\emph{prefix consistency}. The replacement paths $P^*_{i,j}$ and $P^*_{i',j}$ for
$v_{i'} \in P^*_{i,j}$ are prefix-consistent if $P^*_{i',j}=P^*_{i,j}[s, v_{i'}]$.
The advantage of this property is that by adding (missing) last edges of
$P^*_{i',j}$, we also help the longer path $P^*_{i,j}$,
see Fig. \ref{fig:prefix_cons}.
Unfortunately, our construction of the path collection $\mathcal{Q}$ does not
guarantee this property. This is the main motivation for the next step in the
algorithm, where we replace the paths $\mathcal{Q}$ by a new path collection $P^*_{i,j}$ which is somewhat closer to achieving this property (albeit it falls short of doing that.)

\begin{figure}[htbp]
\begin{center}
\includegraphics[scale=0.35]{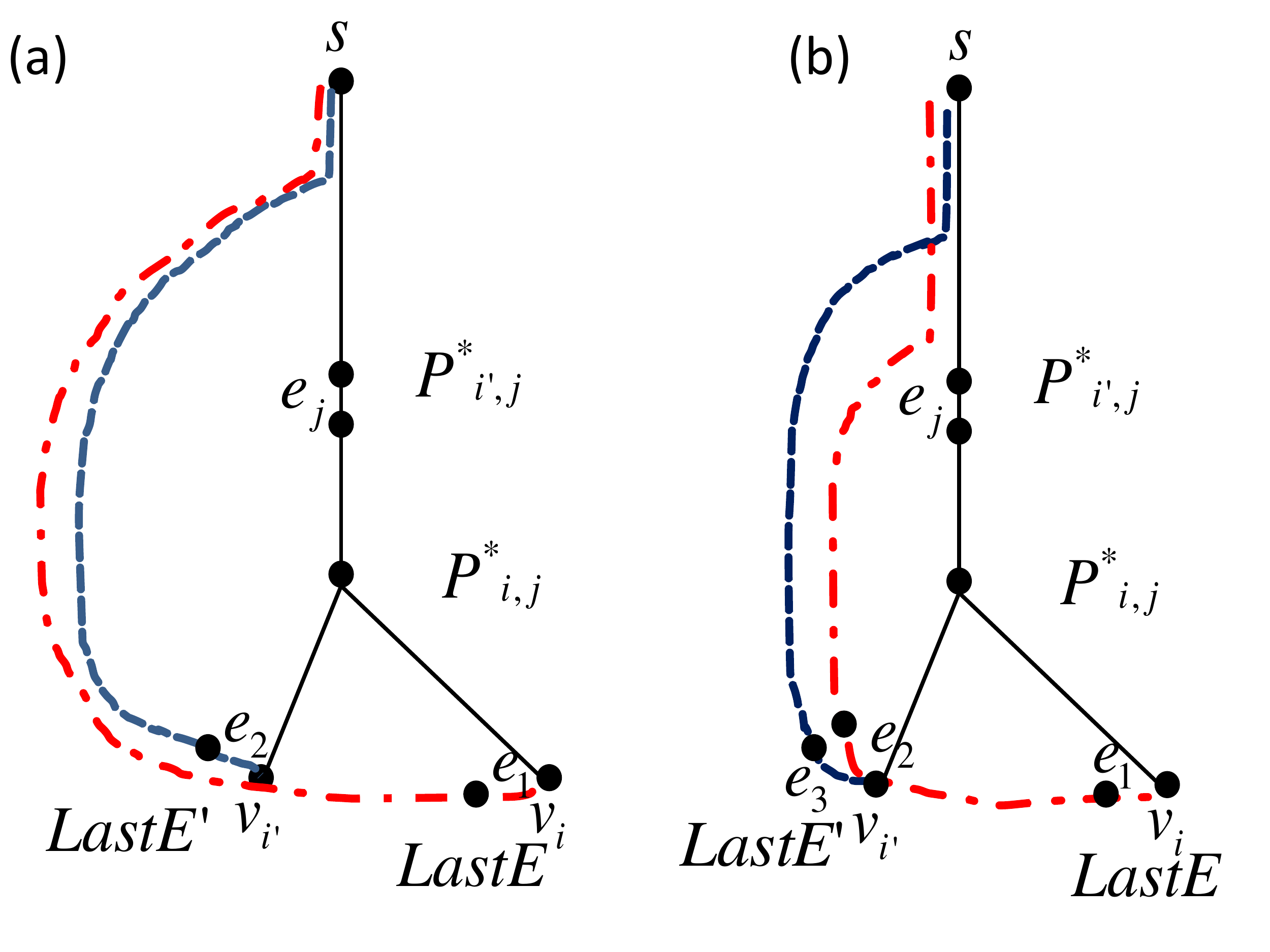}
\caption{ Illustration of the prefix consistency property. (a) The paths $P^*_{i,j}$ and $P^*_{i',j}$ are prefix consistent since $P^*_{i,j}[s, v_{i'}]=P^*_{i',j}$. (b) The paths $P^*_{i,j}$ and $P^*_{i',j}$ are not prefix consistent since $P^*_{i,j}[s, v_{i'}] \neq P^*_{i',j}$ \label{fig:prefix_cons}}
\end{center}
\end{figure}
}
\APPENDPINTRO

%
\def\APPENDNEWREP{
We first describe the algorithm and then prove that these replacement paths satisfy some important properties which are crucial for the efficiency of the subsequent path-buying procedure.

\paragraph{S3.1: Algorithm \mbox{\tt Pcons} for Constructing $P^*_{i,j}$.}
Consider the edges of $T_0$ in \emph{nonincreasing} distance from $s$. For edge $e_j \in T_0$, let the set of vertices sensitive to the failure of $e_j$, $\Sensitive(e_j)=\{v_1, \ldots, v_k\}$, be sorted in \emph{nondecreasing} distance from $s$ in $G \setminus \{e_j\}$.
I.e., $\dist(s, v_1, G \setminus \{e_j\})\leq \ldots \leq \dist(s, v_k, G \setminus \{e_j\})$.
For every $e_j$, the algorithm constructs the replacement path $P^*_{i,j}$ for vertices $v_i \in \Sensitive(e_j)$ according to this order.
Initially set $P^{all}_i=\emptyset$ and $P^*_{i,j}=\emptyset$ for every $v_i \in V(G)$ and $e_j \in \pi(s, v_i)$. The algorithm constructs two intermediate replacement paths,
$\RepOne_{i,j}$ and $\RepTwo_{i,j}$. The final replacement path $P^*_{i,j}$ is obtained from $\RepTwo_{i,j}$.
\par For vertex-edge pair $(i,j)$, let $v_{\ell}$ be the neighbor of $v_i$ on $Q^*_{i,j}$, i.e., $(v_{\ell}, v_i)=\LastE(Q^*_{i,j})$ and do the following.
Consider the following cases.
\\
(R1.1) If $e_j \notin \pi(s, v_{\ell})$, define $\RepOne_{i,j}=\pi(s, v_{\ell})\circ (v_{\ell}, v_i)$. Note that by construction, $\RepOne_{i,j}$ and $Q^*_{i,j}$
share the same last edge.
\\
(R1.2) If $e_j \in \pi(s, v_{\ell})$ (i.e., $v_{\ell} \in \Sensitive(e_j)$) let $\RepOne_{i,j}=P^*_{\ell,j}\circ (v_{\ell}, v_i)$. (Note that $P^*_{\ell,j}$ was already constructed, since $v_{\ell}$ was handled before $v_i$ in our ordering.)
Also note that the path collection $\RepOne_{i,j}$ does achieve the desired property of prefix consistency. Unfortunately, this set no longer enjoys another desirable property that possessed by the original collection $\mathcal{Q}$, namely, the uniqueness of the divergence points $b_{i,j}$. This is why we next modify the paths further, defining the paths $\RepTwo_{i,j}$ which restores the uniqueness property while possibly destroying prefix consistency again.
\\
(R2.1) If $(v_{\ell}, v_i) \in \CurrSpanner_2$ is \emph{not missing},
then let $P^*_{i,j}=\RepTwo_{i,j}=\RepOne_{i,j}$ and proceed to the next vertex-edge pair.
\\
(R2.2) Else, $\LastE(\RepOne_{i,j})=(v_{\ell}, v_i) \notin \CurrSpanner_2$. (Note that this means that $\LastE(\RepOne_{i,j}) \notin E^{near} \cup E^{far}$, and hence by Obs. \ref{obs:close_far},
in this case $v_i$ is clustered and $e_j \in \pi^{mid}(C(v_i))$.)
\par Let $b$ be the first divergence point of $\RepOne_{i,j}$ and $\pi(s, v_i)$. If $\RepOne_{i,j}[b, v_i]$ and $\pi(b, v_i)$ are not edge disjoint ($b$ is not a unique divergence point) then let $w \in \left(\pi(b, v_i) \cap   \RepOne_{i,j}[b, v_i]\right) \setminus \{v_i\}$ be the last mutual vertex in $\pi(b, v_i)$ and $\RepOne_{i,j}[b, v_i]$ which is not $v_i$. Our goal now is to make $w$ the unique divergence point with $\pi(s, v_i)$.\\
(R3) Defining $\RepTwo_{i,j}$: $\RepTwo_{i,j}=\pi(s, w) \circ \RepOne_{i,j}[w, v_i]$, making $b_{i,j}=w$ the unique divergence point of $\RepTwo_{i,j}$ and $\pi(s, v_i)$.
\\(R4) Defining $P^*_{i,j}$: (R4.1) If $b_{i,j} \in \pi^{mid}(v_i)$, then set $P^*_{i,j}=\RepTwo_{i,j}$ as the final replacement path. 
(R4.2) Otherwise, if $b_{i,j} \notin \pi^{mid}(v_i)$, do the following. First, if $P^{all}_i=\emptyset$, set $P^{all}_i \gets \RepTwo_{i,j}$. Finally, let $P^*_{i,j} \gets P^{all}_i$.
This completes the description of the algorithm.

Let
$$\CurrSpanner_3=\CurrSpanner_2 \cup \bigcup\{\LastE(P^{all}_i) ~\mid~ v_i \in V(G)\}.$$
The subgraph $\CurrSpanner_3$ contains, in addition to the edges of $\CurrSpanner_2$, also the last edges of $P^{all}_i$ for every $v_i \in V(G)$. Since each such vertex contributes at most one edge from $P^{all}_i$ to $\CurrSpanner_3$, at most $n$ edges are added, hence $|\CurrSpanner_3|=O(n^{4/3})$.

\paragraph{Analysis of $P^*_{i,j}$ paths.}
We begin by showing that the constructed replacement paths $P^*_{i,j} \in SP(s, v_i, G \setminus \{e_j\})$ are of optimal lengths.
\begin{lemma}
\label{lem:correct}
For every $v_i \in V(G)$ and $e_j \in \pi(s, v_i)$, consider $P^*_{i,j}$ and let $V^+_{i,j}$ be the set of vertices appearing on $P^*_{i,j}$ after the first new edge $e=\FirstEN(P^*_{i,j}) \notin T_0$. Then
\begin{description}
\item{(a)}
$P^*_{i,j} \in SP(s, v_i, G \setminus \{e_j\})$ and
\item{(b)}
$V^+_{i,j} \subseteq \Sensitive(e_j)$.
\end{description}
\end{lemma}
\Proof
We prove this by induction on the iteration in which $P^*_{i,j}$ was constructed by the algorithm. For the induction base, consider the first iteration, when $P^*_{1,1}$ is constructed. By the ordering, $e_1$ is the last edge of some source $s$ to leaf $v$ path, i.e., $e_1=\LastE(\pi(s, v))$, and $v_1$ is the first vertex in the ordered set $\Sensitive(e_1)$.
Let $(v_{\ell}, v_1) =\LastE(Q^*_{i,1})$.
We first prove (a) and show that
$P^*_{1,1} \in SP(s, v_1, G \setminus \{e_1\})$.
By Cl. \ref{cl:q_correct} we have that $Q^*_{1,1}\in SP(s, v_1, G \setminus \{e_1\})$ and therefore $\dist(s, v_{\ell}, G \setminus \{e_\ell\})<\dist(s, v_{1}, G \setminus \{e_1\})$. As $v_1$ is the first vertex in the ordering of $\Sensitive(e_1)$ it implies that $v_{\ell} \notin \Sensitive(e_1)$. Since $\pi(s,v_{\ell})$ has at most one divergence point $b_{i,j}$ with $\pi(s, v_i)$ (in particular, $b_{i,j}=\LCA(v_i, v_{\ell})$), it holds that $\RepTwo_{1,1} =\RepOne_{1,1}$ and since this is that first iteration $P^{all}_i$ was not previously defined. Therefore, $P^*_{1,1}=\RepOne_{1,1}=\RepTwo_{1,1}=\pi(s, v_{\ell})\circ (v_{\ell}, v_1)$ is such that $|P^*_{1,1}|=|Q^*_{1,1}|$ and $P^*_{1,1} \subseteq G \setminus \{e_1\}$. Claim (a) is established. Consider Claim (b). By the definition of $P^*_{1,1}$, it holds that $P^*_{1,1} \setminus \{(v_{\ell},v_1)\} \subseteq T_0$,
hence $V^+_{1,1}=\{v_1\}$,
and since $v_1 \in \Sensitive(e_1)$, the induction base holds.

Assume Claims (a) and (b) of the lemma hold for every replacement path
$P^*_{i',j'}$ constructed up to iteration $t-1$ and
consider the path $P^*_{i,j}$ constructed at iteration $t$.
Let $(v_{\ell},v_{i})=\LastE(P^*_{i,j})$.
We first establish the lemma for $\RepOne_{i,j}$, then for $\RepTwo_{i,j}$,
and finally consider the case where $P^*_{i,j} \neq \RepTwo_{i,j}$,
i.e., where $P^*_{i,j}=P^{all}_i$.
If $v_{\ell} \notin \Sensitive(e_j)$, then Algorithm \mbox{\tt Pcons}
again yields
$P^*_{i,j}=\RepTwo_{i,j}=\RepOne_{i,j}=\pi(s, v_{\ell}) \circ (v_{\ell},v_{i})$,
and similarly to the induction base, by Cl.  \ref{cl:q_correct},
we have that $P^*_{i,j} \in SP(s, v_i, G \setminus \{e_j\})$, yielding Claim (a),
and since $P^*_{i,j} \setminus \{\LastE(P^*_{i,j})\} \subseteq T_0$,
Claim (b) holds as well.
For the rest of the proof, it remains to consider the case where
$v_{\ell} \in \Sensitive(e_j)$. By Cl. \ref{cl:q_correct},
$Q^*_{i,j} \in SP(s, v_i, G \setminus \{e_j\})$, hence
$\dist(s, v_{\ell}, G \setminus \{e_j\})<\dist(s, v_{i}, G \setminus \{e_j\})$ and by the ordering of $\Sensitive(e_j)$, the pair $(\ell,j)$ was considered at iteration $t' <t$ and the induction assumption for $P^*_{\ell,j}$ can be applied.
We then have that $\RepOne_{i,j}=P^*_{\ell,j} \circ \LastE(Q^*_{i,j}) \in SP(s, v_i, G \setminus \{e_j\})$, which satisfies Claim (a).

We now show that Claim (b) holds for the path $\RepOne_{i,j}$.
Note that since $v_{\ell} \in \Sensitive(e_j)$, the path $P^*_{\ell,j}$ must contain a new edge, so
$\FirstEN(\RepOne_{i,j})=\FirstEN(P^*_{\ell,j})$.
Then, by the induction assumption for $P^*_{\ell,j}$ and the fact that $v_i \in \Sensitive(e_j)$, we have that $V(\RepOne_{i,j}[y, v_i]) \subseteq \Sensitive(e_j)$ where $\FirstEN(P^*_{\ell,j})=(x,y)$.
Hence, so far, the lemma holds for $\RepOne_{i,j}$.
It remains to consider the cases where $P^*_{i,j}\neq  \RepOne_{i,j}$, and hence $\LastE(\RepOne_{i,j}) \notin \CurrSpanner_2$.

We next show that the lemma holds for $\RepTwo_{i,j}$.
Since $\LastE(\RepOne_{i,j})=\LastE(Q^*_{i,j})=(v_{\ell}, v_i) \notin T_0$,
Cl. \ref{cl:q_correct}(b) implies that \emph{any} $s-v_i$ replacement path
in $G \setminus \{e_j\}$ must be \emph{missing-ending}.
Let $b$ be the first divergence point of $\RepOne_{i,j}$ and $\pi(s, v_i)$.
If $\RepTwo_{i,j} \neq \RepOne_{i,j}$ then necessarily there exists
another mutual point
$w \in \left(\pi(b, v_i)\cap\RepOne_{i,j}[b, v_i] \right)\setminus \{b, v_{i}\}$
such that $\RepTwo_{i,j}=\pi(s, w) \circ \RepOne_{i,j}[w, v_i]$.
Consider Claim (a). Since $\RepOne_{i,j} \subseteq G \setminus \{e_j\}$ and  $|\RepTwo_{i,j}| \leq |\RepOne_{i,j}|$, to establish Claim (a), it is sufficient to show that $e_j \notin \pi(s, w)$.
\begin{observation}
$e_j \in \pi(w, v_i)$.
\end{observation}
\Proof
Assume, towards contradiction, that $e_j \notin \pi(w, v_i)$.
This implies that the path $P'=\RepOne_{i,j}[s, w] \circ \pi(w, v_i)$
satisfies $|P'|=|\RepOne_{i,j}|$ and $P' \subseteq G \setminus \{e_j\}$,
hence $P' \in SP(s, v_i, G \setminus \{e_j\})$ and it is \emph{not}
missing-ending as $\LastE(P') \in E(T_0)$,
contradicting Cl. \ref{cl:q_correct}(b).
\QED
Hence, $e_j \notin \pi(w, v_i)$, concluding that $\RepTwo_{i,j}\in SP(s, v_i, G \setminus \{e_j\})$ as required, so Claim (a) of the lemma holds for $\RepTwo_{i,j}$. Consider Claim (b).
The prefix $\RepTwo_{i,j}[s,w]$ of $\RepTwo_{i,j}$ is entirely in $T_0$ (since it is equal to $\pi(s,w)$), so $\FirstEN(\RepTwo_{i,j})$ must occur in $\RepTwo_{i,j}[w,v_i]$.
Since $\RepTwo_{i,j}[w,v_i]=\RepOne_{i,j}[w,v_i]$, the validity of Claim (b) for $\RepOne_{i,j}$ implies that Claim (b) holds for $\RepTwo_{i,j}$ as well.
Hence if $P^*_{i,j} = \RepTwo_{i,j}$ then we are done.

It remains to consider the case where $P^*_{i,j}\neq\RepTwo_{i,j}$ or in other words $P^*_{i,j}=P^{all}_i$. Let $t'<t$ be the iteration in which $P^{all}_i$ was defined and let $(i,j')$ be the pair considered at iteration $t'$, hence $P^{all}_i=P^*_{i,j'}=\RepTwo_{i,j'}$. By the induction assumption for $t'<t$, Claims (a) and (b) hold for $P^*_{i,j'}$.
Note that in this case, it holds that $b_{i,j}$ (resp., $b_{i,j'}$), the unique divergence point of $\RepTwo_{i,j}$ (resp., $\RepTwo_{i,j'}$) and $\pi(s, v_i)$, is not in $\pi^{mid}(C(v_i))$.
\begin{observation}
\label{obs:dp_above}
$b_{i,j}, b_{i,j'} \in \pi^{far}(v_i)$.
\end{observation}
\Proof
By the structure of the algorithm (step (R2.2)), it holds that  $\LastE(Q^*_{i,j}), \LastE(Q^*_{i,j'}) \notin \CurrSpanner_2$ and hence by Obs. \ref{obs:close_far} it holds that
\begin{equation}
\label{eq:cor_p}
e_j, e_{j'} \in \pi^{mid}(v_i)~.
\end{equation}
By the correctness established for the paths $\RepTwo_{i,j'}$ and $\RepTwo_{i,j}$ (Claim (a)), we have that $\RepTwo_{i,j} \in SP(s, v_i, G \setminus \{e_j\})$ and $\RepTwo_{i,j'} \in SP(s, v_i, G \setminus \{e_{j'}\})$. Hence, the unique divergence point $b_{i,j}$ (resp.,$b_{i,j'}$) of $\RepTwo_{i,j}$ (resp., $\RepTwo_{i,j'}$) and $\pi(s, v_i)$, appears  on $\pi(s, v_i)$ above the failed edge $e_j$ (resp., $e_{j'}$). Combining Eq. (\ref{eq:cor_p}) with the fact that $b_{i,j'},b_{i,j} \notin  \pi^{mid}(v_i)$, it follows that $b_{i,j'},b_{i,j}$ appear above the vertices of $\pi^{mid}(v_i)$ hence they appear on $\pi^{far}(v_i)$.
\QED
By Claim (a) of the inductive assumption for $P^*_{i,j'}=P^{all}_i$,
it holds that $P^{all}_i \in SP(s, v_i, G \setminus \{e_{j'}\})$.
By the proof of Claim (a) of the lemma for $\RepTwo_{i,j}$, it holds that
$\RepTwo_{i,j} \in SP(s, v_i, G \setminus \{e_j\})$. Combining with
Obs. \ref{obs:dp_above}, it holds that there are two replacement paths $\RepTwo_{i,j}$ and $P^{all}_i$ in $G \setminus \{e_j, e_{j'}\}$.
By the optimality of $P^{all}_i$, it holds that $|P^{all}_i|\leq |\RepTwo_{i,j}|$, and by the optimality of $\RepTwo_{i,j}$, it holds that $|\RepTwo_{i,j}| =|P^{all}_i|$. Claim (a) of the lemma holds.
Consider Claim (b). By Claim (b) of the induction assumption for $P^*_{i,j'}$ it holds that $V^+_{i,j'} \subseteq \Sensitive(e_{j'})$. Since $e_{j'}$ is below $e_j$ (as the edges of $T_0$ are considered in non increasing distance from $s$ and $e_{j},e_{j'} \in \pi(s, v_i)$) it holds that $V^+_{i,j}=V^+_{i,j'} \subseteq \Sensitive(e_{j'}) \subseteq \Sensitive(e_{j})$. Claim (b) follows. The lemma holds.
\QED

For a missing ending replacement path $P^*_{i,j}$, let $V^i_j=\{v^i_1, \ldots, v^i_{r}=v_i\}$ be the end-vertices of missing edges on $P^*_{i,j}$, i.e., $\LastE(P^*_{i,j}[s, v^i_k]) \notin \CurrSpanner_3$ for every $v^i_k \in V^i_j$.
Hereafter, let $b_{i,j}$ be the first divergence point of $P^*_{i,j}$ and $\pi(s, v_i)$.
\begin{lemma}
\label{lem:rep_new}
The following properties hold for every $v^i_k \in V^i_j$:
\begin{description}
\item{(a)}
$e_j \in \pi^{mid}(C(v^i_k))$
(hence $e_j \in \pi(s, v')$ for every $v' \in C(v^i_k)$),
\item{(b)}
$b_{i,j}$ is the \emph{unique} divergence point of $\pi(s, v^i_k)$ and
$P^*_{i,j}[s, v^i_k]$, hence $P^*_{i,j}[b_{i,j}, v^i_k]$ and $\pi(s, v^i_k)$
are edge disjoint, and
\item{(c)}
$b_{i,j} \in \pi^{mid}(C(v^i_k))$.
\end{description}
\end{lemma}
\Proof
First note that since the vertices of $V^i_j$ have missing edges in $\CurrSpanner_3$, it follows by Fact \ref{fc:clustering}(2) that each of them is clustered.
We prove the lemma by induction on the iteration $t$ in which $P^*_{i,j}$ was constructed. For the induction base, consider $t=1$ and let $v_1$ be the first vertex in $\Sensitive(e_1)$ where $e_1$ be the last edge of some $s$ to leaf $v$ path and consider $P^*_{1,1}$. Let $(v_{\ell}, v_1)=\LastE(P^*_{1,1})$. By the induction base of Lemma \ref{lem:correct}, it holds that
\begin{equation}
\label{eq:path_base}
P^*_{1,1}=\pi(s, v_{\ell}) \circ (v_{\ell}, v_1)~.
\end{equation}
Hence the only missing edge on $P^*_{1,1}$ is at most $\LastE(P^*_{1,1})$. I.e., $V^1_1=\emptyset$ or $V^1_1=\{v_1\}$.
If $V^1_1=\emptyset$, the claim holds vacuously. So consider the case where $V^1_1=\{v_1\}$ (i.e., $\LastE(P^*_{1,1})$ is missing).
Note that $\LastE(P^*_{1,1})=\LastE(Q^*_{1,1})$. Hence, by Obs. \ref{obs:close_far}, $e_1 \in \pi^{mid}(C(v_1))$, so Claim (a) holds.

By Eq. (\ref{eq:path_base}), $b_{1,1}=\LCA(v_{\ell}, v_1)$ is the unique divergence point of $P^*_{1,1}$ and $\pi(s,v_1)$. Thus Claim (b) holds.

Consider Claim (c). Recall that in this case $V^1_1=\{v_1\}$ i.e.,
$\LastE(Q^*_{1,1}) \notin \CurrSpanner_3$.
If $b_{1,1} \notin \pi^{mid}(C(v_1))$, then by step (S4.2) of Algorithm
\mbox{\tt Pcons}, $P^*_{1,1}=P^{all}_1$.
Since $\LastE(P^{all}_1) \in \CurrSpanner_3$, we end with contradiction.
The induction base holds.

Assume the claims hold for all replacement paths constructed up to iteration $t-1$ and let $(i,j)$ be the pair considered at iteration $t$. We first prove the lemma for the case where  $P^*_{i,j}=\RepTwo_{i,j}$ and then consider the case where $P^*_{i,j}\neq \RepTwo_{i,j}$.
\par Let $(v_{\ell},v_{i})=\LastE(Q^*_{i,j})$.
If $v_{\ell} \notin \Sensitive(e_j)$, then $P^*_{i,j}=\pi(s, v_\ell) \circ (v_{\ell},v_{i})$. The correctness follows as in the induction base. Thus, it remains to consider the complementary case where $v_{\ell} \in \Sensitive(e_j)$ and by construction, $\RepOne_{i,j}=P^*_{\ell,j} \circ (v_{\ell},v_{i})$. By Cl. \ref{cl:q_correct}, $\dist(s, v_\ell, G \setminus \{e_j\})<\dist(s, v_i, G \setminus \{e_j\})$ and hence the induction assumption for $P^*_{\ell,j}$ can be applied.

Consider the vertices $V^{\ell}_j$ with missing edges on $P^*_{\ell,j}=\RepOne_{i,j} \setminus \{\LastE(\RepOne_{i,j})\}$. Claim (a) holds for $V^{\ell}_j$ by the induction assumption for $P^*_{\ell,j}$.
If $V^{i}_j=V^{\ell}_j$ we are done, else it holds that
$V^{i}_j=V^{\ell}_j \cup \{v_i\}$. Since in this case $\LastE(\RepOne_{i,j})=\LastE(Q^*_{i,j}) \notin \CurrSpanner_3$, it holds by Obs. \ref{obs:close_far} that $e_j \in \pi^{mid}(C(v_i))$. Hence, Claim (a) is established for $V^i_j$.

Consider Claim (b). By Claim (b) of the induction assumption,
the first divergence point of $P^*_{\ell,j}[s, v_k]$ and $\pi(s, v_k)$,
namely, $b_{\ell,j}$ is unique  and common for every $v_k \in V^{\ell}_j$.
Let $b$ be the first divergence point of $\RepOne_{i,j}$ and $\pi(s, v_i)$.
We first claim the following.
\begin{claim}
\label{cl:bk_first}
$b=b_{\ell,j}$.
\end{claim}
\Proof
By Claim (a) of the induction assumption for $P^*_{\ell,j}$ it holds that $e_j=(x,y) \in \pi(s, v_{k})$ for every $v_k \in V^{\ell}_j$. By definition, it also holds that $e_j \in \pi(s, v_i)$.
Hence the first divergence point $b$ satisfies $b \in P^*_{\ell,j}$. Let $b'$ be the vertex that appears
after $b$ on $P^*_{i,j}$ (hence also on $P^*_{\ell,j}$). As $b$ is a divergence point, it must hold that $b' \notin \pi(s, x)$ and therefore $b' \notin \pi(s, v_{k})$ for every $v_k \in V^{i}_j$ and in particular $b' \notin \pi(s, v_i)$. Hence $b_{\ell,j}=b$ as required. The claim follows.
\QED
Let $w \in \left(\pi(b, v_i) \cap \RepOne_{i,j}[b, v_i] \right) \setminus \{v_i\}$ be the last divergence point of $\pi(s, v_i)$ and $\RepOne_{i,j}$. If $w=b$, then $b=b_{i,j}$ is the unique divergence point for every $v_k \in V^i_j$ and Claim (b) holds. It remains to consider the case where $b_{i,j}=w \neq b$, and hence the algorithm takes $\RepTwo_{i,j}=\pi(s, w)\circ \RepOne_{i,j}[w,v_i]$. In this case, we show that $\pi[s, v_{k}]$ is edge disjoint with
$P^*_{i,j}[w, v_{k}]$ for every $v_k \in V^i_j \setminus \{v_i\}=V^{\ell}_j$. I.e., since $w$ is by the definition the unique divergence point for $v_i$ we now want to show that we did not ``ruin" this property for the ``surviving" vertices with missing edges on $P^*_{\ell,j} \cap P^*_{i,j}=P^*_{\ell,j}[w, v_{\ell}]$. Note that since $P^*_{i,j}[s,b]=\pi(s, b)$, it holds that $\depth(w)>\depth(b)$.
By Claim (b) of the induction assumption for $P^*_{\ell,j}$, it holds that for every $v_k \in V^{\ell}_j$, the paths $\pi(s, v_{k})$ and $P^*_{\ell,j}[b, v_k]$ are edge disjoint (where $b=b_{\ell,j}$). In addition, note that by the correctness of $\RepTwo_{i,j}$ in Lemma \ref{lem:correct}(a),
it holds that $w$, the unique divergence point of $\RepTwo_{i,j}$ and $\pi(s, v_i)$ appears on $\pi(s, v_i)$ above the failed edge $e_j$.
By the induction assumption for Claim (a) on $P^*_{\ell,j}$, it holds that  $e_j \in \pi(s, v_{k})$ for every $v_k \in V^{\ell}_j$. Combining the last two observations, it follows that $w \in \pi(s, v_k)$ for every $v_k \in V^{\ell}_j$.
We therefore have that $\pi(w, v_k) \subseteq \pi(s, v_{k})$ and
$P^*_{i,j}[w, v_k]=P^*_{\ell,j}[w, v_k] \subseteq P^*_{\ell,j}[b, v_k]$ are edge disjoint. Finally, since $P^*_{i,j}[s, w]=\pi(s, w)$, it holds that $P^*_{i,j}[w, v_k]$ and $\pi(s, v_k)$ are edge disjoint for every $v_k \in V^{\ell}_j$ as required. Claim (b) is established.
\par Consider Claim (c).
We first show that the claim holds for every $v_k \in V^{\ell}_j \cap P^2_{i,j}$. By induction assumption for $P^*_{\ell,j}$, we have that $b_{\ell,j} \in \pi^{mid}(C(v_{k}))$. By the proof of Claim (b), it holds that $\depth(b_{\ell,j})\leq \depth(b_{i,j})$.
In addition, by Claim (a) of the lemma, $e_j=(x',y')  \in \pi^{mid}(C(v_k))$.
Hence the entire $\pi$ segment from $b$ to $x'$ satisfies $\pi(b_{\ell,j}, x') \subseteq \pi^{mid}(C(v_k))$.
Since $b_{i,j}$ is the unique divergence point (by Claim (b)) it appears above $e_j$ but not above $b_{\ell,j}$, hence $b_{i,j} \in \pi(b_{\ell,j},x') \subseteq \pi^{mid}(C(v_k))$.
We now prove Claim (c) for the case where $V^i_j\setminus V^{\ell}_j=\{v_i\}$.
Assume towards contradiction that $b_{i,j} \notin \pi^{mid}(C(v_i))$.
Then, by construction in this case $P^*_{i,j}=P^{all}_i$. Since $\LastE(P^{all}_i) \in \CurrSpanner_3$, we end with contradiction to the fact that $v_i \in V^i_j$. Hence, the lemma follows for every $(i,j)$ such that $P^*_{i,j}=\RepTwo_{i,j}$.
\par Finally, we consider the complementary case where $P^*_{i,j}=P^{all}_i\neq \RepTwo_{i,j}$.
Let $(i,j')$ be the vertex edge pair considered when $P^{all}_i$ was first defined, thus $P^{all}_i=\RepTwo_{i,j'}=P^*_{i,j'}$. Since $P^*_{i,j'}$ was defined before $P^*_{i,j}$, the induction assumption can be applied.
Claims (b) and (c) for $P^*_{i,j}$ follow immediately by the induction assumption of Claims (b) and (c) for $P^{all}_i$. To see Claim (a), note that by the ordering of the edges, $e_{j'}=(x',y')$ must be below $e_j$ on $\pi(s, v_i)$, in addition, by Lemma \ref{lem:correct}(a) it holds that
$P^{all}_i \in SP(s, v_i, G \setminus \{e_j\})$, hence $b_{i,j'}$ appears above $e_j$. By part (c), $b_{i,j'} \in \pi^{mid}(C(v_i))$, hence $e_j \in \pi(b_{i,j'}, x') \subseteq \pi^{mid}(C(v_i))$, as required.
The lemma follows.
\QED
}
\APPENDNEWREP

\paragraph{S3.2: Path-Buying procedure.}
With the collection of replacement path $P^*_{i,j}$ at hand, we are now ready to present the last step of the algorithm, a modified path-buying procedure, where the replacement path is added entirely to the spanner if it satisfies a particular cost to value balance.

Generally speaking, the high level approach of the path-buying technique is as follows. Recall that in the preliminary clustering sub-stage S1, the graph was condensed into clusters.
There is a collection of $s_i-t_i$ paths, whose distance in the final $H$ is required to approximate the distance in $G$ by an additive factor.
These paths are examined sequentially, where at step $\tau$, a particular candidate path $P_\tau$ is considered to be added to the current spanner $H_\tau$, resulting in $H_{\tau+1}$. The decision is made by assigning each candidate path $P_\tau$ a \emph{cost} $\Cost(P_\tau)$, corresponding to the number of path edges not already contained in the spanner $H_\tau$, and a \emph{value} $\Value(P_\tau)$, measuring how much adding the path would help to satisfy the considered set of constraints on the pairwise distances. The candidate path $P_\tau$ is added to $H_\tau$ if its {\em value to cost ratio} is sufficiently large. Informally, if a path $P_\tau$ is added, then it implies that each of at least some fraction of its new edges $P_\tau \setminus H_\tau$ \emph{contributes} to improving the inter-cluster distances in the current spanner $H_\tau$.
In the context of Stage S3.4 of our algorithm for \FTSPANNERBFS\ structures, we are given a collection of replacement paths $P^*_{i,j}$ where $e_j \in \pi^{mid}(v_i)$ and some preliminary sparse subgraph $\widehat{E}$ consisting of (at least) the edges of the BFS tree $T_0$, the edges of clustering graph $G_C$ and the set $\bigcup_i E_{i}^{far} \cup \bigcup_{i} E_i^{near}$, containing the last edges of replacement paths $P^*_{i',j'}$ protecting against the failure of $e_{j'} \in \pi^{near}(v_{i'}) \cup \pi^{far}(v_{i'})$.
By the preliminary explanation above (see Obs. \ref{obs:only_new_h}) it is sufficient to consider only replacement paths $P^*_{i,j}$ whose last edge is missing in $\widehat{E}$. These paths have a special structure. In particular, there is a unique divergence point $b_{i,j}$ where $P^*_{i,j}$ diverges from $\pi(s, v_i)$ and does not meet it again (i.e., $P^*_{i,j}[b_{i,j}, v_i]$ and $\pi(b_{i,j}, v_i)$ are edge disjoint). Since the common prefix $P^*_{i,j}[s, b_{i,j}]=\pi(s, b_{i,j})$ is contained in $T_0$, the buying procedure restricts attention only to the ``detour" segment
$P^-_{i,j}=P^*_{i,j}[b_{i,j}, v_i]$.
The properties of the partial spanner $\CurrSpanner_3$ constructed so far guarantee that this detour $P^-_{i,j}$ is restricted to $G$, i.e., $P^-_{i,j} \subseteq G$, hence the size of the resulting construct would be bounded as a function of $n=|V(G)|$ as desired.

To gain some intuition regarding our modified path-buying technique, we review some of its principles and draw some differences between our setting and that of \cite{BSADD10} and \cite{CGK13}.
The analysis of the path-buying technique has two main ingredients. The first is the correctness ingredient (A1), where it is required to show that if an $s_i-t_i$ path $P^*$ was \emph{not} added to the current spanner $H_t$ at time $t$, then there exists an alternative $s_i-t_i$ path $P'$ in $H_t$ satisfying that $|P'| \leq |P^*|+\beta$ for some constant integer $\beta>0$. The second is the size ingredient (A2), where it is required to show that the number of edges added due to the paths that were bought by the procedure is bounded.
\par In the 6-additive construction of \cite{BSADD10}, the correctness ingredient (A1) was based upon the fact that if a path $P^*$ was not added, then there exists a vertex $v_q \in P^*$ such that the pairwise $C(s_i)-C(v_q)$ and $C(t_i)-C(v_q)$ distances between clusters in $H_t$ is smaller than that in $P^*$, namely, than $\dist(s_i, v_q, P^*)$ and $\dist(t_i, v_q, P^*)$ respectively.
Similarly, in the subsetwise construction of \cite{CGK13}, (A1) was established by noting that if a path $P^*$ was not added, then there exists a vertex $v_q \in P^*$ such that the vertex to cluster $s_i-C(v_q)$ distance as well as the $t_i-C(v_q)$ distance in $H_t$ are smaller than $\dist(s_i, v_q, P^*)$ and $\dist(t_i, v_q, P^*)$ respectively.
In our setting, this argument becomes more delicate due to the possibility of failures which might render the existing bypasses already available in $H_t$ useless. Hence, when considering a detour $P^-_{i,j}$ of a  replacement path $P^*_{i,j}$ that was not added to the current spanner $H_t$, it is required to show that the inter-cluster bypasses in $H_t$ do not contain the failed edge $e_j$ and hence can safely be used in the surviving structure $H_t \setminus \{e_j\}$.
\par We now consider the second ingredient of the analysis (A2).
Let $\mathcal{B}$ be the set of paths added to the spanner by the path-buying procedure. In both \cite{BSADD10} and \cite{CGK13}, (A2) is established based on the fact that if $P^* \in \mathcal{B}$, then its value satisfies $\Value(P^*)\geq c \cdot \Cost(P^*)$ for some constant $c\geq 1$. This implies that each of at least a constant fraction of the newly added edges on $P^*$ contributes by decreasing some specific inter-cluster distances in the given spanner. Specifically, the value of $P^*$ is the number of pairs $(x, C)$ where $C$ is a cluster and there exists a vertex $v \in C \cap P^*$ such that by adding $P^*$ to the current spanner $H_t$, $\dist(x, C, P^*)$ is improved compared to that in $H_t$.
In the setting of \cite{BSADD10}, $x$ is a cluster (i.e., $x \in \{C(s_i), C(t_i)\}$) and in the setting of  \cite{CGK13}, $x$ is a vertex (i.e., $x \in \{s_i, t_i\}$).
For every $P^* \in \mathcal{B}$, its value (total number of $(x, C)$ pairs) is proportional to its cost $\Cost(P^*)$. Therefore, to bound the number of edges, it is sufficient to bound the number of $(x, C)$ pairs. This involves two steps: (A2.1) showing that the contribution due to a fixed pair $(x, C)$ is bounded (or in other words, that a given pair $(x, C)$ can contribute only a bounded number of times to the value of the paths in $\mathcal{B}$) and (A2.2) showing that the number of distinct $(x, C)$ pairs is bounded. The combination of (A2.1) and (A2.2) bounds the size of the spanner. We now consider (A2.1) and (A2.2) separately.  In both \cite{BSADD10} and \cite{CGK13}, (A2.1) is established by the cluster diameter property of Fact \ref{fc:clustering}, which implies in this context, that every given pair $(x, C)$ can contribute at most a constant number of times to the values of the paths in $\mathcal{B}$. Turning to (A2.2), in \cite{BSADD10}, the total number of distinct pairs corresponds to $|\mathcal{C}| \times |\mathcal{C}|$ since $x \in \mathcal{C}$. In comparison, in \cite{CGK13}, since $x \in S$, there are a total of $O(|\mathcal{C}| \times |S|)$ pairs. Overall, the spanner size is bounded since the number of clusters $\mathcal{C}$, as well as the
cardinality of $S$ in the subsetwise variant of \cite{CGK13}, are bounded.
\par We now contrast this with the situation in our setting of \FTSPANNERBFS\ structures.
Part (A2.1) is no longer straightforward, since every replacement path $P^*_{i,j}$ added to the current spanner exists in a \emph{different} graph $G \setminus \{e_j\}$, and therefore, in contrast to \cite{BSADD10} and \cite{CGK13}, the bounded diameter of the clusters is not sufficient, by itself, to establish (A2.1). Considering (A2.2), the approach of \cite{BSADD10} can be adopted to bound to number of distinct pairs $(x, C)$, by letting  $x \in \{C(b_{i,j}), C(v_i)\}$, but this would result in an additive stretch of 6. To improve the additive stretch to $4$, it is necessary to employ some \emph{intermediate} compromise. We first impose a \emph{direction} on the candidate paths, hence breaking the symmetry between the path endpoints. The direction is imposed by the source $s$. In particular, the two endpoints $b_{i,j}$ and $v_i$ of each detour $P^-_{i,j}$ are treated in an asymmetric manner.
The endpoint $v_i$ is treated as a \emph{cluster} in a similar manner to that of \cite{BSADD10}, that is, the value of $P^-_{i,j}$ counts the improvement of the inter-cluster distances between $C(v_i)$ and some $C(v_k)$ for $v_k \in P^-_{i,j}$.
In contrast, the $b_{i,j}$ endpoint is treated as a \emph{vertex}, as in \cite{CGK13}. Overall, the pairs $(x, C)$ that contribute to the value of $P^-_{i,j}$ are of two types, where $x \in \{b_{i,j}, C(v_i)\}$. In the analysis section, it is shown that the number of contributions of each fixed pair $(x, C)$ is bounded, and moreover, that for every cluster $C \in \mathcal{C}$, the number of distinct pairs in which it appears (of both types) is bounded by $O(n^{4/3})$. The main challenge in this context is to bound the number of distinct contributions of the type $(b_{i,j}, C)$.
(Note that the second type of contribution, where $x \in \mathcal{C}$ is a cluster, is easily bounded, since there are only $O(n^{2/3})$ clusters).
The replacement paths constructed earlier
are designed so that for every cluster $C$ there are at most $O(n^{2/3})$ distinct divergence point $b_{i,j}$ that can be paired with $C$ and contribute to the value of some detour $P^-_{i,j} \in \mathcal{B}$.
This establishes the sparsity of our \FTSPANNERBFS\ structure.

\begin{figure}[t!] 
\begin{center}
\includegraphics[scale=0.45]{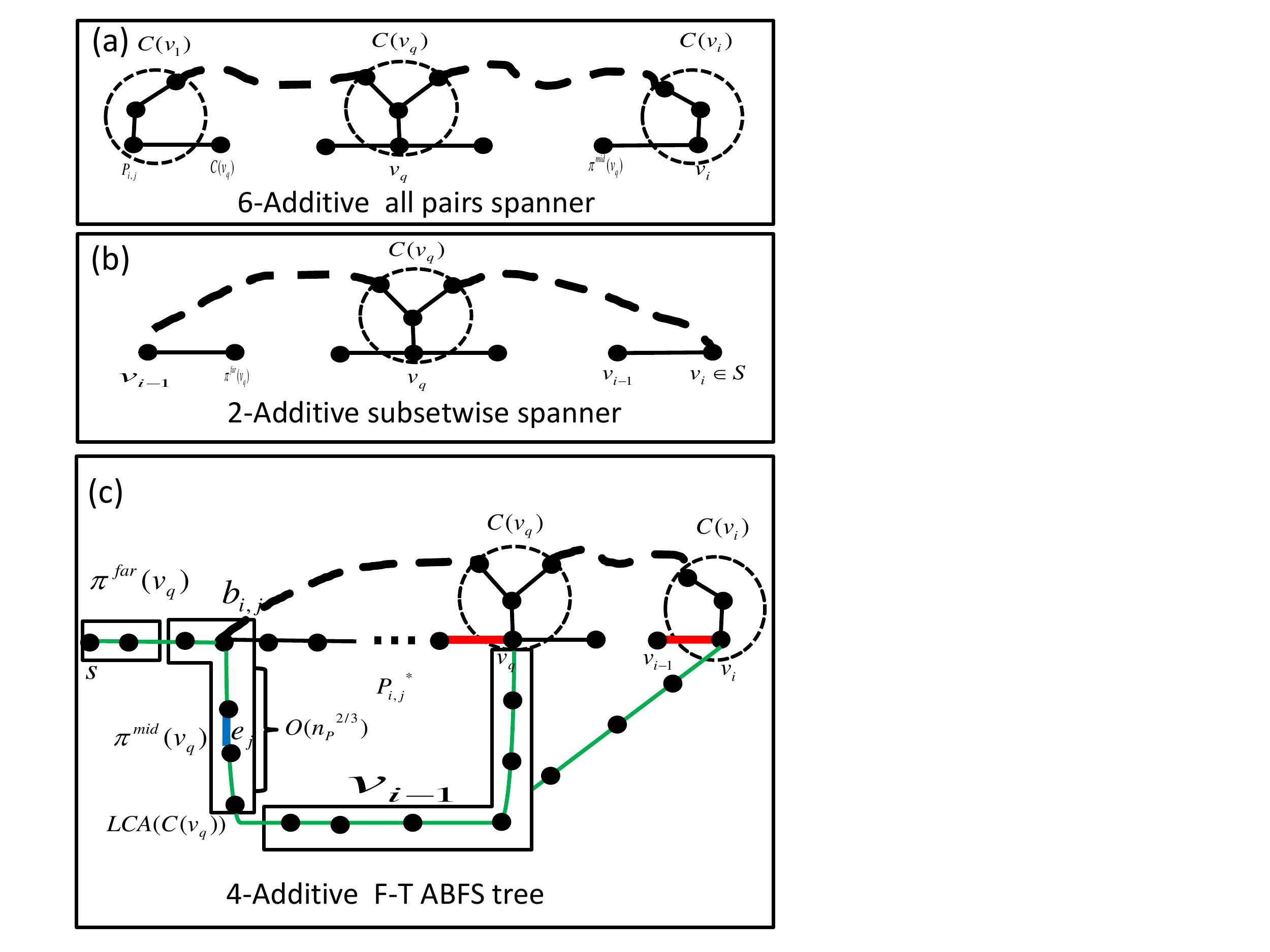}
\caption{
The green line represents the BFS edges. The edge $e_j$ is the failed edge.
Missing edges on the missing-ending path $P^*_{i,j}$ are drawn in red.
Only the detour  segment $P^-_{i,j}=P^*_{i,j}[b_{i,j}, v_i]$ of the path
is a candidate for buying. The edges of $v_i$ appearing on $P^*_{i,j'}$
for $e_{j'} \in \pi^{far}(v_i) \cup \pi^{near}(v_i)$ are already included
in the preliminary subgraph constructed prior to the path-buying procedure.
The shortest inter-cluster paths in $H$ are indicated by dashed curves,
where $H$ is the current subgraph at the time when $P^-_{i,j}$ is considered to be bought. For each cluster $C_q \in \mathcal{C}$, with a vertex $v_q \in C(v_q)$ ending with a missing edge, the branching point (first vertex on the detour) belongs to $\pi^{mid}(v_q)$. Since the size of $\pi^{mid}(v_q)$ is bounded by $O(n^{2/3})$, the contribution of pairs $(b_{i,j}, C(v_q))$ is bounded in a similar manner to the subsetwise case of \cite{CGK13}. In addition, since the endpoint $v_i$ of the detour is treated as a cluster, $C(v_i)$, the contribution of pairs $(C(v_q), C(v_i))$ is bounded by the number of clusters, $O(n^{2/3})$.
\label{fig:pb}}
\end{center}
\end{figure}

Let us note that in our application of the path-buying procedure,
paths with a sufficiently large value to cost ratio are added \emph{entirely}
to the current spanner, and adding just the last edge of each of these paths
will not suffice. This does not contradict Obs. \ref{obs:only_new_h},
for the following reason. Consider step $\tau$ and the current candidate path
$P_{\tau}=P^-_{i,j}$ with a sufficiently large value to cost ratio
with respect to $H_\tau$. If one adds solely the last edge of $P_\tau$
to $H_\tau$, then although the pair $(i,j)$ would be satisfied in
$H_{\tau+1}=H_{\tau} \cup \{\LastE(P_{\tau})\}$,
this would distort the value and cost functions of the subsequent
candidate paths in a way that would force us to add many new last edges
to the spanner. Specifically, since only the last edge of $P_{\tau}$ was added
to $H_{\tau+1}$, the subgraph $H_{\tau+1}$ enjoys almost none of the value
of $P_{\tau}$ in improving the pairwise distances, and as a result,
many subsequent paths would have a larger value (and hence might be suitable
for purchasing) in comparison to their value when the path $P_{\tau}$ is added
in its entirety to $H_{\tau+1}$.
%
%
Fig. \ref{fig:pb} illustrates the high level distinctions between the three
constructions of additive spanners based on the path-buying technique:
(a) The 6-additive construction of \cite{BSADD10},
(b) The 2-additive subsetwise construction of \cite{CGK13},
where the stretch constraint is imposed only on a set of vertex pairs
$S \times S$ for a given $S \subseteq V$, and
(c) The 4-additive \FTSPANNERBFS\ structure presented here.

\par Recall that $\pi(C_k)=\pi(s, \LCA(C_k))$ is the maximal shortest path segment shared by all the members of the cluster. For every $e_j \in T_0$, define
$V_j(\CurrSpanner_3)=\{ v_i \mid \LastE(P^*_{i,j}) \notin \CurrSpanner_3\}.$
By Lemma \ref{lem:rep_new}(c), it then holds that
$b_{i,j} \in \pi^{mid}(C(v_i))$ for every $v_i \in V_j(\CurrSpanner_3)$.
For every missing ending path $P^*_{i,j}$ (i.e., $\LastE(P^*_{i,j}) \notin \CurrSpanner_3$) let $P^{-}_{i,j}=P^*_{i,j}[b_{i,j}, v_i]$ be the detour segment starting from the unique divergence point $b_{i,j}$.
Let \begin{equation}
\label{eq:paths_to_buy}
\mathcal{P}=\{P^-_{i,j}~\mid~ e_j \in T_0, \LastE(P^*_{i,j}) \notin \CurrSpanner_3\}~
\end{equation}
be the candidate paths to be bought and added to $\CurrSpanner_3$.
The benefit of considering these detours is that by the definition of $\CurrSpanner_3$ , we are guaranteed that $P^-_{i,j} \subseteq G$ and in addition, each of these detours $P^-_{i,j}$ is edge disjoint with $\pi(s, \LCA(C_i))$ for every cluster $C_i$ that has a vertex $v' \in C_i \cap P^-_{i,j}$ with missing edge $\LastE(P^-_{i,j}[b_{i,j}, v']) \notin \CurrSpanner_3$. The property is heavily exploited in both the size and correctness analysis of the path-buying procedure.
%

\paragraph{The Scheme.}
The path-buying scheme is as follows.
Starting with $H_0=\CurrSpanner_3 \setminus T_0$, the paths of $\mathcal{P}=\{P^-_{i,j}\}$ are considered in an arbitrary order. At step $t$, we are given $H_t \subseteq G$ and consider the $b_{i,j}-v_{i}$ detour $P^-_{i,j} \in G \setminus \{e_j\}$. To decide whether $P^-_{i,j}$ should be added to $H_t$, the cost and value of $P^-_{i,j}$ are computed as follows.
Let $\Cost(P^-_{i,j})=|P^-_{i,j} \setminus E(H_t)|$ be the number of edges of $P^-_{i,j}$ that are missing in the current subgraph $H_t$.
$\Cost(P^-_{i,j})$ thus represents the increase in the size of the current \FTSPANNERBFS\ structure $H_t$ if the procedure adds $P^-_{i,j}$.
Let $\kappa=\Cost(P^-_{i,j})$ and let $\widetilde{e}_1, \ldots, \widetilde{e}_\kappa$ be the edges in $P^-_{i,j} \setminus E(H_t)$ directed away from $b_{i,j}$ where $\widetilde{e}_\ell=(y_{\ell},z_{\ell})$.

Define
$Z_{i,j}=\{z_{3\ell+1} ~\mid~ \ell \in \{0, \ldots, \lfloor (\kappa-1)/3 \rfloor\} \subseteq  P^-_{i,j}$ be the endpoints of the missing edges on $P^-_{i,j}$.
Hence, it holds that
\begin{eqnarray}
\label{eq:z_up_add}
\LastE(P^-_{i,j}[b_{i,j}, z_{\ell}]) \notin H_t \mbox{~for every~} \nonumber z_{\ell} \in Z_{i,j}
\mbox{~and~}
\dist(z_{\ell}, z_{\ell'}, P^-_{i,j})&\geq& 3~
\end{eqnarray}
for every $z_\ell, z_{\ell'} \in Z_{i,j}$.
Let
\begin{align}
\label{eq:contx}
& Cont_b(P^-_{i,j})=\{(b_{i,j}, C_\ell) \mid \exists z_{\ell} \in C_\ell \cap Z_{i,j} \mbox{~s.t~}
\\ &
\dist(b_{i,j},z_{\ell}, P^-_{i,j})<\dist(b_{i,j},C_{\ell}, H_{t} \setminus \pi(C_{\ell}))\}\nonumber
\end{align}
be the set of pairs $(b_{i,j}, C_\ell)$ such that the distance between $b_{i,j}$ and $C_\ell$ is improved by adding $P^-_{i,j}$ to $H_t \setminus \pi(C_{\ell})$.
Similarly, let $C_{\ell'}=C(v_i)$ be the cluster of $v_{i}$ and define
\begin{align}
\label{eq:conty}
& Cont_v(P^-_{i,j})=\{(v_{i}, C_\ell) ~\mid~ \exists z_{\ell} \in C_\ell \cap Z_{i,j} \mbox{~s.t~}
\\ &
\dist(v_{i},z_{\ell}, P^-_{i,j}) <
\dist(C_{\ell'},C_{\ell}, H_{t} \setminus \pi(C_{\ell'}))\} \nonumber
\end{align}
as the set of pairs $(v_{i}, C_\ell)$ such that the distance between $C_{\ell}$ and $C_{\ell'}$ is improved by adding $P^-_{i,j}$ to $H_t$.
Define
\begin{equation}
\label{eq:value}
\Value(P^-_{i,j})=|Cont_b(P^-_{i,j})|+|Cont_v(P^-_{i,j})|~.
\end{equation}
The path-buying strategy is as follows. If
\begin{equation}
\label{eq:cost_value}
\Cost(P^-_{i,j}) \leq 4 \cdot \Value(P^-_{i,j})
\end{equation}
then we ``buy'' the path $P^-_{i,j}$, namely, set $H_{t+1}=H_{t}\cup P^-_{i,j}$.
Otherwise, we do not buy $P^-_{i,j}$ and set $H_{t+1}=H_{t}$.
The final spanner is given by
$$H=\left(H_{t'} \cup \CurrSpanner_3 \right) \mbox{~where~} t'=|\mathcal{P}|+1.$$

\subsubsection{Analysis}
\begin{figure}[htbp]
\begin{center}
\includegraphics[scale=0.35]{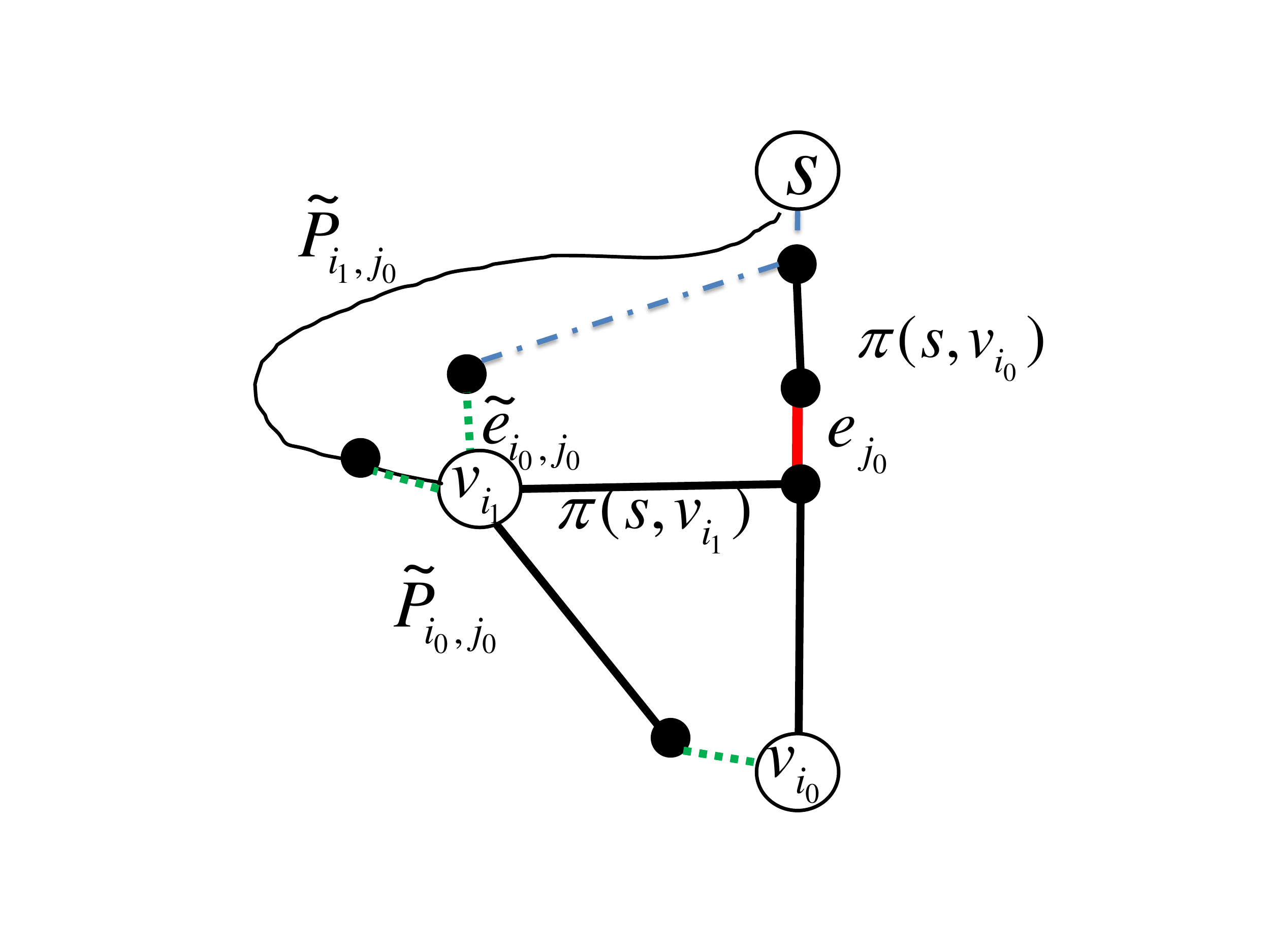}
\caption{ The dashed edges are missing in $H$. Since $(i_0,j_0)$ is a bad pair, it holds that $(i_1,j_0)$ is a bad pair as well and in addition, the last edge of $\widetilde{P}_{i_1,j_0}$ is missing in $H$.
\label{fig:addcorrect}}
\end{center}
\end{figure}

We begin with the following observation.
\begin{observation}
\label{obs:z_in_diffclus}
For every $P^-_{i,j} \in \mathcal{P}$ and for every $z_\ell, z_{\ell'} \in Z_{i,j}$, it holds that $C(z_{\ell}) \neq C(z_{\ell'})$.
\end{observation}
\Proof
Recall that by Fact \ref{fc:clustering}(2), the diameter of each cluster is
at most $2$. As the distance between different $z_{\ell}$'s in $Z_{i,j}$
is at least $3$, it follows that $z_{\ell}$ and $z_{\ell'}$ belong to
distinct clusters.
\QED
To establish the correctness of the construction, we show the following.
\begin{lemma}
\label{lem:correct_add_up_clust}
For every $e_j \in T_0$ and $v_i \in V(G)$ it holds that
$\dist(s, v_i, H \setminus \{e_j\})\leq \dist(s, v_i, G \setminus \{e_j\})+4$.
\end{lemma}
\Proof
The proof is by contradiction.
Let $BP=\{(i,j) \mid
v_i \in V(G), e_j \in \pi(s, v_i)$ and $\dist(s, v_i, H \setminus \{e_j\})> \dist(s, v_i, G \setminus \{e_j\})+4\}$
be the set of ``bad pairs," namely, vertex-edge pairs $(i,j)$ whose additive stretch in $H$ is greater than 4.
Assume, towards contradiction, that $BP\ne \emptyset$.
For each bad pair $(i,j) \in BP$, it holds that $e_j \in \pi(s, v_i)$ (since $T_0 \subseteq H$).
For every bad pair $(i,j)\in BP$ define
$\widetilde{P}_{i,j} \in SP(s, v_i, G \setminus \{e_j\})$ to be
the replacement path whose last missing edge in $H$, $\widetilde{e}_{i,j}$,
is the shallowest among all $s-v_i$ replacement paths in $G \setminus \{e_j\}$.
Let $d(i,j)=\dist(s,\widetilde{e}_{i,j},\widetilde{P}_{i,j})$.
Finally, let $(i_0,j_0) \in BP$ be the pair that minimizes $d(i,j)$, and let $\widetilde{e}_{i_0,j_0}=(v_{i'}, v_{i_1})$.
Note that $\widetilde{e}_{i_0,j_0}$ is the {\em shallowest} ``deepest missing edge'' over all bad pairs $(i,j) \in BP$.
See Fig. \ref{fig:addcorrect} for an illustration.
\begin{claim}
\label{cl:isbad}
$(i_1,j_0) \in BP$~.
\end{claim}
\Proof
Assume towards contradiction that  $(i_1,j_0) \notin BP$ and let
$P'' \in SP(s, v_{i_1}, H \setminus \{e_{j_0}\})$.
Hence, since  $(i_1,j_0) \notin BP$, it holds that
\begin{eqnarray}
\label{eq:up_cor_add}
|P''| &\leq& \dist(s, v_{i_1}, G \setminus \{e_{j_0}\})+4
\\ &=& |P^*_{i_0,j_0}[s, v_{i_1}]|+4. \nonumber
\end{eqnarray}
We now consider the following $s-v_{i_0}$ replacement path
$Q=P'' \circ \widetilde{P}_{i_0,j_0}[v_{i_1}, v_{i_0}]$.
By definition of $(i_1,j_0)$, $Q \subseteq G \setminus \{e_{j'}\}$. In addition,
\begin{eqnarray*}
|Q|&=&|P''|+|\widetilde{P}_{i_0,j_0}(v_{i_1}, v_{i_0})|
\leq
|\widetilde{P}_{i_0,j_0}[s, v_{i_1}]|+4+|\widetilde{P}_{i_0,j_0}(v_{i_1}, v_{i_0})|
\\&=&
|\widetilde{P}_{i_0,j_0}|+4 ~=~ \dist(s, v_{i_0}, G \setminus \{e_{j_0}\})+4~,
\end{eqnarray*}
where the inequality follows by Eq. (\ref{eq:up_cor_add}).
This contradicts the fact that $(i_1, j_0) \in BP$.
\QED
In particular, we have that $e_{j_0} \in \pi(s, v_{i_1})$, hence $\widetilde{P}_{i_1,j_0}$ is defined.
\begin{claim}
$\LastE(\widetilde{P}_{i_1,j_0}) \notin H$.
\end{claim}
\Proof
Assume towards contradiction that $\LastE(\widetilde{P}_{i_1,j_0}) \in H$.
Then since $(i_1, j_0) \in BP$ it holds that there exists at least one missing edge in $\widetilde{P}_{i_1,j_0}$ which is strictly above $\LastE(\widetilde{P}_{i_1,j_0})$. Since $|\widetilde{P}_{i_1,j_0}|=|\widetilde{P}_{i_0,j_0}[s, v_{i_1}]|$, we get that $d(i_1,j_0)<d(i_0,j_0)$, contradiction to the definition of $(i_0,j_0)$.
\QED
Since the last edge of $\widetilde{P}_{i_1,j_0}$ is missing, it holds that (a) $v_{i_1}$ is clustered (by Fact \ref{fc:clustering}) and (b) that the last edge of every $s-v_{i_1}$ replacement path is missing in $H$ as well. In particular, for the replacement path $P^*_{i_1,j_0}$ constructed by Algorithm \mbox{\tt Pcons}, $\LastE(P^*_{i_1,j_0})$ is missing in $H$.
Hence, it holds that $e_{j_0} \in \pi^{mid}(v_{i_1})$.
It then holds by Eq. (\ref{eq:paths_to_buy}), that
$P^{-}_{i_1,j_0}=P^*_{i_1, j_0}[b_{i_1,j_0}, v_{i_1}] \in \mathcal{P}$, i.e., $P^{-}_{i_1,j_0}$ is in the collection of detours considered to be purchased in the path-buying Procedure.
Let $P'=P^{-}_{i_1,j_0}$ and $t$ be the iteration where $P'$ was considered to be added to $H_{t}$.
We now consider two cases. If $P'$ was bought, then $P^*_{i_1,j_0}=\pi(s, b_{i_1,j_0}) \circ P' \subseteq  H$ hence we get a contradiction to the fact that $(i_1,j_0) \in BP$. Hence, it remains to consider the case where $P'$ was not bought.

Let $\kappa=\Cost(P')$ and $Z_{i_1,j_0}=\{z_{1}, \ldots, z_{\kappa'}\}$ for $\kappa'=\lfloor \kappa/3 \rfloor$ be the corresponding vertices in $P'$ with a missing edges that satisfy Eq. (\ref{eq:z_up_add}). By Obs. \ref{obs:z_in_diffclus}, each $z_{\ell}\in Z_{i_1,j_0}$ belongs to a distinct cluster $C_{\ell}$. Hence there are at least $\kappa'$
distinct clusters on $P'$. Let $C_{r}$ be the cluster of $v_{i_1}$ (since $v_{i_1}$ is incident to a missing edge in $G_C \subseteq H$, by Fact \ref{fc:clustering}(1), $v_{i_1}$ is indeed clustered).

A cluster $C_{\ell}$ with $z_{\ell} \in Z_{i_1,j_0}$ is a \emph{contributor}
if adding $P'$ to $H_t$ improves either the $b_{i_1,j_0}-C_{\ell}$ distance or the $v_{i_1}-C_{\ell}$ distance in $H_t$.
I.e., if it satisfies either
$\dist(b_{i_1,j_0}, z_{\ell},P')< \dist(b_{i_1,j_0}, C_\ell,H_{t} \setminus \pi(C_\ell))$ (hence $(b_{i_1,j_0}, C_{\ell}) \in Cont_b(P')$), or $\dist(v_{i_1},z_\ell ,P')<\dist(C_{r}, C_\ell, H_{t} \setminus \pi(C_{r}))$ (hence $(b_{i_1,j_0}, C_{\ell}) \in Cont_v(P')$).
Otherwise, $C_{\ell}$ is \emph{neutral}. That is, $C_{\ell}$ is neutral if $(b_{i_1,j_0},C_{\ell}) \notin Cont_b(P')$ and in addition, $(v_{i_1},C_{\ell}) \notin Cont_v(P')$.
There are two cases to consider.
If all clusters are contributors (i.e., there is no neutral cluster) then all the $\kappa'$ clusters contribute to $\Value(P')$ (either with $b_{i_1,j_0}$ or with $v_{i_1}$ or both).
It then holds that $\Value(P')\geq \kappa' \geq \Cost(P')/4$.  Hence, by Eq. (\ref{eq:cost_value}), we get a contradiction to the fact that $P'$ was not added to $H_t$. In the other case, there exists at least one neutral cluster $C_{\ell}$ having a unique vertex $z_{\ell}$ in $P'$ such that
\begin{eqnarray}
\label{eq:clust_correct}
\dist(b_{i_1,j_0}, C_{\ell} , H_{t} \setminus \pi(C_\ell)) &\leq&
\dist(b_{i_1,j_0}, z_{\ell} ,P') \mbox{~~and~~}
\\
\dist(C_{r}, C_{\ell}, H_{t} \setminus \pi(C_{r})) &\leq&
\dist(v_{i_1}, z_{\ell},P')~.
\end{eqnarray}
Let $u \in C_{\ell}$ be the closest vertex in the cluster $C_{\ell}$ to the divergence point $b_{i_1,j_0}$ in the graph $H_{t} \setminus \pi(C_{\ell})$ and define
$Q_1 \in SP(b_{i_1,j_0}, u, H_{t} \setminus \pi(C_{\ell}))$ such that $u \in C_{\ell}$, hence $|Q_1|=\dist(b_{i_1,j_0}, C_{\ell}, H_{t} \setminus \pi(C_{\ell}))$.
Let $u' \in C_{\ell}$, $y' \in C_{r}$ be the closest pair of vertices in the cluster $C_{\ell}$ and $C_{r}$ repetitively in the graph $H_{t} \setminus \pi(C_{r})$. Define $Q_2 \in SP(u', y', H_{t} \setminus \pi(C_{r}))$, then $|Q_2|=\dist(C_{r}, C_{\ell}, H_{t} \setminus \pi(C_{r}))$.

Let $\widehat{Q}_1\in SP(u',u, G_C)$ and $\widehat{Q}_2\in SP(y',v_{i_1}, G_C)$. Since $y', v_{i_1} \in C_{r}$ and $u,u' \in C_{\ell}$, it holds that $|\widehat{Q}_1|,|\widehat{Q}_2| \leq 2$. Consider the following $b_{i_1,j_0}-v_{i_1}$ replacement path $P_4=Q_1 \circ \widehat{Q}_1 \circ Q_2 \circ \widehat{Q}_2$.
We first claim that $P_4 \subseteq H \setminus \{e_{j_0}\}$. First note that since $G_C \subseteq \left(G \setminus T_0 \right)$, it holds that the intra cluster paths $\widehat{Q}_1$ and $\widehat{Q}_2$ are free of the failed edge $e_{j_0}$.
In addition, since $\LastE(P^*_{i_1,j_0}), \LastE(P^*_{i_1,j_0}[s, z_{\ell}])\notin \CurrSpanner_3$, by Lemma \ref{lem:rep_new}(c), it holds that
$e_{j_0} \in \pi(C_{\ell}) \cap \pi(C_{r})$. In addition, by the definition of $P_4$, it holds that $P_4 \subseteq H_{t} \setminus \left(\pi(C_{\ell}) \cap \pi(C_{r}) \right)$.
\par Finally, we bound the length of $P_4$.
\begin{eqnarray*}
|P_4|&=& |Q_1| +|Q_2|+4
\leq
\dist(b_{i_1,j_0}, z_{\ell}, P')+\dist(v_{i_1}, z_{\ell}, P')+4
=
|P'|+4~,
\end{eqnarray*}
where the first inequality follows by Eq. (\ref{eq:clust_correct}).
We therefore have that the path $P_5=\pi(s, b_{i_1,j_0}) \circ P_4$ exists in $H\setminus \{e_{j_0}\}$ and in addition,
$$|P_5|\leq |P^*_{i_1,j_0}[s, b_{i_1,j_0}]|+|P^{-}_{i_1,j_0}|+4=|P^*_{i_1,j_0}|+4.$$
This contradicts the assumption that $(i_1,j_0) \in BP$. The Lemma follows.
\QED

Finally, we bound the size of $H$.
\begin{lemma}
\label{lem:size_add_up_clust}
$|E(H)| \leq n^{4/3}$~.
\end{lemma}
\Proof
Let $\mathcal{B} \subseteq \mathcal{P}$ be the set of paths that were bought in the path-buying procedure. Since $|E(\CurrSpanner_3) \setminus T_0|=O(n^{4/3})$, it remains to bound the number of edges added due to the paths in $\mathcal{B}$.
Note that for $\gamma=1/3$, it holds by Fact \ref{fc:clustering}, that there are $O(n^{2/3})$ clusters. In addition, recall that for every cluster $C_i$, it holds that $|\pi^{mid}(C_i)|=O(n^{2/3})$.
We have the following.
\begin{eqnarray}
\label{eq:nmiss_val}
|H \setminus \CurrSpanner_3|&=&\sum_{P^{-}_{i,j} \in \mathcal{B}}\Cost(P^{-}_{i,j}) \nonumber
\leq
\sum_{P^{-}_{i,j} \in \mathcal{B}} 4 \cdot \Value(P^{-}_{i,j})
\\&=&
4 \sum_{P^{-}_{i,j} \in \mathcal{B}} \left(|Cont_b(P^{-}_{i,j})|+|Cont_v(P^{-}_{i,j})| \right)~,
\end{eqnarray}
where the inequality follows by Eq. (\ref{eq:cost_value}), and the last equality follows by Eq. (\ref{eq:value}).

Our counting strategy is as follows. We fix a cluster $C_{\ell}$, and bound the
number of times it appears in either $Cont_b(P')$ or $Cont_v(P')$ over all $P' \in \mathcal{B}$.
The following definition is useful in our analysis.
Define the multisets
\begin{eqnarray*}
Cont_b(C_{\ell}) &=& \{(b_{i,j}, C_{\ell}) ~\mid~
P^{-}_{i,j} \in \mathcal{B} \mbox{~and~}
(b_{i,j}, C_{\ell}) \in Cont_b(P^{-}_{i,j})\},
\\
Cont_v(C_{\ell}) &=& \{(v_{i}, C_{\ell}) ~\mid~
\exists P^{-}_{i,j} \in \mathcal{B} \mbox{~and~}
(v_i, C_{\ell}) \in Cont_v(P^{-}_{i,j})\}.
\end{eqnarray*}
Note that certain pairs might appear several times in both $Cont_b(C_{\ell})$ and $Cont_v(C_{\ell})$. Hence, we do not bound the number of unique pairs but also take into account the possible reappearance of the same pair in these expressions.
Eq. (\ref{eq:nmiss_val}) can now be written equivalently as
\begin{eqnarray}
\label{eq:up_add_cluster_cont}
|H \setminus \CurrSpanner_3|&\leq& \nonumber
4 \sum_{P^{-}_{i,j} \in \mathcal{B}} \left(|Cont_b(P^{-}_{i,j})|+|Cont_v(P^{-}_{i,j})| \right)=
4 \sum_{C_\ell}|Cont_b(C_{\ell})|+|Cont_v(C_{\ell})|~.
\end{eqnarray}
We now show the following.
\begin{lemma}
\label{contb_clust}
$|Cont_b(C_{\ell})|=O(n^{2/3})$ for every $C_{\ell}$.
\end{lemma}
\Proof
We begin by showing that for $P^-_{i,j} \in \mathcal{B}$, if $(b_{i,j}, C_{\ell}) \in Cont_b(P^-_{i,j})$ then $b_{i,j} \in \pi^{mid}(C_{\ell})$.
To see that, let $t$ be the iteration in which $P^-_{i,j}$ was added to $H_{t}$. It then holds that there exists $z_{\ell} \in C_{\ell} \cap Z_{i,j}$ that satisfies Eq. (\ref{eq:contx}). By the definition of $Z_{i,j}$, we have that $\LastE(P^*_{i,j}[s, z_{\ell}]) \notin \CurrSpanner_3$. Hence, by Lemma \ref{lem:rep_new}(c) it holds that $b_{i,j} \in \pi^{mid}(C_{\ell})$.
\par We proceed by showing that for every fixed $b_{i,j} \in \pi^{mid}(C_{\ell})$, the pair $(b_{i,j}, C_{\ell})$ can appear at most 3 times in the multiset $Cont_{b}(C_{\ell})$.
Formally, define the set of paths in which $b_{i,j}$ contributes with the cluster $C_{\ell}$ as
\begin{eqnarray*}
A(i,j)&=&\{P^{-}_{i',j'} \in \mathcal{B} ~\mid~ b_{i',j'} = b_{i,j} \mbox{~~and~~}
(b_{i',j'}, C_{\ell}) \in Cont_b(P^{-}_{i',j'})\}~.
\end{eqnarray*}

\begin{claim}
\label{cl:bound_aij}
$|A(i,j)| \leq 3$ for every $i,j$ such that $b_{i,j} \in \pi^{mid}(C_{\ell})$.
\end{claim}
\Proof
Let $A(i,j)=\{Q_1=P^{-}_{i_1,j_1}, \ldots, Q_N=P^{-}_{i_N,j_N}\}$ be sorted according to the time $t_k$ they were bought and added to the graph $H_{t_k}$. Hence, $b_{i_1,j_1}=\ldots=b_{i_N,j_N}=b_{i,j}$.
We claim that $N \leq 3$.
For each $k \in\{1, \ldots, N\}$, considered at time $t_k$ where $t_{k}<t_{k+1}$, there exists a vertex $z_{k} \in Z_{i_k,j_k} \cap C_{\ell}$ such that $\LastE(Q_k[b_{i_k,j_k}, z_k]) \notin H_{t_k}$ and $\dist(b_{i,j}, z_k, Q_k) = \dist(b_{i_k,j_k}, z_k, Q_k)< \dist(b_{i,j}, C_\ell, H_{t_k} \setminus \pi(C_{\ell}))$. This holds as by definition, $b_{i,j}=b_{i_k,j_k}$ for every $k \in \{1, \ldots, N\}$ and by Eq. (\ref{eq:contx}).
We now show that, denoting
$$Y_k=\dist(b_{i,j}, z_k, H_{t_{k+1}} \setminus \pi(C_\ell)),$$
we have $Y_k<Y_{k-1}$
for every $k \in \{2, \ldots ,N\}$.

At each time $t_k$, since a contribution is made,
and by the fact that $z_{k}, z_{k-1} \in C_{\ell}$, we have that
\begin{eqnarray}
\label{eq:pb_zineq}
Y_k
&\leq& \dist(b_{i,j}, z_{k}, Q_k \setminus \pi(C_\ell)) =
\dist(b_{i,j}, z_k, Q_k)
\\&<& \nonumber
\dist(b_{i,j}, C_{\ell}, H_{t_k} \setminus \pi(C_\ell))
\leq Y_{k-1}~,
\end{eqnarray}
where the first inequality holds as $Q_k$ was bought at time $t_k$ and hence $Q_k \subseteq H_{t_{k+1}}$.
By Obs. \ref{obs:z_in_diffclus}, $z_{k-1}, z_k \in C_{\ell}$ are the unique vertices in $C_{\ell}$ in the sets $Z_{i_{k-1}, j_{k-1}}$ and $Z_{i_{k}, j_{k}}$ respectively.  The equality of (\ref{eq:pb_zineq}) holds since $\LastE(P^*_{i_k,j_k}[s, z_{k}]) \notin \CurrSpanner_3$  and hence by \ref{lem:rep_new}(b), $Q_k$ is edge disjoint with $\pi(s, z_k)$.
Since by Obs. \ref{obs:path_pi_cluster}, $\pi(C_{\ell}) \subseteq \pi(s, z_k)$, it holds that $Q_k$ and  $\pi(C_{\ell})$ are edge disjoint as well.  The strict inequality in (\ref{eq:pb_zineq}) follows by Eq. (\ref{eq:contx}) and by the fact that the pair $(z_{k}, C_{\ell})$ contributes to the value of $Q_k$   i.e., $(b_{i,j}, C_{\ell}) \in Cont_b(Q_k)$. The last inequality follows by the fact that $z_{k-1} \in C_{\ell}$. Hence, letting
$$Y=\dist(b_{i,j}, z_N, H_{t_{N+1}} \setminus \pi(C_\ell)),$$
we have
\begin{equation}
\label{eq:ad_up_u1}
Y\leq \dist(b_{i,j}, z_{1}, H_{t_{2}} \setminus \pi(C_\ell))-(N-1).
\end{equation}
Conversely, we also have the following.
\begin{eqnarray}
Y
& \geq & \dist(b_{i,j}, z_N, H_{t_{N+1}} \setminus \{e_{j_1}\}) \label{eq:cor11}
\\& \geq &
\dist(b_{i,j}, z_1, H_{t_{N+1}} \setminus \{e_{j_1}\})-2 \label{eq:cor12}
\\& \geq &
\dist(b_{i,j}, z_1, G \setminus \{e_{j_1}\})-2 \label{eq:cor13}
\\& = &
\dist(b_{i,j}, z_1, Q_1)-2 \label{eq:cor14}
\\& = &
\dist(b_{i,j}, z_1, Q_1 \setminus \pi(C_\ell))-2 \label{eq:cor15}
\\& \geq &
\dist(b_{i,j}, z_1, H_{t_{2}} \setminus \pi(C_\ell))-2~, \label{eq:cor16}
\end{eqnarray}
where Eq. (\ref{eq:cor11}) follows by the fact that $z_{1} \in C_{\ell}$, $\LastE(P^*_{i_1,j_1}[s, z_1])\notin \CurrSpanner_3$ and hence by
Lemma \ref{lem:rep_new}(a), $e_{j_1} \in \pi(C_{\ell})$.
Eq. (\ref{eq:cor12}) follows by the fact that $z_1, z_{N} \in C_{\ell}$ and by Fact \ref{fc:clustering}(2).
Eq. (\ref{eq:cor13}) follows by the fact that $H_{t_{N+1}} \subseteq G$.
Eq. (\ref{eq:cor14}) follows by the fact that $P^*_{i_1,j_1} \in SP(s, v_{i_1}, G \setminus \{e_{j_1}\})$ and
$Q_1[b_{i,j}, z_1]=P^*_{i_1,j_1}[b_{i,j}, z_1]$.
To see Eq. (\ref{eq:cor15}), note that by Lemma \ref{lem:rep_new}(b), it holds that $Q_1[b_{i,j}, z_1]$ and $\pi(s, z_1)$ are edge disjoint. In addition, by Obs. \ref{obs:path_pi_cluster}, $\pi(C_\ell) \subseteq \pi(s, z_1)$, hence $Q_1[b_{i,j}, z_1]$ and $\pi(C_\ell) \subseteq \pi(s, z_1)$ are edge disjoint as well.
Finally, Eq. (\ref{eq:cor16}) follows as $Q_1 \subseteq H_{t_{2}}$ since $Q_1 \in \mathcal{B}$ was bought at time $t_1$.

We therefore have that $\dist(b_{i,j}, z_N, H_{t_{N+1}}\setminus \pi(C_\ell))\geq \dist(b_{i,j}, z_1, H_{t_{2}} \setminus \pi(C_\ell))-2$, combining with Eq. (\ref{eq:ad_up_u1}),
we have $|A(i,j)|=N\leq 3$.
\QED
Since $A(i,j)\neq \emptyset$ only if $b_{i,j} \in \pi^{mid}(C_{\ell})$, it follows from Cl. \ref{cl:bound_aij} that
$$|Cont_b(C_{\ell})| ~=~ \sum_{b_{i,j} \in \pi(s, v_i)} |A(i,j)|
~=~ \sum_{b_{i,j} \in \pi^{mid}(C_{\ell})} |A(i,j)|=O(n^{2/3}).$$
The lemma follows.
\QED
We now turn to consider the second type of contribution of the form $(v_{i}, C_{\ell})$.
\begin{lemma}
\label{contv_clust}
$|Cont_v(C_{\ell})|=O(n^{2/3})$ for every $C_{\ell}$.
\end{lemma}
\Proof
For every cluster $C_{r} \neq C_{\ell}$ define the multiset
$$D_{r}=\{(v_i, C_{\ell}) \in Cont_v(C_{\ell})  \mid v_i \in C_{r}\}.$$
Note that the pair $(v_i, C_{\ell})$ can contribute several times to $D_{r}$ with the same $v_i$. Hence, we do not count only unique pairs as the same pair might contribute several times. In fact, we count the number of all times in which a path $P' \in \mathcal{B}$ was bought, and the pair $(v_{i}, C_{\ell}) \in Cont_{v}(P')$ such that $v_i \in C_{r}$ contributes to the value of the path $P'$.
Since missing edges in $G_C$ are between vertices of different clusters, we have that
$Cont_v(C_{\ell})=\bigcup_{r \neq \ell}D_{r}$.
We now show that each $|D_{r}|\leq 5$ which concludes the proof since there are overall $|\mathcal{C}|=O(n^{2/3})$ clusters in $G_C$.

Let $Q_1=P^{-}_{i_1,j_1}, \ldots, Q_N=P^{-}_{i_N,j_N}$ be such that $(v_{i_k}, C_{\ell}) \in D_{r}$, where the paths are sorted according to the time $t_k$ they were considered, for every $k \in \{1, \ldots, N\}$ and $N=|D_{r}|$.

We do not assume that the $v_{i,k'}$'s are distinct.
Let $z_k \in Z_{i_k,j_k} \cap C_{\ell}$ be the unique vertex of $C_{\ell}$ in $Z_{i_k,j_k}$ that contributes by adding its path $Q_k$ to $v_{i_k}$ to $H_{t_k}$.
We then have that $\LastE(Q_k[b_{i_k, j_k}, z_k]) \notin H_{t_k}$
and hence by Lemma \ref{lem:rep_new}(a), $e_{j_k} \in \pi(C_{\ell})$.
In addition, since $\LastE(Q_k) \notin \CurrSpanner_3$, by Lemma \ref{lem:rep_new}(a) it also holds that $e_{j_k} \cap \pi(C_{r})$ as $v_{i_k} \in C_{r}$. Hence,
\begin{equation}
\label{eq:edge_in_intersection}
e_{j_k} \in \pi(C_{\ell}) \cap \pi(C_{r}) \mbox{~for every~} k \in \{1, \ldots, N\}~.
\end{equation}
We next show that, denoting
$$X_k=\dist(v_{i_k}, z_k, H_{t_{k+1}} \setminus \pi(C_{r})),$$
we have $X_k<X_{k-1}$
for every $k \in \{2, \ldots N\}$.
Note the each $v_{i_{k}}$ belongs to the same cluster $C_{r}$ and every $z_{k}$ for $k \in \{1, \ldots, N\}$ belongs to the same cluster $C_{\ell}$.
Each time a contribution is made at time $t_k$, it implies that
\begin{eqnarray}
X_k &\leq&
\dist(v_{i_k}, z_k, Q_k \setminus \pi(C_{r})) \label{eq:adub_1}
\\&=&
\dist(v_{i_k}, z_k, Q_k) \label{eq:adub_2}
\\&<&
\dist(C_{r}, C_{\ell}, H_{t_{k}} \setminus \pi(C_{r})) \label{eq:adub_3}
\\&\leq&
X_{k-1}~,
\label{eq:adub_4}
\end{eqnarray}
where Eq. (\ref{eq:adub_1}) follows by the fact that $Q_k \subseteq H_{t_{k+1}}$, Eq. (\ref{eq:adub_2}) follows by the fact that $\LastE(P^*_{i_k,j_k}) \notin \CurrSpanner_3$, and hence by Lemma \ref{lem:rep_new}, $Q_k=P^*_{i_k,j_k}[b_{i_k,j_k}, v_{i_k}]$ and $\pi(s, v_{i_k})$ are edge disjoint. In addition, by Obs. \ref{obs:path_pi_cluster}, $\pi(C_{r}) \subseteq \pi(s, v_{i_k})$, hence $\pi(C_{r})$ and $Q_k$ are edge disjoint as well.
Eq. (\ref{eq:adub_3}) follows by Eq. (\ref{eq:conty}) and the fact that $(v_{i_k}, C_{\ell}) \in Cont_v(Q_k)$ and finally, Eq. (\ref{eq:adub_4})
follows by the fact that $v_{i_{k-1}} \in C_{r}$ and $z_{k-1} \in C_{\ell}$.
Therefore, letting
$$X=\dist(v_{i_N}, z_N, H_{t_{N+1}} \setminus \pi(C_{r})),$$
we have that
\begin{equation}
\label{eq:up_srtect1}
X \leq  \dist(v_{i_{1}}, z_{1}, H_{t_{2}} \setminus \pi(C_{r}))-(N-1)~.
\end{equation}
Conversely, we have that
\begin{eqnarray}
X
&\geq&
\dist(v_{i_N}, z_N, G \setminus \pi(C_{r})) \label{eq:up_t01}
\\
&\geq&
\dist(v_{i_1}, z_1, G \setminus \pi(C_{r}))-4 \label{eq:up_t02}
\\
& =&
\dist(v_{i_1}, z_1, Q_1)-4 \label{eq:up_t03}
= \dist(v_{i_1}, z_1, Q_1 \setminus \pi(C_{r}))-4
\\& \geq &
\dist(v_{i_1}, z_1, H_{t_{2}}\setminus \pi(C_{r})\})-4~, \label{eq:up_t04}
\end{eqnarray}
where Eq. (\ref{eq:up_t01}) follows as $H_{t_{N+1}} \subseteq G$.
Eq. (\ref{eq:up_t02}) follows  by the  fact that $v_{i_1},v_{i_N} \in C_{r}$
and $z_1,z_N \in C_{\ell}$ and by Fact \ref{fc:clustering}(b).
Eq. (\ref{eq:up_t03}) follows  by the fact that
$P^*_{i_1,j_1}[z_1,v_{i_1}]=Q_1[z_1,v_{i_1}]$ and
$P^*_{i_1,j_1} \in SP(s, v_{i_1}, G \setminus \{e_{j_1}\})$.
In addition, by Lemma \ref{lem:rep_new}, since
$\LastE(P^*_{i_1, j_1}) \notin \CurrSpanner_3$, it holds that $Q_1$ and
$\pi(s, v_{i_1})$ are edge disjoint.
By Obs. \ref{obs:path_pi_cluster}, $\pi(C_{r}) \subseteq \pi(s, v_{i_1})$,
hence $Q_1$ and $\pi(C_{r})$ are edge disjoint as well.
Finally, Eq. (\ref{eq:up_t04}), follows as $Q_1 \subseteq H_{t_2}$ since $Q_1 \in \mathcal{B}$ was added to $H_{t_1}$ at time $t_1$.
\par Combining with Eq. (\ref{eq:up_srtect1}), we get that $N \leq 5$ and hence $|Cont_v(C_{\ell})|=O(n^{2/3})$. The claim follows.
\QED
Overall, by Lemma \ref{contb_clust} and Lemma \ref{contv_clust}, we have that $|Cont_b(C_{\ell})|+|Cont_v(C_{\ell})|=O(n^{2/3})$ for every $C_{\ell}$. Hence, by plugging into Eq. (\ref{eq:up_add_cluster_cont}),  as there are $O(n^{2/3})$ clusters, we have that $|H| \leq 4\sum_{C_{\ell}}\left( \left|Cont_b(C_{\ell})|+|Cont_v(C_{\ell})\right|\right)=O(n^{4/3})$ as required. The Lemma follows.
\QED




\end{document}